\RequirePackage{fix-cm} 
\documentclass[fleqn,usenatbib]{mnras}

\usepackage{amsmath}
\usepackage{amssymb}
\usepackage{amsfonts}
\usepackage{textcomp}
\usepackage{newtxtext}
\usepackage{bm}
\usepackage{lmodern}

\usepackage[T1]{fontenc}

\DeclareRobustCommand{\VAN}[3]{#2}
\let\VANthebibliography\thebibliography
\def\thebibliography{\DeclareRobustCommand{\VAN}[3]{##3}\VANthebibliography}

\usepackage{graphicx}	
\usepackage{longtable}
\usepackage{booktabs}
\usepackage{amsmath}	
\usepackage{pdflscape} 

\newcommand{\Zsun}{\, \mathrm{Z}_{\odot}}
\newcommand{\logOH}{\ensuremath{12+\log(\mathrm{O/H})}} 
\newcommand{\ratioZZsun}{\ensuremath{Z/Z_{\odot}}}     
\newcommand{\Lx}{L_{\mathrm{X}}}

\def\cm2{cm$^2$}
\def\se1{s$^{-1}$}

\newcommand{\angstrom}{\textup{\AA}}

\usepackage{xfrac}
\newcommand{\altfrac}[2]{\ifmmode\def\tmp{$}\else\def\tmp{}\fi\mbox{%
    {\raisebox{.34\ht\strutbox}{\tmp#1\tmp}}%
    \kern-2.2pt\scalebox{1.5}[1.5]{/}\kern-0.0pt%
    {\tmp#2\tmp}%
    }}

\usepackage{threeparttable}
\usepackage{tablefootnote}
\usepackage{multirow}
\usepackage{booktabs}
\usepackage{wrapfig}
\usepackage{float}
\newcommand{\ii}{~\textsc{ii}}
\newcommand{\iii}{~\textsc{iii}}
\newcommand{\iv}{~\textsc{iv}}

\newcommand{\kmsmpc} {$\rm {km~s^{-1}}~Mpc^{-1}$}

\newcommand{\hii}{H\,{\sc ii }\rm}

\newcommand{\oiii}{[O\,{\sc iii}]}
\newcommand{\oii}{[O\,{\sc ii}]}

\newcommand{\nii}{[N\,{\sc ii}]}
\newcommand{\sii}{[S\,{\sc ii}]}
\newcommand{\siii}{[S\,{\sc iii}]}

\newcommand{\ariv}{[Ar\,{\sc iv}]}

\def\hbeta {H$\beta$}
\def\halpha {H$\alpha$}

\newcommand{\NIIHa}{\ensuremath{[\text{N\,{\sc ii}}]/ \text{H}\alpha}}
\newcommand{\OIIINII}{\ensuremath{[\text{O\,{\sc iii}}]/[\text{N\,{\sc ii}}]}}
\newcommand{\OIIIOII}{\ensuremath{[\text{O\,{\sc iii}}]/[\text{O\,{\sc ii}}]}}
\newcommand{\NIIOII}{\ensuremath{[\text{N\,{\sc ii}}]/[\text{O\,{\sc ii}}]}}

\newcommand{\NIIHA}{\ensuremath{[\text{N\,{\sc ii}}]\lambda6584/\text{H}\alpha\lambda6563}}

\newcommand{\OIIIHB}{\ensuremath{[\text{O\,{\sc iii}}]\lambda5007/\text{H}\beta\lambda4861}}

\newcommand{\aox}{\alpha_{\mathrm{ox}}}
\newcommand{\Ne}{$N_{\rm e}$}
\newcommand{\Te}{$T_{\rm e}$}

\newcommand{\ntwo}{\mathrm{N_2}}
\newcommand{\otnt}{\mathrm{O_3 N_2}}
\newcommand{\ns}{\mathrm{N_2 S_2}}
\newcommand{\no}{\mathrm{N_2 O_2}}

\usepackage{orcidlink}
\defcitealias{StorchiBergmann1998}{SB98}
\newcommand{\SBcalib}[1]{\citetalias{StorchiBergmann1998}\textcolor{blue}{#1}}
\defcitealias{Castro2017}{C17}  
\defcitealias{Carvalho2020}{C20}
\defcitealias{Dors2021b}{D21}
\defcitealias{GRAVITY2024}{GRAVITY}

\usepackage{tikz}
\usetikzlibrary{shadows, shapes.geometric, arrows.meta, positioning, calc, fit, backgrounds}

\title[Identifying AGNs from X-ray detections---I]{Identifying AGNs from X-ray detections---I: Metallicity calibrations in AGNs with X-ray luminosity as the primary input parameter}

\author[M.~Armah et al.]{Mark Armah$^{{\orcidlink{0000-0002-9746-3938}}1}$\thanks{E-mail: mrkrmh@gmail.com (MA)},
O. L. Dors$^{\orcidlink{0000-0003-4782-1570}1}$, 
Rog\'erio Riffel$^{\orcidlink{0000-0002-1321-1320}2}$,
M.~V. Cardaci$^{\orcidlink{0000-0002-8856-602X}3,4}$, G.~F. H\"agele$^{\orcidlink{0000-0002-9011-8517}3,4}$,
\newauthor{Rogemar A. Riffel$^{\orcidlink{0000-0003-0483-3723}5}$ and
J. M. V\'ilchez$^{\orcidlink{0000-0001-7299-8373}6}$}
\\
$^{1}$Universidade do Vale do Para\'iba, Instituto de Pesquisa \& Desenvolvimento, Av. Shishima Hifumi, 2911, CEP: 12244-000, São José dos Campos, SP, Brazil\\
$^2$Departamento de Astronomia, Instituto de F\'isica, Universidade Federal do Rio Grande do Sul, CP 15051, 91501-970, Porto Alegre, RS, Brazil \\
$^3$ Facultad de Ciencias Astron\'omicas y Geof\'{\i}sicas, Universidad Nacional de La Plata, Paseo del Bosque s/n, 1900 La Plata, Argentina\\
$^4$ Instituto de Astrofísica de La Plata (CONICET-UNLP), La Plata, Avenida Centenario (Paseo del Bosque) S/N, B1900FWA, Argentina\\
$^5$Departamento de F\'isica, CCNE, Universidade Federal de Santa Maria, 97105-900, Santa Maria, RS, Brazil\\
$^{6}$Instituto de Astrof\'isica de Andaluc\'ia, CSIC, Apartado de correos 3004, E-18080 Granada, Spain
}

\date{Accepted XXX. Received YYY; in original form ZZZ}

\pubyear{\the\year{}}

\begin{document}
\label{firstpage}
\pagerange{\pageref{firstpage}--\pageref{lastpage}}
\maketitle

\begin{abstract}
We present the first semi-empirical strong-line calibrations to determine metallicity in active galactic nuclei (AGNs) that use the directly observable X-ray luminosity ($\Lx$) instead of the dimensionless ionization parameter ($U$). The calibrations are derived from an extensive grid of photoionization models computed with the {\sc cloudy} code, which are compared with observational data of Seyfert~2 nuclei from the Burst Alert Telescope (BAT) AGN Spectroscopic Survey (BASS). In this first paper, we develop new calibrations for two key optical metallicity diagnostics based on the $\ntwo$ and $\otnt$ indices, which are valid in a metallicity range of $8.0 \lesssim \logOH \lesssim 9.1\, {\rm or}\, 0.2 \lesssim (Z/Z_{\odot}) \lesssim 2.6$, with precision of $1\sigma \approx 0.22$ dex ($\ntwo$) and $\approx0.20$ dex ($\otnt$). We systematically investigate the influence of the AGN spectral index $(\aox)$, narrow-line region (NLR) gas density (\Ne), the characteristic peak temperature of the Big Blue Bump $(T_{\rm BB})$, and $\Lx$. 
We find a strong, opposing secondary dependence on $\Lx$ for both indices. We demonstrate that neglecting this parameter overlooks systematic offsets intrinsic to the diagnostics, leading to metallicity errors of up to $\sim 1.0 Z_{\odot}$ ($\sim 0.50$ dex), particularly for the least and most luminous sources. This framework offers a more precise characterization of chemical enrichment in the NLRs of AGNs by leveraging their intrinsic X-ray emission to mitigate these systematic biases.
\end{abstract}

\begin{keywords}
galaxies: active -- ISM: abundances -- galaxies: ISM -- galaxies: Seyfert -- X-rays: galaxies.
\end{keywords}



\section{Introduction}
\label{sec:introduction}
Active Galactic Nuclei (AGNs) represent some of the most energetic phenomena in the Universe, characterized by strong emission lines in their spectra that serve as fundamental diagnostics of their physical conditions and chemical composition \citep[e.g.][]{Osterbrock2006, Netzer2015, Kewley2019, Maiolino2019Rev}. The metallicity of the gas phase in AGNs, particularly within their narrow-line regions (NLRs), provides a powerful tool for tracing the chemical enrichment history of their host galaxies and understanding galaxy evolution across cosmic time \citep[e.g.][]{1999ARA&A..37..487H,Maiolino2019Rev,Armah2023, 2024MNRAS.527.8193D}.

Oxygen is the primary element used to trace the gas-phase metallicity ($Z$) in various astrophysical environments, including star-forming galaxies, \hii regions and AGNs. This is due to the prominence of emission lines from its dominant ions (i.e. O$^{+}$, O$^{2+}$) in optical spectra \citep[e.g.][]{Hagele2008,Peimbert1969,Groves2004a,PerezMontero2017}. In ionized nebulae, the “direct method” or  electron temperature method (\Te-method) is considered the gold standard for chemical abundance measurements, since a consonance between oxygen abundance  estimates in stars (via absorption lines)
and  in \hii regions (via emission lines)  have been derived    \citep[e.g.][]{2003A&A...399.1003P, 2017MNRAS.467.3759T}. 
This technique requires estimates of \Te, which is derived using the flux of the weak ( typically $\sim 100$ times weaker than H$\beta$) temperature-sensitive auroral lines, such as \oiii$\lambda$4363, \nii$\lambda$5755, \siii$\lambda$6312, and the \oii$\lambda\lambda$7319, 7330 doublet. However, for AGN samples in the local universe ($z \: < \: 0.4$), such as those drawn from the Sloan Digital Sky Survey \citep[SDSS;][]{2000AJ....120.1579Y}, typically fewer than 30 per cent of objects meet the aforementioned criteria \citep[e.g.][]{2020MNRAS.496.2191F,2020MNRAS.492..468D}.

The application of the \Te-method in AGNs and in \hii regions has been shown to yield systematically lower oxygen abundances ($\sim$ 0.2~dex) compared to strong-line methods
\citep[e.g.][]{Dors2020c}.
This discrepancy, often termed the “abundance discrepancy factor”  \citep[ADF;][]{GarciaRojas2007,Peimbert2017} or part of the broader “\Te--problem” \citep[e.g.][]{Peimbert1967,Peimbert1969,Garnett1992,Binette1996},  may arise from several factors in AGNs, including temperature fluctuations \citep{2021MNRAS.506L..11R}, unaccounted heating mechanisms like shocks \citep[e.g.][]{Contini1998,Kewley2013, Dors2020c,Dors2021a} and/or inadequate ionization correction factors (ICFs) for oxygen \citep[e.g.][]{Peimbert2017,PerezMonteiro2019,Dors2020c}. Spatially resolved studies often reveal complex kinematics and shock signatures in NLRs of AGNs \citep[e.g.][]{Riffel2017,  2018ApJ...856...46R, 2021ApJ...910..139R, 2022ApJ...930...14R, DAgostino2019b,Riffel2021,Shimizu2019,RuschelDutra2021,Kakkad2022,Riffel2023}, further complicating direct abundance determinations.

The inherent limitations of the direct method highlight the need for complementary techniques to estimate gas-phase metallicities, especially for large samples and at cosmological distances. Consequently, “strong-line” methods, which use (semi-)empirical or theoretical calibrations between easily measurable bright emission line ratios (ELRs) and metallicity, are predominantly used for AGNs \citep[see][]{Kewley2019,Maiolino2019Rev,2020MNRAS.492..468D}. These methods often rely on photoionization models and/or the \Te-method to interpret the observed LRs. Over the past five decades, photoionization models such as {\sc cloudy} \citep{Ferland1998,Ferland2013,Chatzikos2023} and {\sc mappings} \citep{Binette1985,Sutherland1993,Sutherland2017} have been extensively used to circumvent the limitations of the $T_{\rm e}$-method.

In particular, earlier strong-line calibrations based on AGNs photoionization models, such as those by \citet[][hereafter \citetalias{StorchiBergmann1998}]{StorchiBergmann1998}, used combinations of emission LRs like $\ntwo$ = \NIIHA\  and $O3$=\OIIIHB.  
The $\ntwo$ index remains a readily accessible and widely used metallicity indicator for H{\ii} regions 
\citep[e.g.][]{Pettini2004, Marino2013, Dopita2016, Curti2017}  and AGNs (\citealp[e.g.][hereafter \citetalias{Carvalho2020}]{Carvalho2020}; \citealp{Oliveira2022, 2024PASA...41...99O}). Similarly, the $\otnt$ = (\OIIIHB)$/$(\NIIHA)  has proven useful, particularly in separating star-forming galaxies from AGNs and as a potential metallicity diagnostic \citep[e.g.][]{Alloin1979,Kewley2001,Maiolino2008}.

The accuracy of strong-line methods depends on the underlying photoionization models, which must adequately represent the complex physical conditions within AGN photoionization regions. These conditions are shaped by the AGN’s ionizing continuum (characterized by its $\Lx$, and spectral energy distribution (SED), often parameterized by a spectral index like $\alpha_{\mathrm{ox}}$, \citealt{Tananbaum1979, 1981ApJ...245..357Z}), metallicity ($Z$), electron density ($N_{\rm e}$), and ionization parameter ($U$). Although many studies have explored these dependencies, the focus has largely remained on star-forming regions. However, because these methods rely on empirical correlations observed in local {\hii}regions and star-forming galaxies, their application to objects with different physical conditions such as the harder radiation fields of AGNs or the extreme environments of high-redshift galaxies remains an active debate in the literature. It is well established that secondary dependencies on the $U$ and $N_{\rm e}$ can systematically bias emission line diagnostics in star-forming galaxies \citep[e.g.][]{Kewley2002, Nagao2006b, Sanders2016} and in AGNs \citep[e.g.][]{Nagao2006a, 2014MNRAS.443.1291D}. Consequently, comprehensive grids varying all these key parameters simultaneously are essential for developing robust, strong-line metallicity calibrations.

In this paper, our goal is to develop new semi-empirical strong-line metallicity calibrations for AGNs using the $\ntwo$ and $\otnt$ indices together with the X-ray luminosity of the AGN. In this regard, we combined photoionization model results with a large sample of Seyfert~2 AGN observational data from the Burst Alert Telescope (BAT) AGN Spectroscopic Survey (BASS) Data Release 2 (DR2). Metallicity calibrations for the $\no$ and $\ns$ indices will be presented in a companion paper Armah et al. (in preparation, hereafter Paper II). 

This paper is organized as follows. In \S~\ref{bass_data}, we describe the BASS DR2 observational sample used to test our models. In \S~\ref{method}, we detail the methodology for constructing our extensive grid of photoionization models using {\sc cloudy} code, focusing on the implementation of $\Lx$ as an input parameter.  \S~\ref{results} presents the results, where we compare the photoionization model results with observational data and derive our new luminosity-based metallicity calibrations for the $\ntwo$ and $\otnt$ indices.  In \S~\ref{smetal}, we detail the quantitative analysis used to derive the final metallicity calibrations from the diagnostic diagrams and present the new calibrations. In \S~\ref{discussion}, we discuss the implications of our findings and a detailed comparison with similar results from the literature.  Finally, we summarize our main conclusions in \S~\ref{conclusions}.

Throughout this paper, we adopt a solar oxygen abundance of $\log({\rm O/H})_{\odot} = -3.31$ \citep{Asplund2021} and use the following expression to convert the
oxygen abundance into metallicity and vice versa: $12+\log({\rm O/H})= 12+\log[(Z/\rm Z_{\odot}) \times 10^{-3.31}]$. For our cosmological framework, we use a spatially flat, six-parameter $\Lambda$CDM model with parameters from the \citet{Planck2021}: $\Omega_{\rm m}= 0.315\pm0.007$ and $H_{0} = 67.4\pm 0.5$ \kmsmpc.

\section{Observational sample}
\label{bass_data}
The emission-line observational data for Seyfert galaxies used in this study were obtained from \citet{Armah2023}, who selected AGNs from the publicly available optical spectroscopic data set of the BASS DR2 \citep{Koss2017,Oh2017,Ricci2017BASS,Koss2022a,Oh2022}. The spectra generally span a wavelength range adequate for measuring the principal optical nebular lines from \oii$\lambda 3727$ to \sii$\lambda 6731$, with spectral resolutions better than 7 \angstrom, which are typically sufficient for deblending adjacent features and enabling basic kinematic analyses 
(\citealp[for details on the selection and measurements see ][]{Armah2023} and \citealp[]{Oh2022}).  

An object is retained in our final AGN sample if it satisfies the following two criteria:
(i) it lies simultaneously above both the theoretical maximum starburst demarcation of \citet{Kewley2001} and the Seyfert–LI(N)ER division proposed by \citet{CidFernandes2011}, thereby primarily selecting robust Seyfert nuclei; and (ii) for objects located in the region where the \citet{Kewley2001} curve is not defined (i.e. $\log([\mathrm{N}\,\textsc{ii}]/\mathrm{H}\alpha) > 0.47$), which falls above the \citet{CidFernandes2011} line alone. This second requirement explicitly includes Seyfert-type AGNs situated to the right of the \citet{Kewley2001} asymptote.

Implementation of these selection criteria yields an initial sample of 617 X-ray luminous, optically confirmed Seyfert 2 AGNs, predominantly at low redshift ($z \approx 0.001$--0.31) and moderate to high bolometric luminosity ($\log L_{\rm bol} \approx 41.68$--46.59). The relative lack of low-$z$, high-$\Lx$ systems is largely due to survey volume limitations. 
Stellar mass is commonly used as an indirect proxy for gas-phase metallicity via the mass-metallicity relation (MZR). However, for studying the intrinsic properties of AGNs, $\Lx$ provides a direct measure of the power of the central engine. Unlike optical or UV metrics, X-ray luminosity is largely free from contamination by stellar emissions of the the host galaxy and is substantially less affected by the obscuration discussed by \citet{Davies2014}.
However, fixed-aperture effects in the nearest sources may still contribute to underestimation of the intrinsic X-ray flux.

For the purposes of our analysis, we further restrict the sample to 426 objects. We define reliable, high-quality measurements as those where all emission lines required to construct the diagnostic diagrams presented in \S~\ref{results} (specifically \oii$\lambda3727$, H$\beta$, \oiii$\lambda5007$, \nii$\lambda6584$, and H$\alpha$) are robustly detected. While explicit flux uncertainty estimates are not available for every individual emission line across all objects, the fluxes utilized here were derived from the uniform BASS DR2 spectral fitting procedure \citep{Oh2022}. This procedure strictly modeled and subtracted the host galaxy stellar continuum, ensuring that reported fluxes represent statistically converged, robust line detections free from stellar absorption biases, inherently acting as a strict quality baseline. Furthermore, requiring simultaneous, valid flux measurements to successfully place an object on these multiple diagnostic diagrams effectively acts as an independent filter against marginal or spurious noise detections. Finally, an analysis of the subset of our final sample that does include explicit estimated errors indicates a representative median signal-to-noise ratio (SNR) > 10 for the line fluxes. We therefore adopt this robust multi-line detection, underpinned by the rigorous extraction procedure outlined by \citet{Oh2022}, as our standard for data quality.

\subsection{Electron density}
\begin{figure}
\centering
\includegraphics[angle=0, width=1\columnwidth]{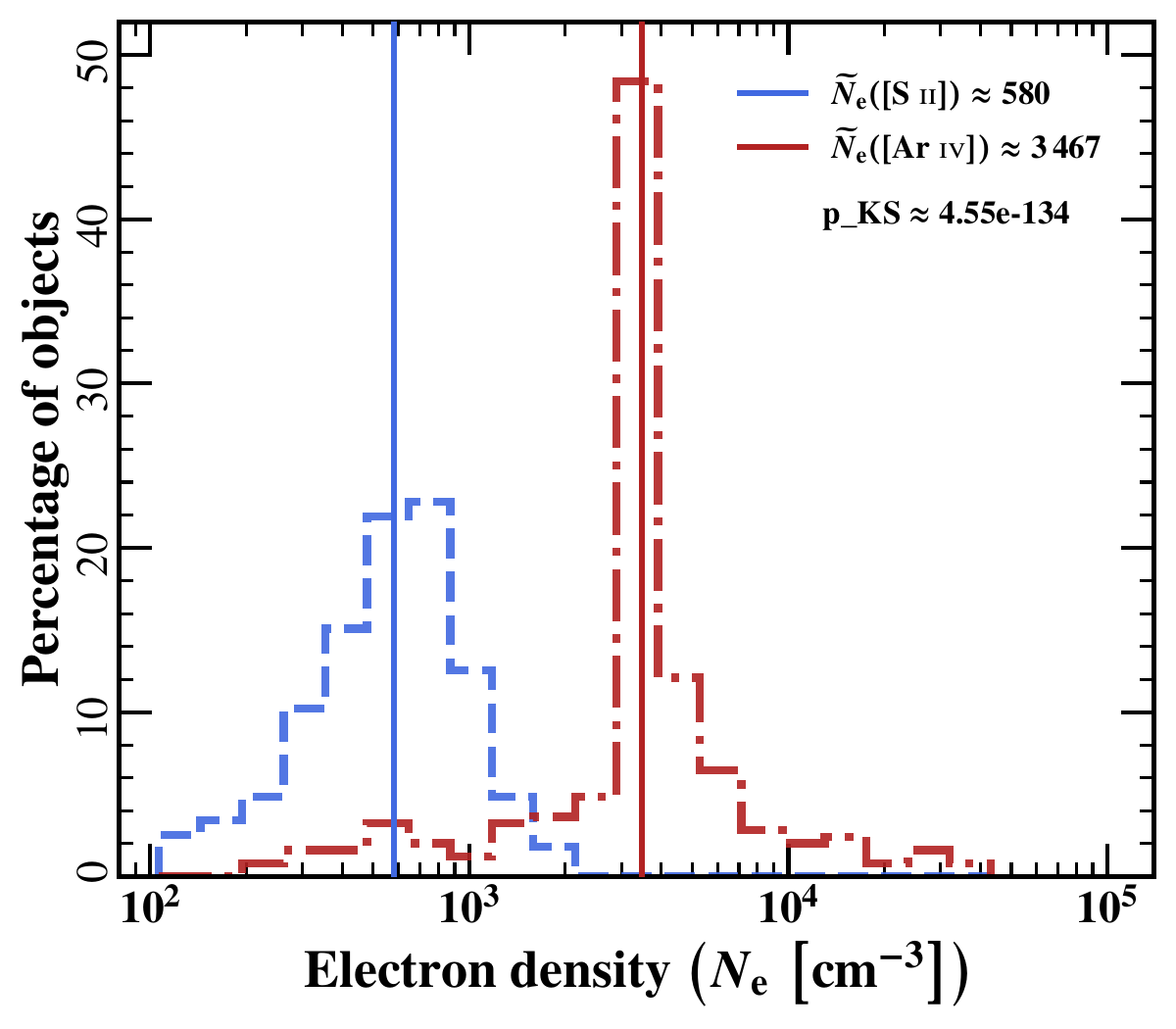}
\caption{Electron density distribution derived from our observational sample. The dashed line represents densities, $N_\mathrm{e}(\text{[S~\textsc{ii}]})$, derived from the \sii$\lambda\lambda6716,6731$ doublet (557 objects; median $\approx 580 \pm 413\,\mathrm{cm}^{-3}$). The dash-dotted line corresponds to densities, $N_\mathrm{e}(\text{[Ar~\textsc{iv}]})$, from the \ariv$\lambda\lambda4711,4740$ doublet (248 objects; median $\approx 3467 \pm 864\,\mathrm{cm}^{-3}$). Vertical lines indicate the median values. The $p$-value from the Kolmogorov-Smirnov (KS) test ($p_{\mathrm{KS}}$) is $4.55 \times 10^{-134}$, confirming that the distributions are statistically distinct and trace stratified NLRs in the AGNs.}
\label{fig_1}
\end{figure}

We derived electron density for the selected sample to characterize the physical environment of the observed NLRs.
For each object, two electron density sensitive emission-line ratios were used: the lower-ionization \sii$\lambda\lambda6716,6731$ lines and the higher-ionization \ariv$\lambda\lambda4711,4740$ lines.  We used the {\sc PyNeb} package \citep{Luridiana2015} to estimate electron densities and their associated uncertainties, propagating flux errors via a Monte Carlo simulation of 1000 iterations and assuming a constant electron temperature of 10\,000 K. Although the observed electron temperature from Seyfert~2 samples show mean or median values of $\sim$13\,000 K  \citep[see][]{2018ApJ...856...46R,Binette2024}, a robust analysis reveals distinct trends for the two density diagnostics. We find that the \sii\ diagnostic is robust against temperature variations \citep[e.g.][]{Osterbrock2006,2014A&A...561A..10P,Sanders2016,2018ApJ...856...46R}, whereas \ariv\ exhibits significant sensitivity \citep[e.g.][]{Keenan1997}, with systematic shifts exceeding estimated errors at the extremes of the observed temperature range (7\,000 K and 22\,000 K). However, even accounting for this dependence, the densities derived from \ariv\ remain consistently and significantly higher than those from \sii, and thus the temperature sensitivity does not qualitatively alter the stratification results.
Figure~\ref{fig_1} presents the resulting distributions.  It is important to note, however, that the \sii\ doublet primarily traces the lower-ionization regions of the NLRs \citep[e.g.][]{Shimizu2019, Davies2020}. For the high-excitation plasma where lines such as \oiii\ originate, other diagnostics such as the \ariv$\lambda\lambda4711,4740$ doublet may provide a more representative  electron density estimate \citep[e.g.][]{Binette2024}.

Figure~\ref{fig_1}  clearly illustrates the density stratification within the $\mathrm{NLR}$. The stark difference between the density distributions derived from the \sii\ and \ariv\ lines is statistically significant, as confirmed by a two-sample Kolmogorov-Smirnov ($\mathrm{KS}$) test, which yields a $p\text{-value}$ of $\mathrm{p\_KS} \approx 4.56 \times 10^{-134}$, indicating that \sii\ and \ariv\ LRs are tracing regions with distinct electron densities. This extremely low value allows us to reject the null hypothesis that the two samples are drawn from the same underlying objects. The \sii\ diagnostic, tracing lower-ionization gas, yields a median density of $\sim 580 \pm 413$ $\mathrm{cm}^{-3}$ (range: $107\lesssim N_{\rm e}\left[{\rm cm}^{-3}\right]\lesssim 1\,984$). In contrast, the \ariv\ lines, which probe the high-excitation plasma where lines like \oiii\ originate \citep[e.g.][]{Binette2024}, reveal a significantly higher and broader range of values. This distribution is strongly right-skewed, with a median density of $\sim 3\,467 \pm 864$ $\mathrm{cm}^{-3}$ (range: $201\lesssim N_{\rm e}\left[{\rm cm}^{-3}\right]\lesssim 43\,206$), nearly an order of magnitude higher than that traced by \sii. This large divergence arises from a significant tail of high-density gas, with approximately 8\,\% of objects exhibiting densities above $10\,000\ \mathrm{cm}^{-3}$.

\section{Methodology}
\label{method}
Our study combines high-quality observational data from hard X-ray selected AGNs with a comprehensive suite of photoionization models. A key novelty of our model is the direct use of $\Lx$ of AGN as an input parameter defining the strength of the ionizing radiation field, rather than relying on the dimensionless ionization parameter ($U$) at a specific radius of the nebula. This allows for a more direct comparison with the observable properties of the AGN.

\subsection{Photoionization models with \texorpdfstring{\boldmath $\Lx$}{Lx}}
\label{models}
We computed an extensive grid of photoionization models using version C23.01 of the spectral synthesis code {\sc cloudy} \citep{Ferland2017, Chatzikos2023}. A primary innovation in our modelling  approach is the direct specification of the total ionizing luminosity, connected by the observable X-ray luminosity, $\Lx$, instead of the dimensionless ionization parameter, $U$.

Traditionally, grids of photoionization models have been constructed using $U$ as input parameter to reproduce UV \citep[e.g.][]{Groves2004b,Nagao2006a, Matsuoka2009,Matsuoka2011,2014MNRAS.443.1291D,Feltre2016,2024ApJ...977..187Z}, 
optical \citep[e.g.][]{Ferguson1997,StorchiBergmann1998,Carvalho2020,Richardson2022,Carr2023,2024ApJ...977..187Z}  and
infrared \citep[e.g.][]{Groves2004b,Groves2006b,2017MNRAS.470.1218P,Richardson2022,Calabro2023} ELRs. However, $U$ is fundamentally defined as the dimensionless ratio of the flux of hydrogen ionizing photons to the total hydrogen density: $\left[U = Q(\mathrm{H})/4\pi r^2 n_{{\rm H}} \rm c \right]$, where $Q(\mathrm{H})$ is the  emission rate of ionizing photons of the source, $r$ is the distance from the source to the illuminated gas cloud, $n_{\rm H}$ is the hydrogen number density and $\rm c$ is the speed of light. Although $U$ theoretically describes the local physics, it is not directly observable. Comparison of traditional models based on $U$ with observations of an AGN with a given luminosity requires assuming a distance, $r$, effectively creating a degeneracy between the luminosity of the source and the physical location of the gas \citep[e.g][]{Netzer2015}.  

Applying a $U$--based calibration to an observed object implicitly assumes a specific geometry that may not match the physical reality of the source. Our approach circumvents this by using the directly observable X-ray luminosity ($\Lx$) as the primary input. To construct the grid, we use a fixed radius ($r_{\rm in} = 0.3$ pc), which replaces a completely free parameter with a measurable one, allowing us to calibrate metallicity against a quantity that can be directly constrained from X-ray surveys. This assumed radius is consistent with reverberation mapping and interferometric constraints on the inner NLR/torus region \citep[e.g.][]{Audibert2017,Shimizu2019,GRAVITY2024}. By fixing $r_{\rm in}$, we effectively translate the variation in $\Lx$ directly into the ionization conditions of the gas. This allows us to estimate metallicity using two observables ($\Lx$ and LRs) without solving for the intermediate parameter $U$.

\begin{figure*}
    \centering
\includegraphics[width=1.0\textwidth]{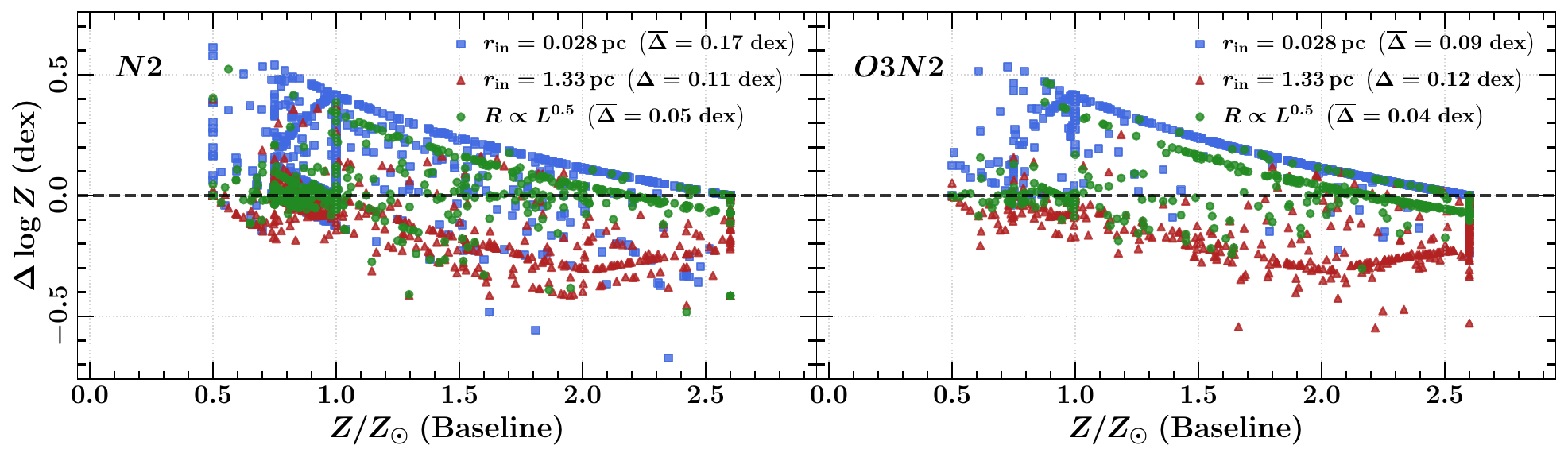}
\caption{Logarithmic residuals of derived gas-phase metallicity ($\Delta \log Z$ in dex) resulting from variations in the inner radius ($r_{\mathrm{in}}$) of the photoionized cloud for the $\ntwo$ (left panel) and $\otnt$  (right panel) diagnostics. The residuals are plotted as a function of the baseline metallicity derivation ($r_{\mathrm{in}} = 0.3$ pc). 
To evaluate the extreme spatial boundaries of the \citet{GRAVITY2024} sample, the derived metallicities were analytically recalculated assuming fixed constant radii of $r_{\mathrm{in}} = 0.028$ pc (blue squares) and $r_{\mathrm{in}} = 1.33$ pc (red triangles).
The green circles represent the scenario where $r_{\mathrm{in}}$ is continuously coupled to the AGN luminosity (i.e. $R \propto L^{0.5}$), effectively evaluating the grid at a constant ionization parameter ($U$) corresponding to a pivot luminosity of $\Lx = 10^{43.5}$ erg s$^{-1}$. The dashed horizontal line indicates zero residual. The mean absolute deviations $\left(\overline{\Delta}\right)$ provided in the legend demonstrate that incorporating the $R \propto L^{0.5}$ correlation significantly reduces systematic bias across the sample compared to assuming unscaled extreme spatial boundaries.
}
\label{fig_1b}
\end{figure*}

Interferometric measurements by the \citet{GRAVITY2024} demonstrate that the dust sublimation radius spans a wide physical range (i.e. $0.028 \lesssim r_{\rm in} \lesssim 1.33$ pc) and correlates strongly with the AGN bolometric luminosity, scaling as $R \propto L^{0.5}$. Because the ionization parameter scales as $U \propto \Lx/r_{\rm in}^2$, coupling the inner radius to the luminosity via this empirical relation yields a scenario where $U$ remains entirely independent of the varying luminosity at the sublimation boundary.
To rigorously evaluate how this physical scaling impacts our calibrations utilizing our fixed $r_{\rm in} = 0.3$ pc model grid, we exploit the fact that evaluating a luminosity $L$ at radius $r$ produces the exact same ionization parameter (and resulting line ratios) as a model with luminosity $L^{\prime}$ at radius $r^{\prime}$ if $L/r^2 = L^{\prime}/(r^{\prime})^2$. Analytically, we can simulate this by applying a shift to the $\Lx$ of the observational targets prior to mapping them onto our model grids to interpolate their metallicities (see \S~\ref{smetal} for details). This shift is given by:
\begin{equation}
    \Delta \log \Lx = -2 \log \left(\frac{r^{\prime}}{r_0}\right),
 \label{Lx_shiftt}
\end{equation}
where $r_0 = 0.3$ pc is our baseline radius and $r^{\prime}$ is the varied inner radius. For comparison, as shown in Figure~\ref{fig_1b}, we first evaluated the extreme spatial boundaries by analytically recalculating the metallicities under the assumption of fixed, unscaled radii of 0.028 pc and 1.33 pc.

However, when coupling the radius to the luminosity, we require a fiducial pivot point to normalize the $R \propto L^{0.5}$ scaling. According to standard theoretical dust sublimation relations (e.g. $R_{\mathrm{sub}} \approx 0.4 \times (L_{\mathrm{bol}}/10^{45})^{0.5}$ pc; \citealt{2008ApJ...685..160N}), an inner radius of 0.3 pc corresponds to a bolometric luminosity of $L_{\rm bol} \approx 10^{44.75} \text{ erg s}^{-1}$. Applying a typical bolometric correction factor of $k_{\rm bol} \approx 17$ \citep[e.g.][]{2019MNRAS.488.5185N,2020A&A...636A..73D}, this equates to an X-ray luminosity of $\Lx \approx 10^{43.5} \text{ erg s}^{-1}$.
The $R \propto L^{0.5}$ scaling  dictates that $r^{\prime} = r_0 (\Lx / 10^{43.5})^{0.5}$. Substituting this into Equation~\ref{Lx_shiftt} reduces the shifted luminosity for every object to exactly $10^{43.5} \text{ erg s}^{-1}$. Therefore, rather than calculating individual spatial shifts, we project the entire sample onto this single pivot coordinate on our fixed 0.3 pc grid. This evaluates the model gas under a constant ionization parameter consistent with the spatial scaling established by the GRAVITY Collaboration.

It is important to note that this exact pivot luminosity is inherently dependent on the assumed normalization of the dust sublimation relation and the applied bolometric correction. For instance, our derived pivot of $\Lx = 10^{43.5} \text{ erg s}^{-1}$ aligns perfectly with the silicate dust sublimation boundary (e.g. $R_{\rm sub,Si} \simeq 1.3 \times (L_{\rm bol}/10^{46})^{0.5}$ pc; \citealt{Netzer2015}). In contrast, highly refractory carbon (graphite) grains can survive at higher temperatures closer to the central engine. Adopting a carbon-dominated sublimation relation would require nearly an order of magnitude higher luminosity (e.g. $\log \Lx \approx 44.3$) to push the boundary out to 0.3 pc. Therefore, assuming different dust grain properties or bolometric corrections would shift this pivot coordinate, effectively evaluating the sample at a different constant ionization parameter and potentially introducing a global systematic offset in the derived metallicities.

If we adopt the extreme constant boundaries of the GRAVITY sample without scaling them with luminosity, we find large individual deviations in the derived metallicity spanning from $-0.67$ dex to $+0.61$ dex, although the majority of the data points fall within a more restricted range of $\pm 0.5$ dex in both panels of Figure~\ref{fig_1b}. These extreme values align with maximum absolute deviations of $0.53 - 0.67$ dex at the extreme inner boundary ($r_{\mathrm{in}} = 0.028$ pc), yielding mean systematic biases of 0.17 dex for $\ntwo$ and 0.09 dex for $\otnt$. At the outer boundary ($r_{\mathrm{in}} = 1.33$ pc), the maximum absolute deviations are $0.45 - 0.55$ dex, with mean systematic biases of 0.11 dex for $\ntwo$ and 0.12 dex for $\otnt$.  However, when $r_{\mathrm{in}}$ is self-consistently coupled to $\Lx$ following the $R \propto L^{0.5}$ relation (implemented via our projection to the pivot coordinate), this geometric dependency is substantially mitigated. This continuous scaling diminishes the averaged systematic deviations, yielding a mean (median) absolute difference of only $0.05$ ($0.02$) dex for $\ntwo$ and $0.04$ ($0.01$) dex for $\otnt$. Furthermore, additional analytical tests assuming extreme, unscaled spatial limits well beyond the GRAVITY sample boundaries (i.e. $0.005 \lesssim r_{\mathrm{in}} \lesssim 3.0$ pc) yield similar systematic biases.

These deviations are substantially smaller than the intrinsic scatter typical of empirical strong-line metallicity calibrations ($\sim 0.20$ dex). Therefore, utilizing a fixed baseline radius of $r_{\mathrm{in}} = 0.3$ pc when recognized as a stable pivot point representing the primary ionization parameter behaviour does not alter the underlying physics of the model. Rather, it eliminates the need to treat $r_{\mathrm{in}}$ as an unconstrained free parameter while introducing negligible systematic uncertainty to the final metallicity derivations.

\subsubsection{AGN ionizing continuum}
The generic SED is characterized by a prominent BBB and a connecting X-ray power law. The BBB, which peaks in the extreme ultraviolet (EUV), is understood to be the thermal emission from the optically thick, geometrically thin accretion disk surrounding the supermassive black hole \citep[e.g.][]{Shakura1973, Marconi2004, Brandt2015, Netzer2015, Lusso2017}. For our models, we adopt a characteristic peak temperature for this component of $T_{\mathrm{BB}} = 1.0 \times 10^5$\,K, a representative value for Seyfert-luminosity. Although our primary analysis uses this value, we also explorea wide range of temperatures, from $T_{\mathrm{BB}}=1.0\times10^{4}\,\mathrm{K}$ to $T_{\mathrm{BB}}=3.0\times10^{5}\,\mathrm{K}$   \citep[see also][]{Carr2023}, to test the robustness of our calibrations against this key assumption. This broader exploration allows us to account for the intrinsic variations in the accretion disc temperatures across the AGN population.

The shape of the SED is further refined by two spectral indices. The first is the index of the non--thermal power law at energies below the Big Bump, for which we adopt a typical value of $\alpha_{\mathrm{uv}} = -1.0$ (where $F_{\nu} \propto \nu^{\alpha_{\mathrm{uv}}}$). The second parameter is the two-point spectral index, $\aox$, which connects the optical/UV emission to the X-ray emission. It is defined as the slope of a power law between the monochromatic luminosities at 2500\,\AA\ and 2\,keV \citep{Tananbaum1979}.
The $\aox$ parameter is varied across our grid (see \S~\ref{grid_params}), reflecting the observed range and its known correlation with luminosity \citep[e.g.][]{Lusso2017}.

Finally, the total luminosity of this continuum is specified and scaled by the input 2--10 keV X-ray luminosity. The hard X-ray luminosity is less susceptible to dust extinction than UV/optical emission and serves as a robust observational proxy for the total ionizing output of the AGN \citep{Marconi2004}. While this work explores a range of standard continuum shapes, alternative SEDs have been proposed to resolve discrepancies between observed and modeled LRs. For instance, double-bump SEDs, with a second peak near 200 eV, have been shown to increase the plasma heating rate and can potentially reproduce the high NLR temperatures that standard models fail to achieve \citep[][]{Binette2024}.

\subsubsection{Model grid parameters}
\label{grid_params}
Our grid of photoionization models is designed to span the wide range of physical conditions observed in the NLRs of AGNs. The principal parameters defining the grid are the ionizing luminosity, the shape of the ionizing continuum, the gas density, and the gas-phase metallicity. Each parameter is varied to encompass values representative of the full range of Seyfert galaxies.

\begin{itemize}

\item Ionization spectral index ($\aox$): 
This index governs the hardness of the ionizing radiation field; a less negative value indicates a relatively stronger X-ray component, which enhances the heating and ionization of the gas, favoring higher-ionization species. However, a more negative value corresponds to a softer ionizing continuum. We consider five values, $\aox = [-2.0, -1.7, -1.4, -1.1, -0.8]$, which span and extend the typical observed range for Seyfert galaxies and quasars to include hard ionizing continua (e.g. \citealt{1994ApJS...92...53W, 2005AJ....130..387S, 2006AJ....131.2826S, Just2007, 2011ApJ...726...20M, 2019A&A...630A.118V, 2021RNAAS...5..101T, 2021RAA....21....4Z}). It is well-established that the $\aox$ parameter correlates with AGN luminosity, whereby more luminous sources typically exhibit softer spectra, corresponding to more negative $\alpha_{\rm ox}$ values  \citep[e.g.][]{Tananbaum1979, 2006AJ....131.2826S, Lusso2010}.
Moreover, $\aox$ has an indirect relation with the ionization parameter \citep{2025A&A...696A.229P}. While $\Lx$ sets the normalization of the ionizing photon rate (total power), $\aox$ determines the spectral shape (hardness). Both are necessary to fully describe the ionizing field; $\Lx$ drives the overall intensity, while $\alpha_{ox}$ influences the heating efficiency and relative ion fractions.

\item X-ray luminosity: The 2--10 keV X-ray luminosity, $\Lx$, serves as an observational proxy for the bolometric luminosity of the AGN \citep[e.g.][]{Ichikawa2017} and sets the overall power of the ionizing source in the models. We explore six values for its logarithm, that is, $\log \Lx ({\rm erg \: s^{-1}})= [38, 40, 42, 44, 46, 48]$. This extensive range covers the observed luminosities from low-luminosity AGNs (LLAGNs) and local Seyfert galaxies ($\log \Lx \sim 40-44$) up to the most luminous Seyfert galaxies, ensuring our models are applicable across the full range of Seyfert activity \citep[e.g.][]{Brandt2015}.

\item Electron density:  We vary the hydrogen number density across the grids corresponding to electron densities of $N_{\rm e} = 100$, $500$, $1\,000$, $2\,000$, $3\,000$, $5\,000$, and $10\,000$\,cm$^{-3}$. This range corresponds to typical conditions in the NLR, as inferred from diagnostic line ratios such as \sii$\lambda6716/\lambda6731$ and \ariv$\lambda\lambda4711,4740$ doublets  \citep[e.g.][]{Osterbrock2006, Netzer2015, Binette2024}. The upper end of this range approaches the critical densities of several key auroral lines, making the models sensitive to density-driven changes in the cooling curve. While the \sii\ doublet typically traces lower-density gas in the partially ionized zone, higher electron densities ($N_{\rm e} \sim 10^4 - 10^6$\,cm$^{-3}$) are frequently found in the fully ionized, inner regions of the NLR. These higher densities are derived from auroral and trans-auroral lines \citep[e.g.][]{Rose2018, Baron2019, Davies2020, 2021ApJ...910..139R, 2024ApJ...960..108Z}.

\item Gas-phase metallicity ($Z/\rm Z_{\odot}$): The abundance of heavy elements relative to solar, $Z/\rm Z_{\odot}$, is a primary regulator of the gas thermal balance, as metal-line emission is the dominant cooling mechanism. We explore five metallicities, $(Z/\Zsun) = 0.2, 0.5, 0.75, 1.0,$ and $ 2.6$. This range accommodates the diverse environments of AGN host galaxies \citep[e.g.][]{2020MNRAS.492..468D, 2025arXiv250805397D, 2024ApJ...977..187Z},
from potentially metal-poor systems to the highly enriched nuclei of massive galaxies, which are expected to host the most luminous AGNs, consistent with the observed luminosity-metallicity relation \citep[e.g.][]{Nagao2006b, Matsuoka2009}. 
\end{itemize}

\subsubsection{Elemental abundances and scaling}
To characterize the chemical composition of the gas, we adopt the baseline solar abundance pattern from the comprehensive compilation by \citet{Asplund2021}. 
We assume that most elemental abundances scale linearly with the overall gas-phase metallicity ($Z/\mathrm{Z}_{\odot}$). The exceptions are nitrogen and helium, for which we adopt specific empirical scaling relations derived using the $T_{\rm e}$-method, expressed as a function of the oxygen abundance $m = 12+\log(\mathrm{O/H})$. For the nitrogen-to-oxygen ratio, we apply the piecewise relation by \citet{2024MNRAS.534.3040D}:
\begin{equation}
    \log(\mathrm{N/O}) = 
    \begin{cases} 
        0.86 \times m - 8.39 & \text{for } m > 8.0, \\
        -1.42 & \text{for } m < 8.0.
    \end{cases}
    \label{eqn_1}
\end{equation}
For the helium abundance, we utilize the relation by \citet{2022MNRAS.514.5506D}:
\begin{equation}
    12+\log(\mathrm{He/H}) = 0.1215 \times m^{2} - 1.8183 \times m + 17.6732.
    \label{eqn_2}
\end{equation}

\subsubsection{Other nebular parameters}
For all models, we adopt a set of fixed nebular parameters that are representative of the typical physical conditions within the NLRs of Seyfert galaxies.

The NLR is known to be a clumpy, inhomogeneous medium rather than a uniform shell of gas. We account for this by adopting a low volume filling factor ($f$) of 1 per cent, which includes the ionized gas filling factor itself plus the (geometrical) volume cloud covering factor. Such a small filling factor is required to reconcile the observed line luminosities with the large volume of the region and is a canonical value used in many AGN photoionization models \citep[e.g.][]{Groves2004b}.

Observed NLR line profiles, with typical widths of several hundred km\,s$^{-1}$, are significantly broader than the thermal width of the photoionized gas ($\sim 10$\,km\,s$^{-1}$), indicating the presence of substantial non-thermal motions. We include a microturbulent velocity of $100 \, \mathrm{km\,s}^{-1}$ (default value in the {\sc cloudy} code). This parameter accounts for internal, non-thermal motions within individual gas clouds, contributing to the broadening of emission lines and providing an additional source of pressure support \citep{1992ApJ...387...95F, Komossa2008}.

We include a secondary ionization source from background cosmic rays with an ionization rate (default value in the {\sc cloudy} code) of
$2\times10^{-16} \: \rm  s^{-1}$  \citep{2007ApJ...671.1736I}.
The influence of cosmic ray ionization on the predictions of the AGN model
was undertaken by \citet{2025A&A...693A.215K}, who found that it produces additional heating of the ISM and ionization of ions
with a low ionization potential (e.g. $\rm N^{+}$).
The geometry of the models is plane-parallel, with the outer radius defined as the location where the electron temperature drops to 4\,000 K, which is the default stopping condition in the {\sc cloudy} code. 
We adopt dust-free photoionization models for the entire NLR. This assumption is motivated by the distinct grain-destruction mechanisms operating across the stratified NLR. In the inner, higher-ionization region, dust grains are likely destroyed by sublimation due to the intense radiation field \citep[e.g.][]{Ferguson1997,Kraemer2000,Stern2014}. In the outer, lower-ionization region--while cooler--the gas dynamics are often dominated by outflows and shocks \citep[e.g.][and references therein]{Riffel2017,  2018ApJ...856...46R, 2021ApJ...910..139R, 2022ApJ...930...14R, DAgostino2019b,Riffel2021,Shimizu2019,RuschelDutra2021,Riffel2023}; theoretical studies indicate that thermal sputtering in fast shocks ($v \gtrsim 200$ km s$^{-1}$) efficiently destroys dust grains, returning refractory elements to the gas phase \citep[e.g.][]{Dopita1996,Contini2004,Micelotta2010}. 

Empirically, adopting a dust-free model resolves the “\Te-problem” noted by \citet{Binette1996} and \citet{Nagao2006a}. Standard dusty models often predict electron temperatures that are too low to reproduce the high observed ionization ratios. Dust-free grids typically occupy the upper region of the  Baldwin-Phillips-Terlevich  \citep[BPT;][]{1981PASP...93....5B} diagnostic diagram (higher \OIIIHB), matching the locations of observed Seyfert galaxies better than dusty models, which often predict ratios that are too soft \citep[e.g.][]{Groves2004b}. 
This choice is reinforced by \citet{Feltre2016}, who, while demonstrating that including dust shifts model grids toward higher \NIIHA\ ratios due to depletion-induced heating (see their Figure 3), they explicitly acknowledge the findings of \citet{Nagao2006a} that high-ionization AGN signatures are best reproduced by dust-free models at $n_{{\rm H}} \lesssim 10^3\, \rm cm^{-3}$. 
Adopting dust-free, we assert that the hard AGN radiation has destroyed the grains, allowing the gas to be hotter and more efficient at emitting these specific diagnostic lines.

Consequently, we do not include dust grain physics (depletion or photoelectric heating) in our model grids. 
While the observational hard X-ray luminosity (14–195 keV) used to scale our models is largely insensitive to obscuration up to Compton-thick levels \citep[$N_{\rm H}$ > 10$^{24}$ cm$^{-2}$,][]{2015ApJ...815L..13R,2016ApJ...825...85K}, the optical emission lines remain subject to reddening by foreground/host galaxy dust. To ensure consistency between the models and data, the observational emission line fluxes were corrected for extinction  following a similar methodology for the dust-extinction correction by \citet{Armah2023} using the Balmer decrement prior to analysis. We corrected emission line fluxes using the \citet{Cardelli1989} extinction curve. We assumed an $R_V = A_V/E(B-V) = 3.1$ and an intrinsic H$\alpha/$H$\beta$ ratio of  2.86. This Balmer decrement is consistent with Case B recombination conditions ($T \approx 10^4$ K, $N_{\rm e} \sim 10^2$ cm$^{-3}$; \citealt{Osterbrock2006}) and is empirically supported by our observational data. The gas ionization degree parametrized by the \oiii$\lambda 5007/$\oii$\lambda 3727$ line ratio versus the H$\alpha$/H$\beta$ ratio
indicate that the vast majority of the models ($\sim$98 per cent) predict $\mathrm{H}\alpha/\mathrm{H}\beta$ values in the range from 2.69 to 3.10 \citep[see][]{Storey1995}. This is consistent with the observational data, where we find that $\sim$98 per cent (418/426) of the reddening-corrected BASS DR2 sample also lies within these theoretical boundaries. For both observational data and model predicted ratios, the most representative value of $\mathrm{H}\alpha/\mathrm{H}\beta$ is $\sim2.80$ with an average value of $\sim2.84$. 
This confirms that our objects reside in the physical regime where this model assumption holds true, typical of regions dominated by the hard ionizing spectra of AGNs \citep[e.g.][]{2015ApJS..217...12D, Feltre2016}.

A suite of 6\,300 photoionization models was generated to cover a broad spectrum of nebula conditions (see Table~\ref{table_1} for a full summary of all parameters).

\begin{table}
\centering
\setlength{\tabcolsep}{1.5pt}
 \caption{Parameter space of the {\sc cloudy} photoionization models.}
 \label{table_1}
 \begin{tabular}{l c l}
  \hline
  Model parameter & Symbol & Values \\
  \hline
  \multicolumn{3}{c}{\textit{Varied Parameters}} \\
  \hline
  Gas-phase metallicity & $Z/Z_{\odot}$ & [0.2, 0.5, 0.75, 1.0, 2.6] \\
  X-ray luminosity (2-10 keV) & $\log \Lx$ & [38, 40, 42, 44, 46, 48] \\
  Electron density\textsuperscript{a} & \Ne\ [cm$^{-3}$] & $10^2 \lesssim N_{\rm e}\lesssim 10^4$ \\
  Ionizing spectral index & $\alpha_{\rm ox}$ & [-2.0, -1.7, -1.4, -1.1, -0.8] \\
  Big Blue Bump temperature\textsuperscript{b} & $T_{\rm BB}$ [K] & $1.0 \times 10^4$ to $3.0 \times 10^5$ \\
  \hline
  \multicolumn{3}{c}{\textit{Fixed Parameters}} \\
  \hline
  Geometry & -- & Plane-parallel \\
  Inner radius of cloud & $r_{\rm in}$ & 0.3 pc \\
  Microturbulent velocity & $v_{\rm turb}$ & 100 km s$^{-1}$ \\
  Volume filling factor & $f$ & 0.01 \\
  Dust content & -- & Dust-free \\
  UV spectral index & $\alpha_{\rm uv}$ & -1.0 \\
  Stopping condition & \Te\ & 4\,000 K \\
  \hline
 \end{tabular}
 \begin{flushleft}
 \textsuperscript{a} Corresponds to the varied hydrogen number density. \\
 \textsuperscript{b}The main results use a fiducial temperature of $T_{BB} = 1.0 \times 10^5$ K, with other values explored from $T_{\mathrm{BB}} = 1.0 \times 10^4\,\mathrm{K}$ to $T_{\mathrm{BB}} = 3.0 \times 10^5\,\mathrm{K}$ shown in the bottom row of Figure~\ref{fig_4}.
 \end{flushleft}
\end{table}

\begin{figure*}
    \includegraphics[angle=0.0,width=0.92\textwidth]{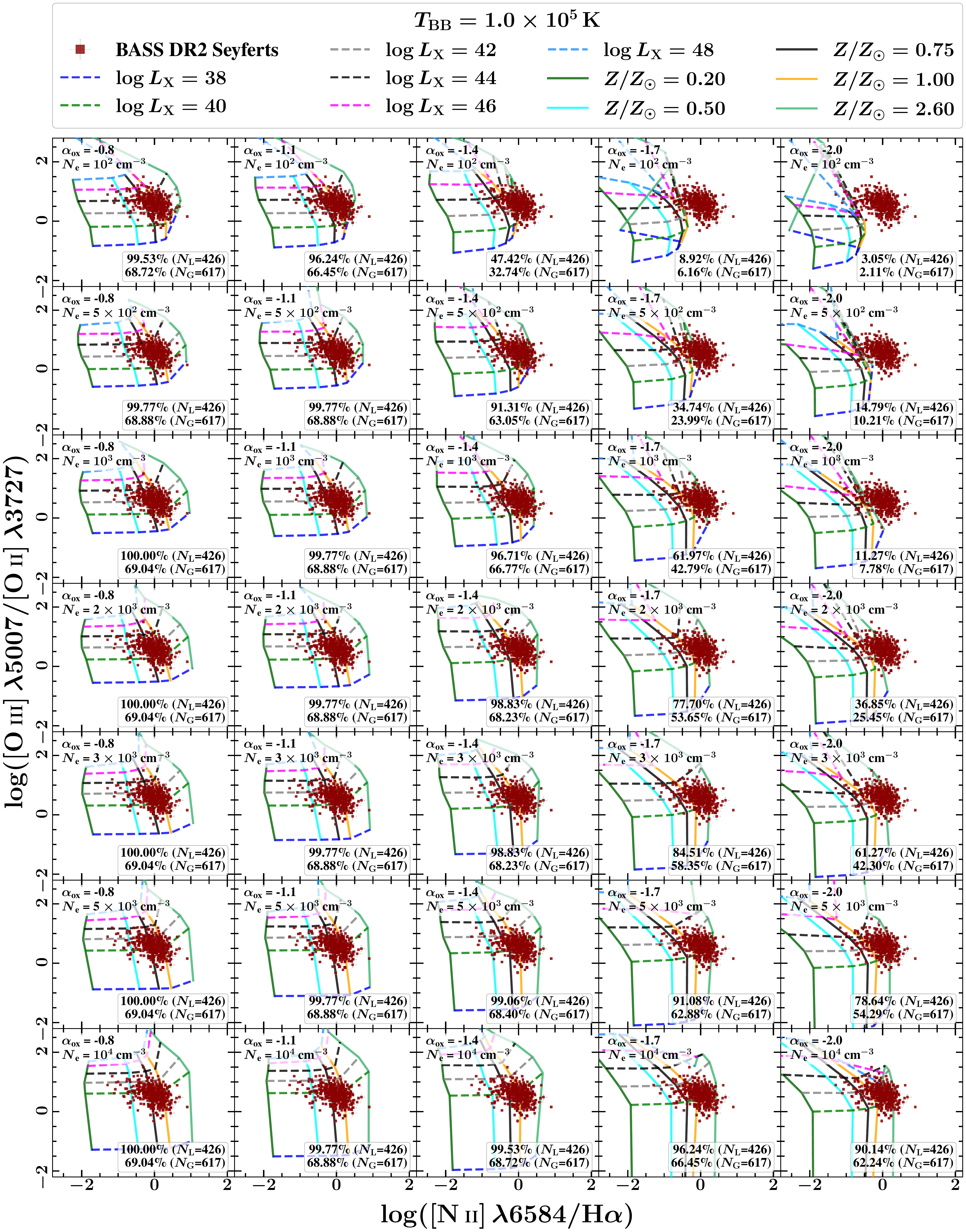}
\vspace{-8pt} 
\caption{Diagnostic diagram of log(\oiii$\lambda 5007$/\oii$\lambda 3727$) versus log(\nii$\lambda6584$/\halpha) from the models and the observational sample of Seyfert 2s. Each panel represents a grid of photoionization models calculated for specific values of the ionizing spectral index ($\aox$) and logarithm of electron density (\Ne\, [cm$^{-3}$]), as indicated in the top-left corner of each subplot. The rows correspond to $\aox$ values of -2.0, -1.7, -1.4, -1.1, and -0.8 (from bottom to top), while columns correspond to \Ne\ values from $\rm 10^2\, to \, 10^4\,cm^{-3}$ (top to bottom). All models assume a blackbody temperature $T_{\rm BB} = 100\,000\, {\rm K}$ for the ionizing continuum. Within each panel, solid lines represent models of constant metallicity ($Z/\Zsun = \rm 0.20, 0.50, 0.75, 1.00, and\, 2.60$, with colors indicated in the main legend at the top of the figure), varying with $\log \Lx$. Dashed lines represent models of constant X-ray luminosity ($\log \Lx = 38, 40, 42, 44, 46, 48$, with colors also indicated in the main legend), varying with metallicity. The red filled squares show the observational data from the BASS DR2.
In the bottom-right corner of each subplot, two coverage percentages are provided, each associated with a total object count ($N_{\rm L}$ or $N_{\rm G}$):
Percentages indicate the coverage fraction, defined as the proportion of the plotted data points ($N_{\rm L}$) and the total BPT-selected AGN sample ($N_{\rm G}$) that fall within the boundaries of the model grid (see \S~\ref{bass_data}). Note that while the total sample sizes $N_{\rm L}$ and $N_{\rm G}$ (denominators) remain constant across all panels, the coverage percentages vary depending on the specific model parameters (\Ne\ and $\aox$) used.
    }
  \label{fig_2}
\end{figure*}

\begin{figure*}
    \centering
    \includegraphics[width=1.0\textwidth]{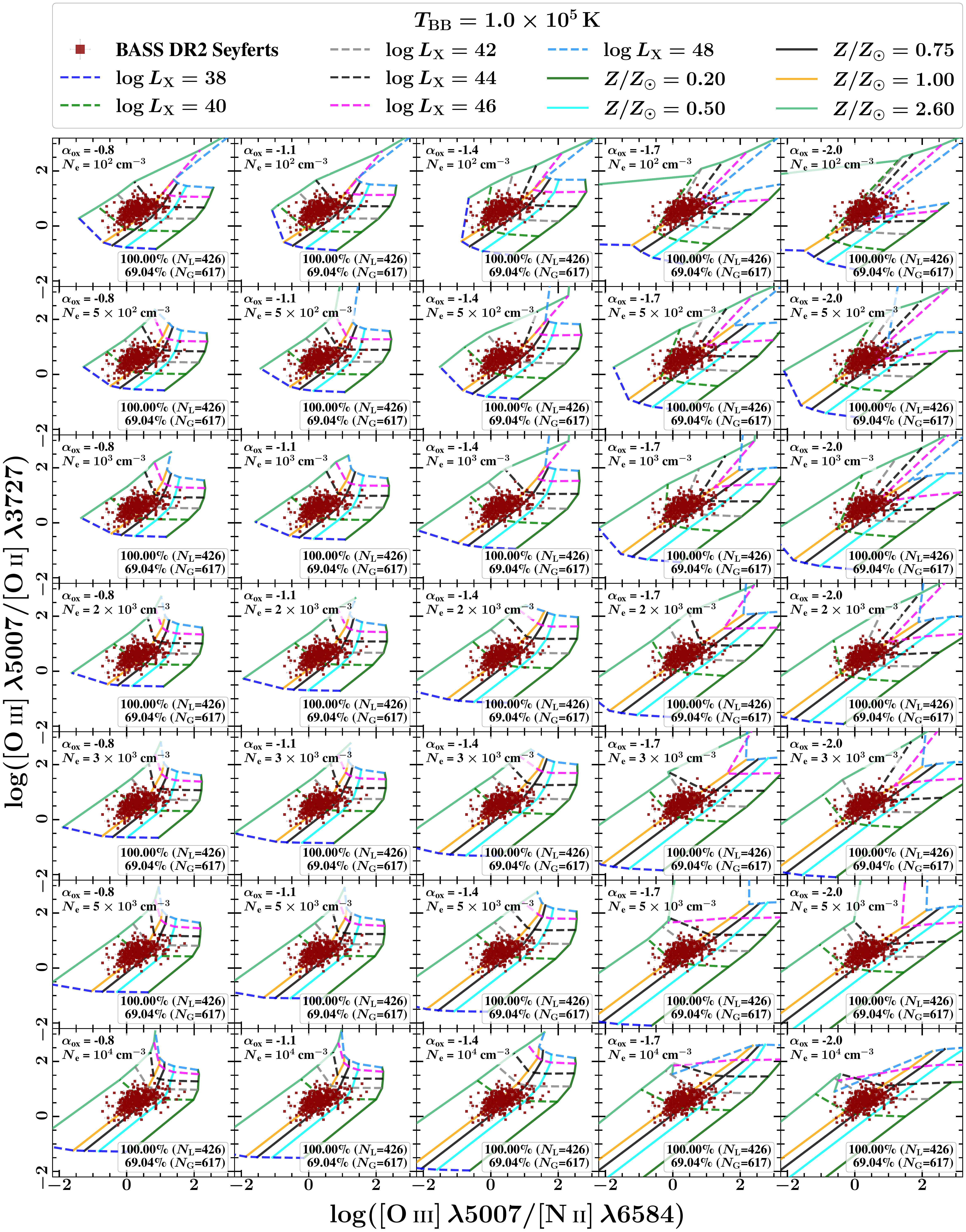}
 \caption{Same as Figure~\ref{fig_2} but for the diagnostic diagram of log(\oiii$\lambda5007$/\oii$\lambda 3727$) versus log(\oiii$\lambda 5007$/ \nii$\lambda6584$).
    }
    \label{fig_3}
\end{figure*}

\section{Results}
\label{results}

We compare the predictions of our  grid of photoionization models with observational data from the BASS DR2 sample, using two distinct diagnostic diagrams, presented in Figures~\ref{fig_2} and  \ref{fig_3}, to determine the range of conditions  in NLRS of our sample. 
The columns vary in $\alpha_{\rm ox}$ from $-0.8$ (left) to $-2.0$ (right), while the rows  vary in \Ne\ from $\rm 10^2\, to \, 10^4\,cm^{-3}$ (top to bottom). The percentages shown quantify the consistency between the photoionization models and the observational data. For each panel, the top percentage indicates the coverage fraction of the plotted data points, defined as the proportion of the total subsample ($N_{\rm L}=426$) that falls within the model grid. The bottom percentage indicates the coverage fraction relative to the entire BPT-selected AGN sample ($N_{\rm G}=617$). We note that $N_{\rm L}$ and $N_{\rm G}$ represent the constant total sample sizes (see \S~\ref{bass_data}), while the displayed percentages vary depending on the specific model parameters used in each panel, providing a measure of the performance of the model against the full observational dataset. The use of diagnostic diagrams to derive our calibrations is discussed in the following.

\subsection{The \texorpdfstring{\OIIIOII\ versus \NIIHa}{[O III]/[O II] versus [N II]/Halpha} diagnostic}

Figure~\ref{fig_2} presents the  log(\oiii$\lambda5007/$\oii$\lambda 3727$) versus log(\nii$\lambda6584$/\halpha) diagnostic diagram.  This diagram utilizes the \NIIHa\ line ratio, a metallicity indicator \citep{1994ApJ...429..572S},
whose sensitivity is enhanced in the high-metallicity environments typical of AGN host galaxies due to the secondary production of nitrogen \citep[e.g.][]{Groves2006a, 2020MNRAS.496.2191F, Carvalho2020, 2024MNRAS.534.3040D}. The \OIIIOII\ ratio, is a powerful tracer of $U$ and \Ne. While a higher ionization state (O$^{2+}$) is favoured over a lower one (O$^{+}$) by a more intense radiation field  \citep[e.g.][]{McGaugh1991}, the ratio is also strongly dependent on \Ne\ due to the low critical density of the \oii\ doublet ($\approx 10^{3.7}$ cm$^{-3}$; \citealt{Vaona2012}).

In each panel, solid lines trace models of constant metallicity, while dashed lines trace models of constant X-ray luminosity. The overlaid BASS DR2 data predominantly cluster in the upper region of the diagram, corresponding to near-solar to super-solar metallicities $(Z/\rm Z_{\odot} \gtrsim0.75)$ and higher densities ($N_{\rm e} \gtrsim 1000$ cm$^{-3}$). However, as shown by the coverage fractions ($N_{\rm L}$ versus $N_{\rm G}$), the data display an extended distribution, with a population of objects scattering downwards into regions of lower \OIIIOII\ ratios. While these lower values align with the predictions of models with softer ionizing continua (which shift the model grids downwards), relying on such soft models causes the grid to detach from the upper region.
This is quantitatively supported by the coverage fractions shown in each panel of Figure~\ref{fig_2}. For instance, with a hard ionizing continuum ($\aox = -0.8$) and high density ($N_{\rm e} = 10^4 \text{ cm}^{-3}$), the model grid extends sufficiently to encompass 100\,\% of the plotted data points ($N_L=426$).  In contrast, a very soft continuum ($\aox = -2.0$) at the same density shifts the grid downwards, losing coverage of the main cluster and enclosing only 90.14\,\% of the sample. 
For this specific diagnostic, models with hard-to-intermediate ionizing continua ($\aox \gtrsim -1.4$) provide the best and most stable calibration framework. This is because these models successfully trace the primary locus of the BASS DR2 data across the full metallicity range, from the sub-solar regime to the high-metallicity cluster.

\subsection{The \texorpdfstring{\OIIIOII\ versus \OIIINII}{[O III]/[O II] versus [O III]/[N II]} diagnostic}
Figure~\ref{fig_3} presents the log(\oiii$\lambda 5007$/\oii$\lambda 3727$) versus log(\oiii$\lambda5007$/\nii$\lambda6584$) diagnostic diagram. In this projection, the \OIIINII\ ratio functions as a sensitive metallicity indicator. At a fixed ionization level, this ratio decreases strongly with increasing metallicity, a trend primarily driven by the secondary nucleosynthesis of nitrogen which enhances \nii\ emission relative to oxygen \citep[e.g.][]{Kewley2002}.

The grids in Figure~\ref{fig_3} clearly show that metallicity and X-ray luminosity (our proxy for the ionization parameter) produce more distinct vectors in this parameter space  compared to Figure~\ref{fig_2}, significantly improving the separation of parameters. While the vectors are not perfectly orthogonal for all spectral indices (particularly in the upper-right region for softer continua, where grid lines become more parallel and overlap), this diagnostic effectively reduces the severe degeneracies that affect Figure~\ref{fig_2}, thereby providing a more reliable method for disentangling metallicity from ionization effects across the majority of the sample. A qualitative inspection demonstrates that the BASS DR2 data are well-reproduced by the model grids across the entire range of densities, metallicities, and X-ray luminosities. However, significant differences arise depending on the ionizing continuum. While models with harder spectra ($\aox \gtrsim -1.4$) provide a dynamic range that encompasses the data distribution, the grids for the soft continua ($\aox = -2.0, -1.7$) are noticeably more overlapping or compressed. This compression is supported by the analysis, which finds that the inherent degeneracies within the soft models prevent the stable interpolation of metallicity, leading to their exclusion from the final calibration (as will be discussed in \S~\ref{smetal}).

Together, these diagrams (Figures~\ref{fig_2} and~\ref{fig_3}) demonstrate that reproducing the full distribution of the observed sample requires varying secondary parameters (such as electron density and spectral index). This is consistent with predictions from comprehensive AGN photoionization models \citep[e.g.][]{Groves2004b, Feltre2016} rather than relying on a single fixed model configuration, for instance, using models that assume a constant $U$ to derive metallicities from observed ELRs. The scatter in the BASS DR2 data points to a genuine diversity in the physical conditions of NLRs, particularly in gas density and metallicity, even among a sample of similar type of AGNs.

\section{Metallicity calibration}
\label{smetal}
The final metallicity calibrations, presented in Figure~\ref{fig_4}, are the result of a quantitative analysis designed to translate the two-dimensional information from the diagnostic diagrams (Figures~\ref{fig_2} and \ref{fig_3}) into direct, one-dimensional relations between strong-line indices and gas-phase metallicity ($Z/\rm Z_{\odot}$). A critical prerequisite to this derivation was an assessment of model grid stability. While models with soft ionizing continua ($\alpha_{\mathrm{ox}} = -2.0, -1.7$) provided partial coverage of the diagnostic planes, they introduced significant degeneracies that precluded robust metallicity interpolation, often yielding unstable fits that extrapolated to unphysical regimes. Consequently, to ensure the reliability of the derived relations, the final calibrations were generated exclusively using models with harder ionizing continua ($\alpha_{\mathrm{ox}} = -1.4, -1.1, -0.8$).

Following the methodology of \citet{2011MNRAS.415.3616D}, we employed the comprehensive grid of {\sc cloudy} photoionization models (described in \S~\ref{models}) as a multi-dimensional reference frame.  Within the diagnostic diagrams (Figures~\ref{fig_2} and \ref{fig_3}), this grid functions as a map, translating the observed two-dimensional line-ratio space into the intrinsic physical parameters of the models, primarily metallicity ($Z/\rm Z_{\odot}$) and X-ray luminosity ($\Lx$). The BASS DR2 observational data were then positioned on these diagrams according to their measured line ratios to facilitate interpolation.

We derive the calibrations using the following three-step procedure:
\begin{itemize}
\item Grid mapping: The {\sc cloudy} model grid is projected onto the observational diagnostic plane (e.g. \OIIIOII\ versus \nii) as a function of both $\Lx$ and $Z$.
\item Interpolation: For each observed object with measured coordinates ($x_{\rm obs}, y_{\rm obs}$) and luminosity $L_{\rm X,obs}$, we interpolate the metallicity ($Z_{\rm est}$) from the surrounding model grid points $(x_{\rm mod}, y_{\rm mod}, Z_{\rm mod})$. This yields the theoretical metallicity required to reproduce the observed position given the specific luminosity of the object.
\item Function fitting: We perform a polynomial fit to the relation between the observed line ratios and the individually interpolated $Z_{\rm est}$ values to generate the final calibration functions.
\end{itemize}
We derive a simple functional form to describe the relation between a given logarithmic line ratio $x$ and the derived metallicity $Z_{\rm est}$. The general functional form of the polynomial fit to the set of interpolated data points ($x$, $Z_{\text{est}}$) is given by
\begin{equation}
    (Z/ {\rm Z_{\odot}}) = \sum_{n=0}^{N} c_n \mathcal{F}_n(x)
  \label{curve_fit}
\end{equation}
where $c_n$ represents the best-fit coefficients and the piecewise basis function $\mathcal{F}_n(x)$ is defined as:
\begin{equation}\mathcal{F}_n(x) =
\begin{cases}
\delta_{n,0} + \delta_{n,1} \cdot c_n^{x-1} & \text{for $\ntwo$ (} N=1 \text{)}\\
x^n & \text{for $\otnt$ (} N=2 \text{)}.
\end{cases}
\label{curve_fit2}
\end{equation}
This is accomplished using the non-linear least squares method, which yields the best-fit coefficients ($c_n$) 
and their corresponding 1$\sigma$ uncertainties. The resulting functions are shown as solid and dashed lines in Figure~\ref{fig_4}, with the derived coefficients listed in Table~\ref{table_2}. For the $\ntwo$ diagnostic, we adopt the exponential parameterization in Equation~\ref{curve_fit2} to explicitly model the steep, monotonic rise associated with secondary nitrogen production at high metallicities. This form ensures robust statistical fits that are free from unphysical inflection points. In contrast, we employ a second-order polynomial for the $\otnt$ diagnostic to accurately capture the curvature and saturation effects driven by electron cooling, providing a more accurate representation of the data compared to linear or higher-order models.

Finally, to investigate the systematic effects of other physical parameters, we repeated the entire process of interpolation and fitting procedure for different subsets of the full model grid. As shown in the rows of Figure~\ref{fig_4}, we performed this analysis independently for subsets binned by the $\aox$, \Ne, $\Lx$, and the $T_{\rm BB}$. The  final step in our script involves a stability check, thus, any fitted curve that extrapolated to physically unrealistic metallicity values outside the valid plot range was automatically discarded to ensure the robustness of the final presented calibrations. A `Global fit', derived using the entire model grid simultaneously, was also calculated and shown as a dashed black line in each panel for reference.

A detailed analysis of Figure~\ref{fig_4} reveals the systematic effects of the different model parameters on the final calibrations. The four rows in the figure  examine the robustness of the calibrations against the ionizing spectral index ($\aox$), the electron density (\Ne), the X-ray luminosity ($\Lx$), and the Big Blue Bump temperature ($T_{\rm BB}$). The $\ntwo$ diagnostic exhibits a weak-to-moderate dependence on both $\aox$ and \Ne, and  a systematic dependence on $T_{\rm BB}$. In contrast, the $\otnt$ diagnostic is remarkably robust against variations in all three of these parameters ($\aox$, \Ne, and $T_{\rm BB}$). However, both diagnostics show a strong, systematic secondary dependence on $\Lx$ as seen in the third row panels, though the trends are opposite and the model tracks remain distinctly separated. The model tracks are distinctly separated. For the $\ntwo$ diagnostic, lower luminosity models (i.e.
$38 \lesssim \log \Lx < 42$, blue points) predict a higher metallicity  for a fixed LR compared to higher luminosity models (i.e. $44 \lesssim \log \Lx \lesssim 48$, red points). For the $\otnt$ diagnostic, this trend is reversed, with higher luminosity models predicting a higher metallicity for a fixed LR. In the following subsections, we present each of the resulting metallicity calibrations.

\subsection{The \texorpdfstring{\boldmath $\ntwo$}{N2} diagnostic: \texorpdfstring{\ratioZZsun\ versus \NIIHa}{Z/Z\_sun versus [N II]/Halpha}}
\label{n2_diag}
\begin{figure*}
    \centering
\includegraphics[width=1.0\textwidth]{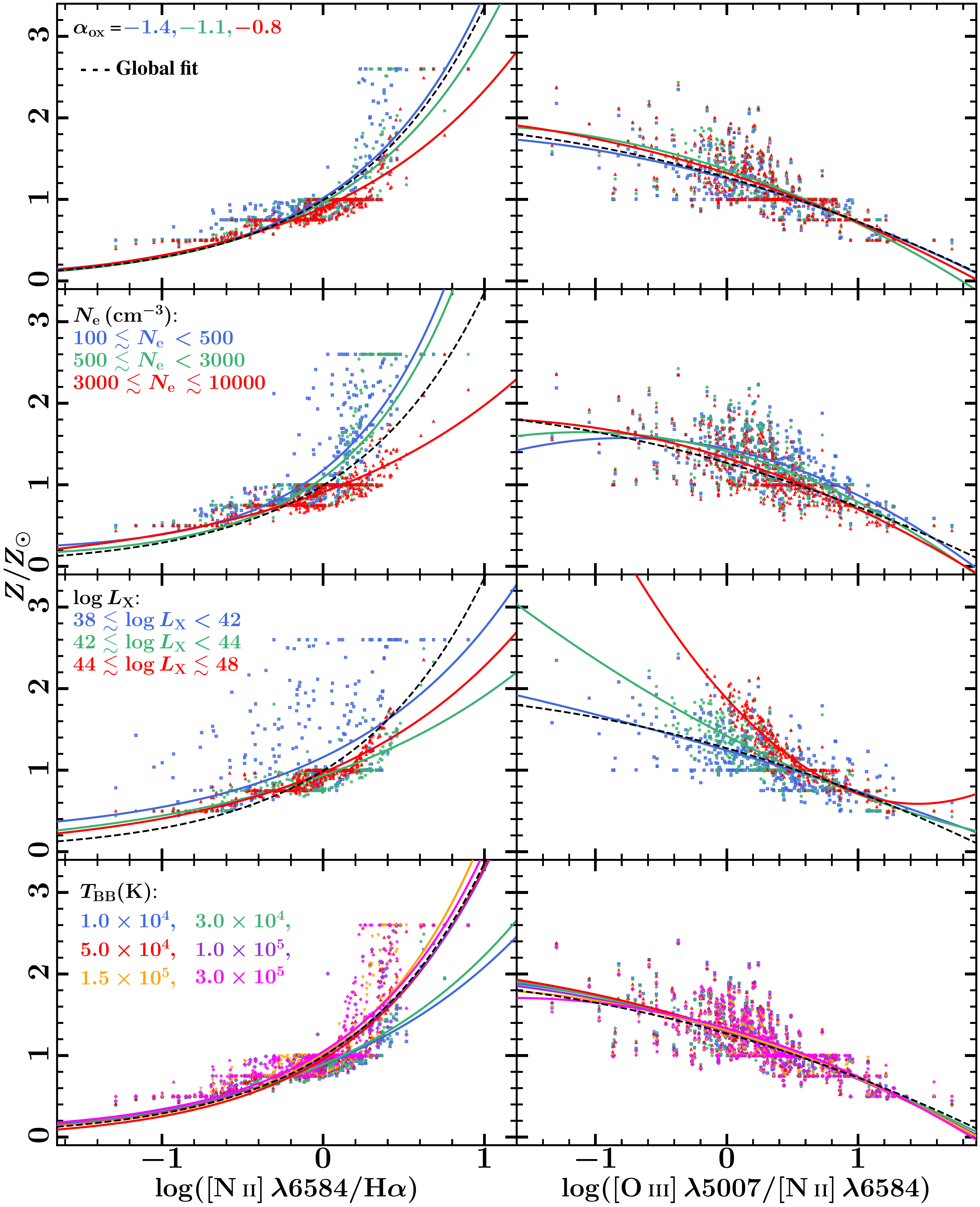}
\vspace{-10pt} 
\caption{Metallicity calibrations based on the $\ntwo$ and $\otnt$ strong-line indices, derived from interpolating photoionization model results onto the observed BASS DR2 data.
{\it Left column:} \ratioZZsun as a function of the log(\nii$\lambda6584$/\halpha) ratio ($\ntwo$).
{\it Right column:} \ratioZZsun\ as a function of the log(\oiii$\lambda 5007$/\nii$\lambda6584$) ratio ($\otnt$).
    In both columns, each panel explores the effect of a different model parameter on the estimated metallicity. The colored markers represent the interpolated metallicity estimates ($Z_{\rm est}$) obtained for each observational data point from a subset of the model grid, while the lines are the corresponding polynomial fits (Equations~\ref{cal_1} and \ref{cal_2}). The dashed black line in each panel represents the global fit using all model data. From top to bottom, the rows vary the $\aox$, \Ne\ in cm$^{-3}$, $\log \Lx$, and the $T_ {\rm BB}$ in K. The coefficients for each fit are presented in Table~\ref{table_2}.}
\label{fig_4}
\end{figure*}

The left column of Figure~\ref{fig_4} shows our first new calibration, based on the widely used
$\ntwo = \log([\rm N\,\textsc{ii}]\lambda6584/\rm H\alpha)$, which plots the estimated metallicity against it. 
The global relation derived from all model points, represented by the dashed black line in the left panels of Figure~\ref{fig_4}, can be expressed as:
 \begin{eqnarray}
       \begin{array}{l@{}l@{}l}
(Z/{\rm Z_{\odot}})\, & = \,  &  c_1^x + c_0,   \\
       \end{array}
\label{cal_1}
\end{eqnarray}
where $ x = \ntwo$.
Consistent with established results for chemically enriched gas \citep[e.g.][]{Kewley2002,Carvalho2020}, our calibration yields a residual dispersion of $1\sigma \approx 0.22$ dex across the entire metallicity range covered by the models. This relation is semi-empirically derived over the index range $-1.29 \lesssim \ntwo \lesssim 0.90$. While the diagnostic serves as a robust tracer for super-solar metallicities where the gradient is steep, we caution that the correlation flattens significantly at lower index values, which reduces sensitivity and restricts the precision of this diagnostic in the low-metallicity regime.

The left column of Figure~\ref{fig_4} explores the robustness of this diagnostic against the model parameters. The $\ntwo$ calibration exhibits a weak-to-moderate dependence on both $\aox$ and \Ne. For intermediate spectral indices $(\aox = -1.4)$, the resulting calibration fit is systematically higher than for harder indices ($\aox = -0.8, -1.1$). This indicates that as the ionizing field becomes less hard, it becomes less efficient at exciting \nii\ via X-ray heating; consequently, a higher intrinsic metallicity is required to produce the same observed line ratio.

The most significant secondary dependence is on $\Lx$, as shown in the third left-row panel of Figure~\ref{fig_4}. This dependence is considered dominant because it induces systematic vertical offsets throughout the entire calibration range ($-1.29 \lesssim \ntwo \lesssim 0.90$), whereas the divergence driven by other model parameters (e.g. $\aox$, \Ne, $T_{\mathrm{BB}}$) become prominent primarily in the high-metallicity regime ($\ntwo > -0.3$).
We observe that the vertical separation between calibration curves for distinct luminosity ranges is regime-dependent. At lower index values ($\ntwo \lesssim -0.3$), the systematic offset is $\sim 0.2 Z_{\odot}$ ($\sim 0.30$ dex). However, at higher index values ($\ntwo > -0.3$), this separation increases substantially, reaching up to $0.5 Z_{\odot}$ ($\sim 0.16$ dex). This magnitude of separation applies not only when contrasting the low-luminosity bin ($38 \lesssim \log \Lx < 42$) versus the high-luminosity bin ($44 \lesssim \log \Lx \lesssim 48$), but also characterizes the deviation of the intermediate-luminosity range ($42 \lesssim \log \Lx < 44$) against either extreme.

At lower index values ($\ntwo \lesssim -0.3$), the data display a clear, monotonic stratification. For a fixed $\ntwo$ value, more luminous AGNs ($44 \lesssim \log \Lx \lesssim 48$) yield systematically lower estimated metallicities compared to their lower-luminosity counterparts ($38 \lesssim \log \Lx < 42$) \citep[e.g.][]{Armah2023,Armah2024}. 
This suggests that the intense ionizing radiation field in high-luminosity sources drives deeper penetration of X-ray photons, altering the ionization structure and increasing the heating rate in the partially ionized zone \citep[e.g.][]{Groves2004a}. This elevates the electron temperature, thereby boosting the collisional excitation rates for the nitrogen transitions \citep[e.g.][]{Osterbrock2006};
thus, a lower chemical abundance is sufficient to produce the same observed $\ntwo$ ratio that would require higher abundance in a lower-luminosity system.
However, at higher index values ($\ntwo > -0.3$), this stratification becomes disordered. The intermediate luminosity curve ($42 \lesssim \log \Lx < 44$) exhibits crossover behaviour, deviating from the monotonic trend and rising above the high-luminosity calibration. This likely reflects the transition into the high-metallicity saturation regime, where increased cooling efficiency and line saturation disrupt the linear response of the diagnostic to the ionizing field \citep[e.g.][]{Maiolino2008,Curti2017}. Consequently, in this high-metallicity regime, the systematic variation driven by luminosity becomes complex relative to other parameters.

Finally, the bottom-left panel of Figure~\ref{fig_4} explores the effect of the $T_{\mathrm{BB}}$. The calibration shows a clear systematic dependence on this parameter, which remains consistent across the full dynamic range. The fits for different temperatures are clearly stratified, with hotter models ($T_{\mathrm{BB}}\gtrsim5.0\times10^{4}$ K) systematically offset from the cooler models ($T_{\mathrm{BB}}\lesssim3.0\times10^{4}$ K). This indicates that the $\ntwo$ diagnostic is persistently sensitive to variations in the accretion disk temperature, even in regimes where the luminosity dependence becomes degenerate.

The $\ntwo$ index is strongly dependent on the $ \rm N/O$ ratio, which can vary independently of $\rm O/H$  due to factors such as the star formation history \citep{VilaCostas1993}. This dependence can introduce scatter and systematic uncertainties in metallicity estimates, particularly when applying a single $\ntwo$ calibration to AGNs,  where the nitrogen has mainly a secondary origin.

\subsection{The \texorpdfstring{\boldmath $\otnt$}{O3N2} diagnostic: \texorpdfstring{\ratioZZsun\ versus \OIIINII}{Z/Zsun versus [O III]/[N II]}}
\label{03n2_diag}
In the right column of Figure~\ref{fig_4}, we present our calibration relating the estimated $Z/\rm Z_{\odot}$ to the $\otnt$ index.   A clear and tight anti-correlation is observed, establishing this ratio as a strong tracer of metallicity. This relation arises because, in a given ionization state, an increase in overall $Z/ \rm Z_{\odot}$ is accompanied by an even faster increase in the $\rm N/O$ ratio, causing the \OIIINII\ ratio to drop. 
The global fit for this diagnostic, shown as the dashed black lines in the right column of Figure~\ref{fig_4}, is given by:
 \begin{eqnarray}
       \begin{array}{l@{}l@{}l}
(Z/ \rm Z_{\odot})\, & = \, &  c_2 y^2 + c_1 y + c_0, \\
       \end{array}
\label{cal_2}
\end{eqnarray}
where $y = \otnt$.
The derived relation for the $\otnt$ index shows a monotonic decrease with increasing metallicity, characterized by a residual dispersion of $1\sigma \approx 0.20$ dex  over the index range $-1.33 \lesssim \otnt \lesssim 1.71$. This scatter is comparable to that of the $\ntwo$ diagnostic, indicating that while $\otnt$ is sensitive to ionization parameter variations, the new calibration effectively mitigates these effects within the modeled parameter space.

The systematic effects of the spectral index, electron density, and Big Blue Bump temperature are investigated in the top two and bottom panels of the right column of Figure~\ref{fig_4}. Notably, within the considered range of harder ionizing continua ($\aox = -1.4, -1.1, -0.8$), as well as across the different bins of electron density and $T_{\rm BB}$, the resulting curve fits are nearly indistinguishable from one another and from the global fit (dashed black line). This stability demonstrates that the $\otnt$ diagnostic index is remarkably robust against variations in the shape of the ionizing continuum, gas density, and accretion disk temperature, offering a significant advantage over many other metallicity indicators.

However, the third panel from the top reveals a significant secondary dependence on X-ray luminosity. The data are clearly stratified by X-ray luminosity, where, for a fixed $\otnt$ value, higher X-ray luminosity values systematically correspond to higher derived metallicities and vice versa \citep[e.g.][]{Nagao2006b, Matsuoka2009}. Similar to the $\ntwo$ diagnostic, the vertical separation between the calibration curves for distinct luminosity ranges is regime-dependent. In the low-metallicity regime ($\otnt \gtrsim 1.2$), the separation spans $\sim 0.5 Z_{\odot}$ ($\sim 0.54$ dex). In the intermediate regime ($0.7 < \otnt < 1.2$), the offset transitions smoothly, increasing substantially to reach up to $\sim 1.0 Z_{\odot}$ ($\sim 0.20$ dex) in the high-metallicity regime ($\otnt \lesssim 0.7$). The distinct separation between the second order fits for each luminosity bin confirms that $\Lx$ must be included as a second parameter to achieve an accurate metallicity estimate with this diagnostic.

\section{Discussion}
\label{discussion}
\subsection{Implications of the \texorpdfstring{\boldmath $\Lx$}{Lx}--based approach}
\label{LX_vs_U}
Departing from standard parameterizations based on the local ionization parameter, $U$, we utilize the approach detailed in \S~\ref{method} to connect our model grids (Figures~\ref{fig_2} and \ref{fig_3}) directly to the observable X-ray luminosity ($\log \Lx = 38-48$), thereby linking the diagnostics to the intrinsic energetic output and fundamental accretion characteristics that define the BASS DR2 sample (see \S~\ref{bass_data}).

The third row panels in Figure~\ref{fig_4} reveal that while both diagnostics have a strong, systematic secondary dependence on X-ray luminosity \citep[also see][]{Matsuoka2011,Oh2017}, their stratification trends are in direct opposition.
For the $\ntwo$ diagnostic (left panel), the models are stratified such that for a fixed $\ntwo$ value, the lower luminosity models (blue) predict a higher metallicity than the higher luminosity models (red and green), with systematic vertical offsets ranging from $\sim 0.2 Z_{\odot}$ ($\sim 0.30$ dex; $\ntwo \lesssim -0.3$) to $\sim 0.5 Z_{\odot}$ ($\sim 0.16$ dex; $\ntwo >  -0.3$). We note that the horizontal “striping” seen in the interpolated data points is an artifact of the discrete metallicity steps [($Z/\rm Z_\odot) = 0.2, 0.5, 0.75, 1.0, 2.6$] in the model grid nodes.
This `reversed' trend is consistent with the classic $\Lx$--$Z$ photoionization
degeneracy, as illustrated in Figure~\ref{fig_5} (top-left panel),
where a higher $\Lx$ is required to produce the same ionization state in a
less-efficiently cooled, lower-metallicity gas \citep[e.g.][]{Groves2004a}.

Conversely, the $\otnt$ diagnostic (right panel) shows the exact opposite trend: a higher luminosity systematically corresponds to a higher derived metallicity, exhibiting pronounced vertical offsets ranging from $\sim 0.5 Z_{\odot}$ ($\sim 0.54$ dex) in the low-metallicity regime to $\sim 1.0 Z_{\odot}$ ($\sim 0.20$ dex) in the high-metallicity regime. 
While this behaviour mimics the well-known empirical MZR where higher luminosity and stellar mass correspond to higher metallicity, it is in direct contrast with the trend exhibited by the $\ntwo$ calibration, which is driven by the classical ionization parameter degeneracy. This fundamental discrepancy is a significant finding.
The fact that two diagnostics applied to the same dataset yield strong but diametrically opposed physical conclusions (i.e. metal-poor inflow versus mass-metallicity co-evolution) strongly suggests that both indices are still significantly contaminated by ionization-parameter-driven effects. 

The $\Lx$ dependency in the $\ntwo$ diagnostic appears to be dominated by the classic $\Lx$--$Z$ photoionization degeneracy. The $\Lx$ dependency in the $\otnt$ diagnostic, on the other hand, concurrently mimics the expected MZR and luminosity-metallicity relation \citep[e.g.][]{Nagao2006b,Ellison2008,Matsuoka2009}.  There is a semi-empirical precedent for this trend; the luminosity-metallicity dependence we observe here in our optical strong lines (specifically the $\otnt$ index) has been independently established in the literature using UV emission line diagnostic diagrams, such as C\iv$\lambda1549/$He\ii$\lambda$1640 versus C\iii]$\lambda1909/$C\iv$\lambda$1549 \citep[e.g.][]{Nagao2006b, Matsuoka2009}.
In this case, elevated ionization parameters (driven by higher nuclear luminosities) increase excitation levels and systematically shift emission line ratios independent of metallicity. Similar systematic offsets have been observed in high-redshift star-forming galaxies, where elevated ionization parameters and densities shift the $\otnt$ ratio relative to local calibrations \citep{Sanders2016}. However, given the contradictory results, neither calibration in isolation can be considered to have successfully disentangled the degeneracies to probe the true underlying metallicity. We therefore note that both the $\ntwo$ and $\otnt$ calibrations are heavily impacted by degeneracies with the ionization state of the gas (proxied by $\Lx$), introducing systematic errors of up to $\sim 1.0$ dex if neglected. In principle, a combined ratio or a differential index using both $\ntwo$ and $\otnt$ could break the degeneracy with $\Lx$. While such a joint calibration falls outside the scope of this current work, it represents a highly promising avenue for future AGN metallicity studies. This dependency highlights the necessity of the multi-parameter approach ($\Lx$ + LR) proposed in this work, which explicitly accounts for such variations in the ionization parameter to recover robust metallicity estimates. Furthermore, it underscores the need for diagnostics constructed to be robust against such variations, such as the calibrations for the $\no$ and $\ns$ indices that will be explored in Paper II.

\begin{table}
\centering
\setlength{\tabcolsep}{1.5pt} 
\caption{The values of the best-fit coefficients ($c_2$, $c_1$, and $c_0$) and their $1\sigma$ uncertainties resulting from the curve fits (Equations~\ref{cal_1} and \ref{cal_2}) obtained from the estimations in Figure~\ref{fig_4}. The fits are provided for the $\ntwo$ and $\otnt$ diagnostics, stratified by different model parameter subsets. The final row, Global fit, lists the coefficients derived from the entire model grid without parameter discrimination.}
\label{table_2}
\resizebox{\columnwidth}{!}{%
\begin{tabular}{@{}l | r@{$\pm$}l r@{$\pm$}l | r@{$\pm$}l r@{$\pm$}l r@{$\pm$}l@{}}
\hline
\multicolumn{1}{c}{} & \multicolumn{4}{c}{\textbf{$\ntwo$}} & \multicolumn{6}{c}{\textbf{$\otnt$}} \\
\hline
Parameter & \multicolumn{2}{c}{$c_1$} & \multicolumn{2}{c|}{$c_0$} & \multicolumn{2}{c}{$c_2$} & \multicolumn{2}{c}{$c_1$} & \multicolumn{2}{c}{$c_0$} \\
\hline
\multicolumn{1}{l}{$\alpha_{\rm ox}$} & \multicolumn{2}{c}{} & \multicolumn{2}{c}{} & \multicolumn{2}{c}{} & \multicolumn{2}{}{} & \multicolumn{2}{c}{} \\
$-1.40$ & 3.50 & 0.15 & 0.02 & 0.01 & -0.09 & 0.03 & -0.44 & 0.03 & 1.25 & 0.01 \\
$-1.10$ & 3.08 & 0.11 & -0.04 & 0.01 & -0.13 & 0.03 & -0.53 & 0.03 & 1.36 & 0.01 \\
$ -0.80$ & 2.43 & 0.06 & -0.10 & 0.01 & -0.09 & 0.03 & -0.51 & 0.03 & 1.32 & 0.01 \\
\\
\multicolumn{1}{l}{$N_{\rm e}\, ({\rm cm^{-3}})$} & \multicolumn{2}{c}{} & \multicolumn{2}{c}{} & \multicolumn{2}{c}{} & \multicolumn{2}{c}{} & \multicolumn{2}{c}{} \\
 $\in [100, 500)$  & 4.76 & 0.27 & 0.18 & 0.02 & -0.22 & 0.04 & -0.35 & 0.03 & 1.44 & 0.02 \\
$\in [500, 3\,000)$ & 4.41 & 0.20 & 0.09 & 0.02 & -0.19 & 0.04 & -0.43 & 0.03 & 1.41 & 0.02 \\
$\in [3\,000, 10\,000]$ & 2.06 & 0.04 & -0.09 & 0.01 & -0.13 & 0.04 & -0.50 & 0.03 & 1.33 & 0.02 \\
\\
\multicolumn{1}{l}{$\log L_{\mathrm{X}}$} & \multicolumn{2}{c}{} & \multicolumn{2}{c}{} & \multicolumn{2}{c}{} & \multicolumn{2}{c}{} & \multicolumn{2}{c}{} \\
$\in [38, 42)$ & 2.58 & 0.20 & 0.16 & 0.02 & -0.03 & 0.04 & -0.47 & 0.03 & 1.24 & 0.02 \\
$\in [42, 44)$  & 1.98 & 0.05 & -0.06 & 0.01 & 0.11 & 0.05 & -0.83 & 0.05 & 1.42 & 0.01 \\
$\in [44, 48]$ & 2.30 & 0.05 & -0.03 & 0.01 & 0.61 & 0.06 & -1.77 & 0.07 & 1.88 & 0.02 \\
\\
\multicolumn{1}{l}{$T_{\rm BB}$ [K]} & \multicolumn{2}{c}{} & \multicolumn{2}{c}{} & \multicolumn{2}{c}{} & \multicolumn{2}{c}{} & \multicolumn{2}{c}{} \\
$1.0 \times 10^{4}$ & 2.20 & 0.04 & -0.11 & 0.01 & -0.10 & 0.03 & -0.51 & 0.03 & 1.32 & 0.01 \\
$3.0 \times 10^{4}$  & 2.34 & 0.05 & -0.11 & 0.01 & -0.08 & 0.03 & -0.50 & 0.03 & 1.30 & 0.01 \\
$5.0 \times 10^{4}$ & 3.36 & 0.11 & -0.04 & 0.01 & -0.08 & 0.03 & -0.52 & 0.03 & 1.31 & 0.01 \\
$1.0 \times 10^{5}$ & 3.28 & 0.12 & -0.01 & 0.01 & -0.09 & 0.03 & -0.49 & 0.03 & 1.29 & 0.01 \\
$1.5 \times 10^{5}$ & 3.71 & 0.14 & 0.02 & 0.01 & -0.11 & 0.03 & -0.48 & 0.03 & 1.31 & 0.01 \\
$3.0 \times 10^{5}$ & 3.46 & 0.14 & 0.05 & 0.01 & -0.14 & 0.03 & -0.46 & 0.03 & 1.33 & 0.01 \\
\hline
Global fit & 3.37 & 0.14 & -0.01 & 0.01 & -0.08 & 0.03 & -0.46 & 0.03 & 1.27 & 0.01 \\
\hline
\end{tabular}%
}
\end{table}

\subsection{Luminosity dependence and the influence of model parameters}
\begin{figure*}
\centering
\includegraphics[width=1.0\textwidth]{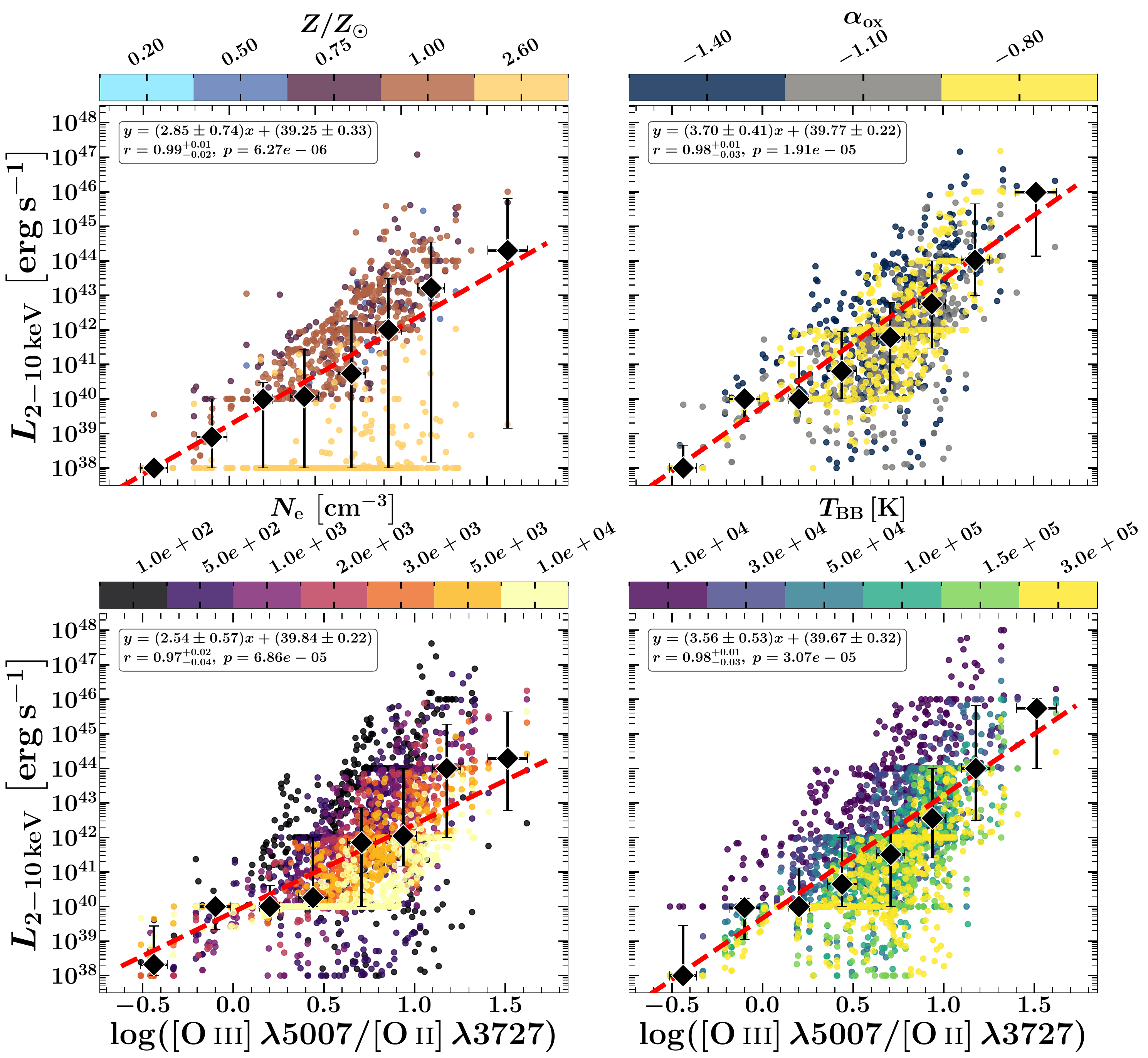}
    \caption{
        The X-ray luminosity versus the $\log ([\mathrm{O}\,\textsc{iii}]\,\lambda 5007/[\mathrm{O}\,\textsc{ii}]\,\lambda 3727)$ ELR. 
        The data points are derived from interpolating photoionization models onto the positions of observed galaxies in a multi-dimensional parameter space. Each panel is color-coded by a different model parameter to show its effect on the distribution. {\it Top-left panel:} points are color-coded by metallicity (\ratioZZsun). {\it Top-right panel:} points are color-coded by the spectral index ($\aox$). {\it Bottom-left panel:} points are color-coded by the electron density (\Ne\ in cm$^{-3}$). {\it Bottom-right panel:} points are color-coded by the characteristic peak temperature of the BBB component ($T_{\rm BB}$
  in K). In each panel, the black diamonds represent the median values in X-ray luminosity bins, with error bars indicating the 16th-84th percentile range. The red dashed line shows the linear fit to these binned median values. The equation for this fit ($y= mx + c$), along with its Pearson correlation coefficient ($r$) and $p$--$value$, are indicated in the top-left corners. For these equations, $y$ is the vertical axis (ordinate) value ($L_{2-10~\text{keV}}~[\text{erg s}^{-1}]$) and $x$ is the horizontal axis (abscissa) value [log(\oiii$\lambda5007/$\oii$\lambda 3727$)], $m$ is the slope, and c represents the y-intercept.}
    \label{fig_5}
\end{figure*}

Figure~\ref{fig_5} presents a semi-empirical diagnostic plane of the X-ray luminosity ($\Lx$) versus the log(\oiii$\lambda 5007/$\oii$\lambda 3727$).
To construct this figure, we applied the same interpolation methodology outlined in \S~\ref{smetal} for the derivation of the metallicity calibrations.
We observe a distinct positive correlation between $\Lx$ and the log(\OIIIOII). This trend is quantified by linear fits to the binned data (black diamonds), which consistently show strong, statistically significant correlations across all panels (with Pearson correlation coefficients $r \approx +0.97^{+0.02}_{-0.04}$ to $r \approx+0.99^{+0.01}_{-0.02}$  with extremely low $p$--$values$, $p\approx10^{-5}$). This implies that sources with higher X-ray luminosities tend to have higher ionization states. However, the considerable scatter in the individual data points is systematically driven by the physical conditions within the NLR, as explored in the four panels of the figure.

The top-left panel, which is color-coded by the gas-phase metallicity, $Z/\rm Z_{\odot}$, reveals a strong linear correlation in the binned data, described by the relation $y=(2.85 \pm 0.74)x+(39.25 \pm 0.33)$ with a Pearson correlation coefficient of $r = 0.99^{+0.01}_{-0.02}$ $\left(p=6.27\times10^{-6}\right)$. Here,  $x$ and $y$ represent the log(\oiii$\lambda5007/$\oii$\lambda 3727$) and the 2-10 keV X-ray luminosity ($L_{2-10~\text{keV}}$ in $\text{erg s}^{-1}$), respectively.  The panel shows that for a fixed log(\OIIIOII) ratio, sources with lower metallicities (blue/purple points) are systematically associated with higher X-ray luminosities, while higher metallicity gas can reach lower X-ray luminosities \citep[e.g.][]{Armah2023, Armah2024}.
This demonstrates the significant impact of metal content on the emergent LRs at a given X-ray luminosity. This stratification is a direct consequence of how metallicity regulates the gas temperature. Higher metallicity gas radiates energy more effectively, settling at a cooler equilibrium temperature  \citep[e.g.][]{Sutherland1993}. The emissivities of the \oii\ and \oiii\  lines respond differently to this temperature change; the \oii\ emission, arising from the ${}^2D^o \to {}^4S^o$ transition which requires an excitation energy of $\approx 3.32$ eV, is suppressed more strongly in this cooler gas than the \oiii\  nebular emission (transition ${}^1D_2 \to {}^3P$), which possesses a lower excitation energy of 2.51 eV \citep[e.g.][]{Osterbrock2006}. This preferentially boosts the log(\OIIIOII) ratio. Therefore, to match a fixed, observed log(\OIIIOII) ratio, this high-$Z$ (cool) gas must be ionized by a weaker radiation field (lower $\Lx$) to reduce the O$^{2+}$/O$^{+}$ fraction \citep[e.g.][]{Kewley2002} and bring the LR back down. Conversely, low-metallicity gas is hotter, which lowers the log(\OIIIOII) ratio, thus requiring a stronger ionizing source (higher $\Lx$) to compensate by creating more O$^{2+}$. This well-documented degeneracy between $Z$ and $U$ (proxied by $\Lx$ in this work) is a fundamental aspect of AGN photoionization modelling \citep[e.g.][]{Groves2004a}.

The top-right panel explores the influence of the spectral index, $\aox$. The binned data again show a robust linear relationship, with a fit of $y=(3.70 \pm 0.41)x+(39.77 \pm 0.22)$ and a correlation coefficient of $r=0.98_{-0.03}^{+0.01}$ $\left(p=1.91\times10^{-5}\right)$. However, the color-coded points reveal a more complex stratification than a simple linear trend. While the full range of $\aox$ values is present across several orders of magnitude of X-ray luminosity, indicating significant intrinsic scatter, the envelopes of the distribution are clearly defined by the hardness of the ionizing continuum. Specifically, the hardest continua (less negative $\alpha_{ox}$ values, shown in yellow/orange) predominantly occupy the higher $L_{2-10 \text{ keV}}$ and higher log(\OIIIOII) regime (upper-right portion) of the parameter space. This signifies that a harder continuum is required to produce a higher ionization state at the highest X-ray luminosities. In contrast, the softest continua (more negative $\aox$ values, shown in purple and blue) extend to the lower $L_{2-10 \text{ keV}}$ and lower log(\OIIIOII) regime (lower-left portion). This is expected, as a softer, less efficient ionizing source requires a lower overall luminosity to achieve a lower degree of ionization. This observation aligns with the well-established correlation between $\Lx$ and $\aox$ \citep[e.g.][]{2006AJ....131.2826S, Lusso2010} and implies that the hardness of the ionizing radiation plays a significant role in determining the observed emission LRs, consistent with findings on X-ray spectral indices in radio-loud (RL) and radio-quiet (RQ) AGNs \citep[e.g.][]{Just2007, Gupta2018, Gupta2020}.

The bottom-left panel, which plots the \OIIIOII\ ELR against X-ray luminosity, serves as a robust diagnostic of the physical conditions within the ionized gas, particularly its electron density and ionization state. The linear fit to the binned data, $y=(2.54 \pm 0.57)x+(39.84 \pm 0.22)$, reinforces the strong positive correlation $\left(r=0.97_{-0.04}^{+0.02}, p=6.86\times10^{-5}\right)$. The color-coding clearly shows that for a given $\Lx$, gas with lower density exhibits higher \OIIIOII\ ratios, while higher-density gas (purple points) systematically populates the lower envelope of the relation.
The relative strength of these two emission lines depends fundamentally on the abundance of doubly-ionized oxygen (O$^{2+}$) compared to singly-ionized oxygen (O$^{+}$). Creating O$^{+}$ from neutral oxygen requires photons with energies exceeding 13.62 eV, while subsequent ionization to O$^{2+}$ requires significantly more energetic photons, with energies above 35.12 eV \citep{Vaona2012}. Consequently, the \oiii$\lambda$5007 line traces regions exposed to a harder ionizing radiation field than the \oii$\lambda\lambda3726,3729$ doublet.
The \oii$\lambda\lambda3726,3729$ doublet has a relatively low critical density of $N_{\rm c}\sim10^{3.7} \: \rm cm^{-3}$, while the \oiii$\lambda$5007 line has a much higher critical density of $N_{\rm c}\sim10^{5.8} \: \rm cm^{-3}$ \citep{Vaona2012}. Above the critical density of the \oii\ doublet, collisional excited ions are more likely to lose energy via collisions with free electrons rather than through radiative decay \citep[e.g.][]{Osterbrock2006}. Therefore, in regions with higher-density gas (\Ne\ $\gtrsim 5000$ cm$^{-3}$), the emission from the \oii\ doublet is progressively suppressed. 
Although the \oiii\ line has a high critical density and is unaffected by collisional de-excitation at these densities, the behaviour of the \OIIIOII\ ratio is dominated by the ionization parameter, $U$. Since $U$ is inversely proportional to the gas density ($U \propto N_{\rm e}^{-1}$), higher-density gas is characterized by a lower ionization state, which reduces the relative abundance of $\rm O^{2+}$. This effect overpowers the suppression of \oii\ emission via collisional de-excitation (which would otherwise tend to increase the \OIIIOII\ ratio), resulting in the systematic decrease in the overall \OIIIOII\ ratio observed in the bottom-left panel of Figure~\ref{fig_5}. The opposite is true for lower-density gas, where the consequently higher $U$ drives a much higher \OIIIOII\ ratio \citep[e.g][]{Osterbrock2006,Dopita2013}.

The bottom-right panel, color-coded by the characteristic peak temperature of the BBB component ($T_{\rm BB}$), reveals a systematic, albeit complex, trend. A strong linear trend is again recovered in the binned data, with a fit of $y=(3.56 \pm 0.53)x+(39.67 \pm 0.32)$ and $r=0.98_{-0.03}^{+0.01}$ $\left(p=3.07\times10^{-5}\right)$.
Similarly to the other model parameters, this plot shows a systematic stratification of the data points based on $T_{\rm BB}$. Contrary to simple expectations, the models show that lower values of $T_{\rm BB}$ preferentially populate the upper envelope of the relation, associated with higher $\Lx$ and higher \OIIIOII\ ratios, while higher values of $T_{\rm BB}$ form the lower envelope, appearing more prominent in the lower-luminosity and lower-ratio regions of the diagram. This effect is expected because a hotter blackbody produces a harder ionizing continuum, which shifts the peak of its thermal emission to higher energies and enhances the production of highly ionized species.  In other words, models with hotter accretion disks (higher $T_{\rm BB}$) require less $\Lx$ to achieve the same ionization state (represented by the \OIIIOII\ ratio), thereby populating the lower envelope of the relation, whereas cooler disks (lower $T_{\rm BB}$) require higher X-ray luminosities to reach the same ratio, populating the upper envelope. 
This change in the ionization structure of the gas alters the emergent LRs in a way that is degenerate with the effect of a higher ionization parameter, making it challenging to separate the two effects using a single diagnostic \citep{2024ApJ...977..187Z}. This fundamental relation between the hardness of the ionizing source and the resulting ionization state of a nebula is a central tenet in nebular astrophysics \citep{Osterbrock2006}. Although this demonstrates that $T_{\rm BB}$ clearly influences the LRs for any given object, a key finding from a broader analysis of the diagnostic diagrams from $T_{\mathrm{BB}} = 1.0 \times 10^4\,\mathrm{K}$ to $T_{\mathrm{BB}} = 3.0 \times 10^5\,\mathrm{K}$ is that the overall morphology and coverage of the model grids are remarkably insensitive to the specific choice of $T_{\rm BB}$ within the explored range (i.e. $\aox = -1.4, -1.1, -0.8$). This is a {key} result because it establishes that while $T_{\rm BB}$ has a discernible effect, it is a second-order effect compared to the dominant drivers of metallicity, $\Lx$, $\aox$, and \Ne. The robustness of our model predictions against this key assumption of the ionizing continuum strengthens the foundation of our luminosity-based calibration method.

Figure~\ref{fig_5} deconstructs the physical drivers behind the observed LRs in AGN. It establishes a strong, positive correlation between the observable X-ray luminosity and the ionization state of the gas, traced by the log(\OIIIOII) ratio. More importantly, it demonstrates conclusively that the significant scatter around this relation is not random noise but is systematically governed by the key physical properties of the NLR. Specifically, a gas with lower metallicity, a softer ionizing continuum (less negative $\aox$), or lower density requires a higher $\Lx$ to produce the same ionization state. This provides the fundamental justification for the fact that luminosity must be treated as an important secondary parameter in metallicity calibrations to account for the diverse physical conditions within AGN hosts.

\subsection{Comparison between the  \texorpdfstring{\boldmath $\ntwo$}{N2} and  \texorpdfstring{\boldmath $\otnt$}{O3N2} calibrations}
\begin{figure}
    \centering
\includegraphics[angle=0.0, width=1.0\columnwidth]{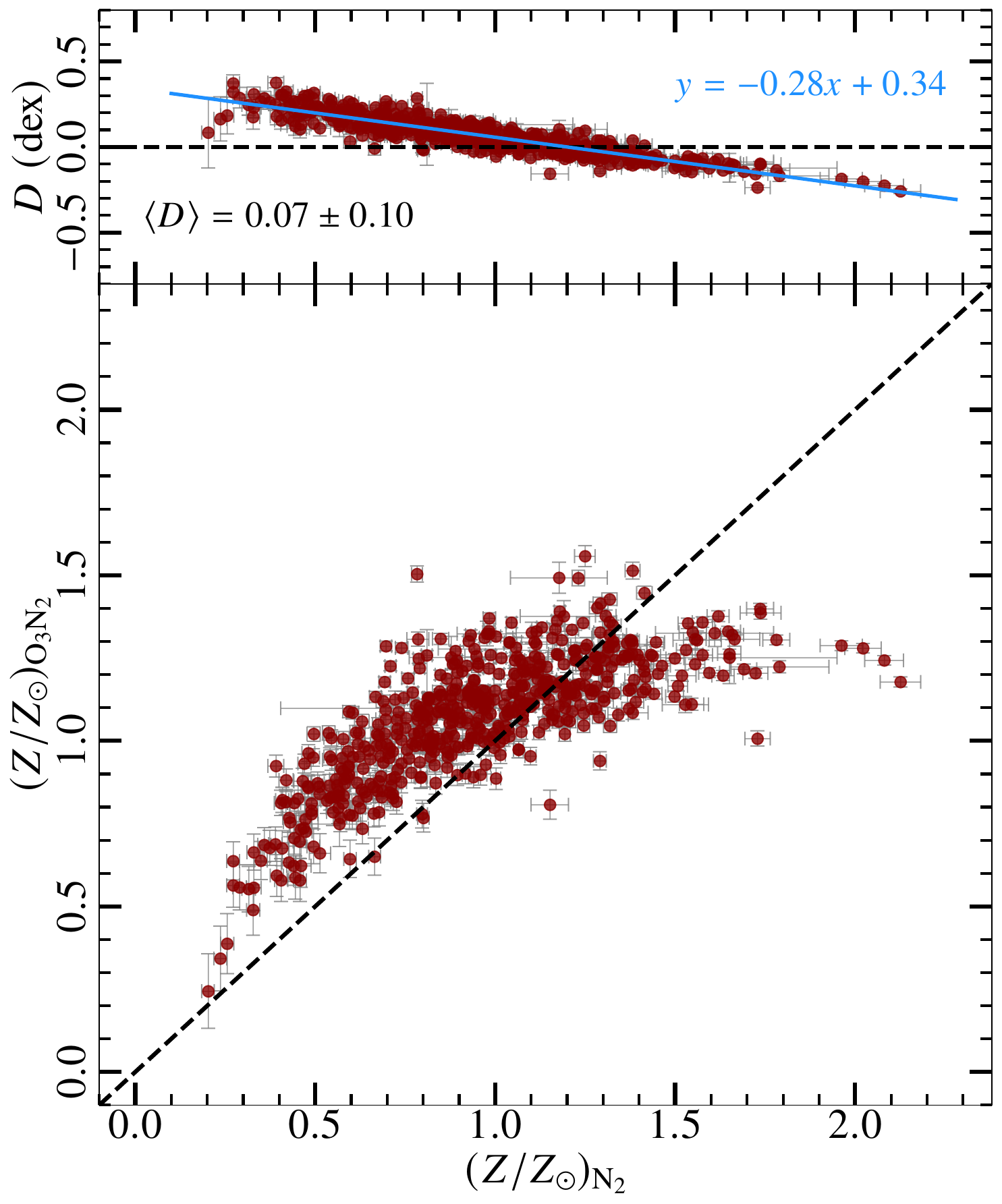}
\caption{{\it Bottom panel}: Comparison between the metallicity estimates derived from the $\ntwo$ and $\otnt$ diagnostics. The dashed black line represents the 1:1 relation. While the two estimates show a strong correlation with a Pearson coefficient of $r=0.76$ ($p = 5.20 \times 10^{-104}$) and a Spearman rank coefficient of $\rho=0.77$ ($p = 8.55 \times 10^{-111}$), calculated for a sample of $N=558$ objects, significant systematic deviations are observed. {\it Top panel}: the logarithmic difference ($D = \log(Z/Z_{\odot})_{\rm ordinate} - \log(Z/Z_{\odot})_{\rm abscissa}$) between the metallicity estimates from $\ntwo$ and $\otnt$. The dashed black line marks a zero difference, and the solid blue  line represents a linear regression to
these differences as indicated. The mean residual  ($\langle D \rangle$), with its value and standard deviation is also indicated. The residuals exhibit a clear metallicity--dependent trend ($y=-0.28x+0.34$), where the $\ntwo$ diagnostic predicts systematically lower metallicities than $\otnt$ in the low--$Z$ regime ($Z/Z_{\odot} \lesssim 1.0$) and higher metallicities in the high--$Z$ regime, with differences reaching up to $\sim 0.5$ dex at the extremes of the distribution.
    }
 \label{fig_6}
\end{figure}

Figure~\ref{fig_6} tests the internal consistency between the two new metallicity calibrations developed in this work: $Z/\rm Z_{\odot}$ derived from the $\ntwo$ and $\otnt$ indices. By applying both calibrations to the same sample of galaxies, we can assess their agreement and identify potential systematic biases. The bottom panel shows a direct comparison between the two metallicity estimates.
While there is a clear correlation, there is also considerable scatter, particularly at the extremes of the distribution; at lower metallicities ($Z/\rm Z_{\odot} \lesssim 1.0$), the $\otnt$ calibration tends to predict a higher metallicity than the $\ntwo$ calibration, whereas at high metallicities ($Z/\rm Z_{\odot} \gtrsim 1.5$), this trend reverses. The top panel quantifies the difference ($D = (Z/{\rm Z_{\odot}})_{\otnt} - (Z/{\rm Z_{\odot}})_{\ntwo}$) as a function of the $\ntwo$--derived metallicity, revealing a negative linear trend defined by $y = -0.28x + 0.34$ and an overall mean offset of $\langle D \rangle = 0.07 \pm 0.10$ dex.

We compare the metallicities derived from the $\ntwo$ and $\otnt$ calibrations to assess their consistency. Although the aforementioned mean difference is technically small, indicating that $\otnt$ yields slightly higher metallicities on average. We note that this difference is in order of the error derived via direct and indirect metallicity estimates, i.e. $\lesssim 0.20$ dex \citep[e.g.][]{Hagele2008, Marino2013, Dors2020c}. To assess the agreement between the two diagnostics, we performed a statistical correlation analysis on the $N=558$ sources with simultaneous metallicity estimates. We find a strong, statistically significant positive correlation between the metallicities derived from the $\ntwo$ and $\otnt$ calibrations. The Pearson correlation coefficient is $r=0.76$ with a 95 per cent confidence interval (CI) of $[0.72, 0.79]$ and a negligible $p$-value ($5.20 \times 10^{-104}$), indicating a robust linear relation. Furthermore, the non-parametric Spearman rank correlation is marginally stronger ($\rho=0.77$; 95 per cent CI $[0.73, 0.80]$; $p < 10^{-100}$). This demonstrates that while intrinsic scatter is present, the monotonic relation is well-preserved, ensuring highly consistent metallicity rankings between the two methods. 
However, these global metrics obscure a strong systematic trend in the residuals (see Figure~\ref{fig_6}, top panel). As derived from the linear fit to the residuals (i.e. $y=-0.28x+0.34$), the $\ntwo$ diagnostic yields systematically lower metallicities than $\otnt$ in the low--$Z$ regime (where $D > 0$) and higher metallicities in the high--$Z$ regime (where $D < 0$), with discrepancies reaching $1.0 Z_{\odot}$ ($\sim 0.50$ dex) at the extremes. We identify a direct, causal correlation between the vertical stratification observed in the calibration planes (Figure~\ref{fig_4}) and these systematic differences, which mirror the trends observed in Figure~\ref{fig_6}. The magnitude of these deviations corresponds precisely to the separation of the outermost luminosity tracks, demonstrating that the systematic error for any individual object is determined by its specific location within the luminosity-stratified diagnostic plane.
This confirms that the two indices have distinct sensitivities to the ionization parameter variations driven by $\Lx$, as presented in \S~\ref{smetal}. Consequently, relying on a single line ratio without accounting for these systematic $\Lx$--dependent offsets can lead to significant abundance errors.

\subsection{Metallicity calibration comparison with the literature} 
\label{sec:comparison_literature} 
\begin{figure}
    \centering
\includegraphics[angle=0.0, width=1.0\columnwidth]{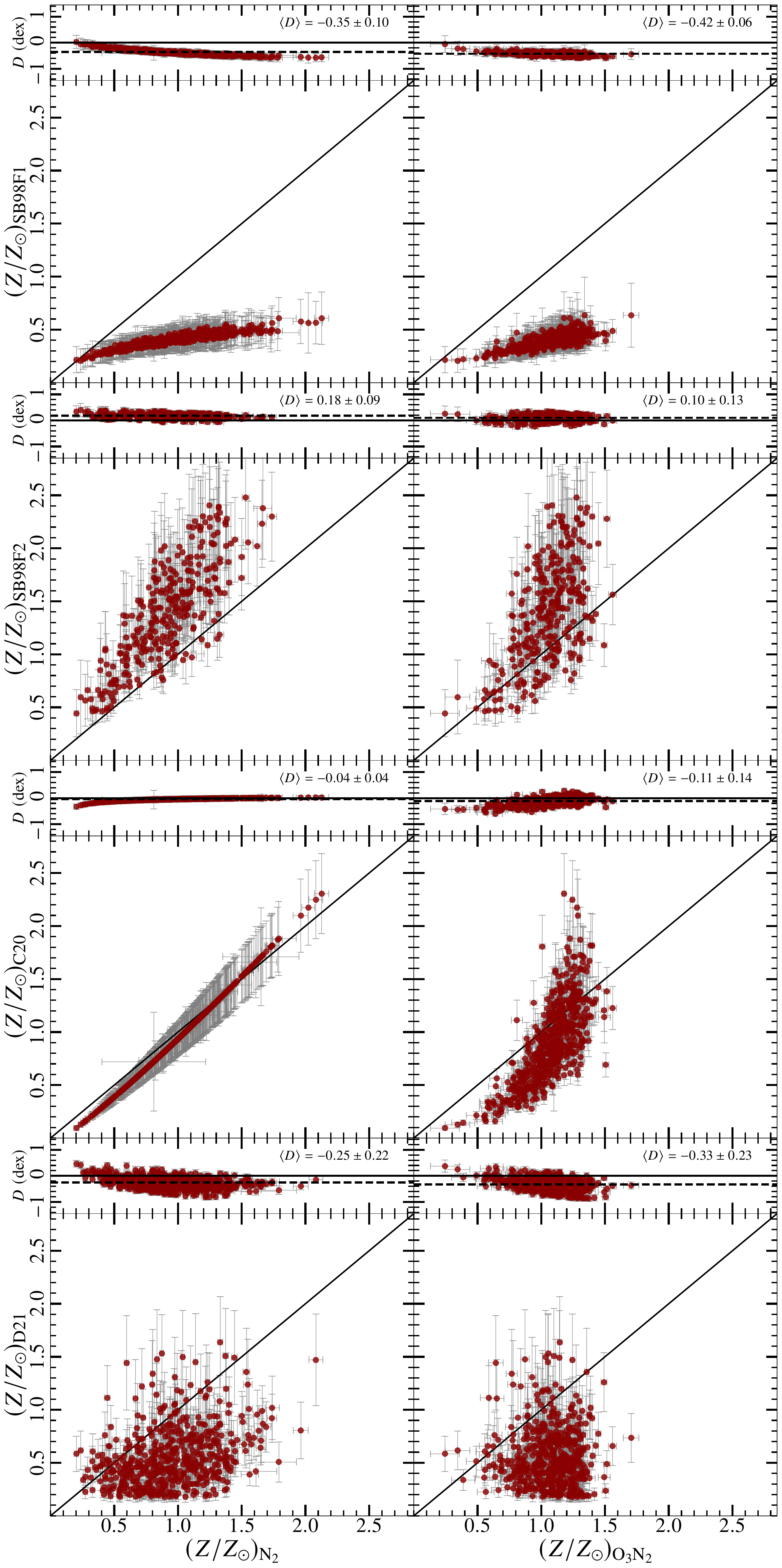}
\caption{Comparison of metallicities derived from our new calibrations ($\ntwo$ and $\otnt$; abscissas) against several literature methods (ordinates). In each column, the bottom panel shows the one-to-one relation (solid black line), while the top panel shows the logarithmic residuals ($D = Z_{\text{Literature}} - Z_{\text{This work}}$).
{\it Top row:} comparison with the \SBcalib{F1} calibration (showing a systematic offset of $\sim -0.4$ dex due to its softer ionizing continuum)
{\it Second row:} comparison with the \SBcalib{F2} calibration (offset $\sim +0.10$ to $\sim +0.18$ dex).
{\it Third row:} comparison with the semi-empirical \citet[][i.e. \citetalias{Carvalho2020}]{Carvalho2020} calibration, which shows the best agreement ($\langle D \rangle \approx -0.04$ dex for $\rm N_2$ and $\approx -0.11$ dex for $\rm O_3 N_2$).
{\it Bottom row:} comparison with the empirical, $T_{\rm e}$-based \citet[][i.e. \citetalias{Dors2021b}]{Dors2021b} calibration (offset $\sim -0.3$ dex, consistent with the known ADF). The mean difference ($\langle D \rangle$) and its standard deviation are indicated in each residual panel.
}
 \label{fig_7}
\end{figure}

To contextualize our new semi-empirical luminosity-based calibrations within existing frameworks, we compared the metallicities derived from our global fits ($\ntwo$ and $\otnt$) with those obtained from several widely adopted methods in the literature: \citetalias{StorchiBergmann1998}, \citetalias{Carvalho2020}, and \citetalias{Dors2021b} (see Figure~\ref{fig_7}). This comparative analysis clarifies the systematic differences that arise when applying calibrations based on distinct underlying assumptions about the ionizing source and its primary input parameters. Specifically, our models utilize the observable X-ray luminosity ($\Lx$) as the primary input, while other calibrations typically rely on the theoretical ionization parameter ($U$). 
We quantify the precision of our semi-empirical calibrations using the $1\sigma$ residual dispersion of the data points around the best-fit polynomial. We find dispersions of $1\sigma = 0.22$ dex for the $\ntwo$ index and $0.20$ dex for the $\otnt$ index. These values represent the intrinsic uncertainty associated with using these line ratios as metallicity proxies across the diverse parameter space in the NLRs of AGNs. It is imperative to compare these dispersions to those found in metallicity calibrations for star-forming galaxies AGNs which relied on $U$. Empirical calibrations derived for {\hii} regions typically achieve tighter constraints, with residual dispersions often in the range of $\sim 0.14 - 0.18$ dex \citep[e.g.][]{Pettini2004, Marino2013, Curti2017}. The slightly larger scatter observed in our AGN calibrations is physical rather than instrumental. Unlike {\hii} regions, which follow a relatively tight sequence in ionization parameter driven by stellar effective temperature, AGNs exhibit a much broader range of ionizing spectral energy distributions, accretion rates, and gas densities. The hardness of the AGN ionizing field introduces secondary dependencies in the emission line ratios that are not fully degenerate with metallicity, a complexity well-documented in previous photoionization studies of the NLR \citep[e.g.][]{Groves2004a, Dors2015, Carvalho2020}. Consequently, the $\approx 0.20$ dex dispersion reflects the irreducible scatter introduced by these intrinsic variations.

However, previous studies deriving semi-empirical or empirical metallicity calibrations for the NLR have reported residual dispersions typically in the range of $\sim 0.05 - 0.5$ dex \citep[e.g.][]{StorchiBergmann1998, Castro2017, Carvalho2020, 2024ApJ...977..187Z}. Our derived values of $1\sigma \approx 0.20$ dex are consistent with these established limits. The slight variance across different calibrations is largely attributable to the methodology used; for instance, calibrations based on stacked spectra \citep[e.g.][]{Curti2017} tend to suppress intrinsic object-to-object variations in density and ionization parameter, yielding nominally lower dispersion values ($\sigma \approx 0.14$ dex). In contrast, our calibration utilizes individual detections across a broad grid of photoionization models, explicitly accounting for the wider intrinsic scatter expected in the diverse BASS DR2 population.

A comparison with the theoretical calibrations of \citetalias{StorchiBergmann1998}, which were developed for Seyfert 2 nuclei, revealed very large systematic offsets. The \SBcalib{F1} calibration typically yields metallicities that are, on average, $\sim$ --0.4 dex lower than our methods. In contrast, the \SBcalib{F2} calibration shows a large but opposing offset, systematically overestimating metallicities by $\sim$ 0.10  to 0.18 dex. This profound discrepancy stems from fundamental differences in the ionizing continuum assumed for the models. The foundational \citetalias{StorchiBergmann1998} models utilized a significantly softer SED, consisting of a power-law combined with a blackbody component characteristic of hot stars, resulting in a cooler gas state compared to our hard, X-ray-rich continuum models.

The softer continuum used by \citetalias{StorchiBergmann1998} is less efficient at heating the gas, resulting in a lower equilibrium temperature. This physical state is degenerate with the effect of high metallicity (which also lowers temperature via enhanced metal cooling). Conversely, our hard, X-ray-rich continuum is more efficient at heating, leading to a hotter gas state, which is degenerate with low metallicity. Therefore, applying the  “cooler” \citetalias{StorchiBergmann1998} model, calibrated to associate a given LR with a comparatively lower metallicity, to gas ionized by a hard AGN source results in a systematic and significant underestimation (for \SBcalib{F1}) or overestimation (for \SBcalib{F2}) of the true metallicity. This underscores the importance of matching the ionizing source of the photoionization model to the dominant physical process in the observed objects.

Beyond the differences in the ionizing continuum, the \citetalias{StorchiBergmann1998} models were constructed assuming a single, fixed electron density of $N_{e} = 300\ \mathrm{cm}^{-3}$, a common simplification at the time. The authors acknowledged that NLRs have a range of densities and that the calibrations, derived using a fixed density of $300\ \mathrm{cm}^{-3}$, are sensitive to this parameter. They found that the dependence is approximately linear with the logarithm of the density and proposed a correction to the calculated oxygen abundance $(\mathrm{O}/\mathrm{H})$. However, real AGN NLRs exhibit a considerable range of electron densities \citep[e.g.][]{Osterbrock2006, Vaona2012, Rose2018, Baron2019, Davies2020, 2021ApJ...910..139R}. The observed BASS DR2 sample, in contrast, exhibits a median density of $580 \pm 413\,\mathrm{cm}^{-3}$ from the lower-ionization \sii$\lambda\lambda$6716,6731 doublet and a much higher median of $3467\pm 864\,\mathrm{cm}^{-3}$ from the higher-ionization \ariv$\lambda\lambda$4711, 4740 doublet. This density distribution is clearly spread over a wide range, from $\sim 100$ to over $10\,000\ \mathrm{cm}^{-3}$. This single, fixed density assumption, inconsistent with observations, represents an additional source of systematic error, contributing to the observed offset, as gas density affects key line ratios via collisional de-excitation.

In the third row of Figure~\ref{fig_7}, we compare the metallicities derived from our calibrations with those from the semi-empirical calibration of Seyfert 2 nuclei by \citetalias{Carvalho2020}. This comparison contrasts two modern, but different, modelling approaches for the same class of objects.
The left panel, which compares our $\ntwo$--based metallicities, shows a very tight correlation with a small scatter and a mean difference of only $\langle D\rangle = -0.04 \pm 0.04$ dex. This agreement is expected, as both our work and \citetalias{Carvalho2020} utilize the $\ntwo$ index, which is a strong function of the $\rm N/O$ abundance ratio.
In contrast, the right panel, comparing our $\otnt$ calibration with the \citetalias{Carvalho2020} method, shows a larger systematic offset of $\langle D\rangle = -0.11 \pm 0.14$ dex. This increased discrepancy is a direct consequence of the differing assumptions about the AGN ionizing continuum used in the two sets of models. The \OIIIOII\ ratio, assumed by \citetalias{Carvalho2020} to derive their calibration,
is highly sensitive to the ionization structure of the gas, which is dictated by the shape of the ionizing SED. Therefore, the distinct continua assumed by each study result in different predicted LRs for a given metallicity, naturally leading to the larger systematic offset seen when comparing these ionization-sensitive diagnostics.

Finally, in the bottom row of Figure~\ref{fig_7}, we extend our comparison to the empirical metallicity calibration from \citet[][hereafter \citetalias{Dors2021b}]{Dors2021b}. This calibration is distinct from the others as it is not based on photoionization models but was derived empirically using a sample of Seyfert 1 and 2 nuclei with oxygen abundances determined via the \Te-method, adapted for AGNs. The \citetalias{Dors2021b} calibration utilizes the $R_{23} =$ (\oii$\lambda 3727 +$ \oiii$\lambda4959+\lambda5007)/$\hbeta\ and $P =\left[([\text{O}\,\textsc{iii}]\lambda4959+\lambda 5007)/\text{H}\beta\right]/R23$  LRs, which differ from the nitrogen-based indices used in our work. Their work represents another important benchmark for AGN metallicity studies. 

The comparison of metallicities from our calibrations with those from \citetalias{Dors2021b}, reveals a systematic offset, with the \citetalias{Dors2021b} calibration yielding moderately lower metallicities on average (the mean differences range from $\langle D \rangle \approx$ --0.3 dex across our two calibrations). This offset is expected and highlights the well-known discrepancy between theoretical (photoionization model-based) and empirical (\Te--based) abundance scales, which can be on the order of 0.2 dex \citep[e.g.][]{Dors2020c} or more. The  \citetalias{Dors2021b} calibration relies on a specific theoretical relation between the temperatures of the low- and high-ionization zones ($t_2$ -- $t_3$) derived from a separate grid of models, introducing a different set of systematic uncertainties compared to our direct modelling approach. This fundamental difference in methodology could be the primary driver of the observed offset. The comparisons with our diagnostics ($\ntwo$ and $\otnt$) show similar offsets and scatter, reinforcing the conclusion that the discrepancy is systematic between the two approaches.
Therefore, the observed offset between our new calibrations and the \citetalias{Dors2021b} method does not necessarily invalidate either approach. Instead, it underscores a known, fundamental difference between two of the primary methodologies used for chemical abundance studies in gaseous nebulae.

It is well-established that the distinct methodologies used to derive gas-phase metallicities, including empirical techniques, which are typically calibrated against \Te-method abundance measurements \citep[e.g.][]{Marino2013, Curti2017}, theoretical approaches, based purely on grids of photoionization models \citep[e.g.][]{McGaugh1991, StorchiBergmann1998, Kewley2002, Tremonti2004, Nagao2006b, Dopita2016, 2024ApJ...977..187Z}, semi-empirical or hybrid methods, combining aspects of the \Te-method with constraints from photoionization models \citep[e.g.][]{ Castro2017,Carvalho2020, Oliveira2022}, and complex Bayesian frameworks (based on simultaneous fits of most strong emission lines with stellar evolutionary synthesis, e.g.  \texttt{IZI}: \citealt{Blanc2015}; \texttt{BOND}: \citealt{ValeAsari2016}; \texttt{BEAGLE}: \citealt{Chevallard2016}; \citealt{VidalGarcia2024}; \texttt{NebulaBayes}: \citealt{Thomas2018};   \texttt{BAGPIPES}: \citealt{Carnall2019} and  \texttt{HCM}: \citealt{PerezMonteiro2019})--are generally not consistent with one another \citep[see][]{2010A&A...517A..85L, 2020MNRAS.492..468D}. Consequently, comparisons between different strong-line calibrations, even those based on the same emission-line diagnostics, are expected to produce systematic offsets.

This issue is particularly pronounced in AGN studies, where discrepancies can be substantial. For instance, a direct comparison of AGN-specific diagnostics by \citet{2024ApJ...977..187Z} found that metallicity estimates using the methods of \citetalias{StorchiBergmann1998} and \citetalias{Castro2017}  can differ by up to $\sim$0.5~dex. The same study highlighted that even when using an identical diagnostic LR (\NIIOII), the calibration from \citet{Thomas2018} yields metallicities that are systematically  $\sim$0.4~dex higher than that of \citetalias{Castro2017}, underscoring the profound impact of the underlying model assumptions. Similarly, \citet{Carr2023} reported that while their results agreed well with the $\ntwo$ method of \citetalias{Carvalho2020}, they diverged significantly from the \Te-based calibration of \citet{2020MNRAS.496.2191F}, with an average offset of 0.34~dex.

Overall, reported systematic differences in derived abundances for both \hii regions and AGNs commonly range from $\sim -0.1$ to $\sim 0.8$ dex \citep[e.g.][]{Denicol2002, Pettini2004, Tremonti2004, 2008ApJ...681.1183K, Maiolino2008, 2010A&A...517A..85L, Marino2013, Blanc2015, Bian2017, Curti2017, Dors2020c}. This systematic range is consistent with the discrepancies found in comparison of our new calibrations with similar calibrations in the literature.

\section{Summary and Conclusions}
\label{conclusions}
In this paper, we have presented a new, comprehensive analysis of the physical conditions in the NLRs of Seyfert galaxies, with a focus on developing robust calibrations for gas-phase metallicity. Our work is based on an extensive grid of photoionization models computed with {\sc cloudy} photoionization code, which we compare with observational data from the BASS DR2 survey. Our primary findings and conclusions are summarized below.
\begin{enumerate}
    \item Through this approach, we found that the $\ntwo$ and $\otnt$ diagnostics exhibit a strong, but opposing, secondary dependence on $\Lx$ (Figure~\ref{fig_4}). Estimates from the $\ntwo$ index decrease with increasing X-ray luminosity, while $\otnt$ estimates increase with increasing X-ray luminosity. This dependence introduces systematic vertical offsets ranging from $\sim 0.2 Z_{\odot}$ ($\sim 0.30$ dex) to $\sim 0.5 Z_{\odot}$ ($\sim 0.16$ dex) for the $\ntwo$ diagnostic and from $\sim 0.5 Z_{\odot}$ ($\sim 0.54$ dex) to $\sim 1.0 Z_{\odot}$ ($\sim 0.20$ dex) for the $\otnt$ diagnostic. 
    
    \item Despite the luminosity dependence, we confirmed that both indices are good tracers of metallicity in the high-metallicity regime ($Z/Z_{\odot} \approx 0.2-2.6$) typical of AGN hosts. We found that the $\otnt$ diagnostic is robust against variations in gas density, spectral index, and Big Blue Bump temperature compared to the $\ntwo$ diagnostic. Our analysis is consistent with previous studies that favour photoionization models with relatively hard ionizing continua (i.e. $\aox = -1.4, -1.1, -0.8$).

    \item We derived new semi-empirical calibrations for estimating gas-phase metallicity in the NLRs using these two strong-line ratio diagnostics ($\ntwo$ and $\otnt$). These calibrations achieve residual dispersions of $\sigma \approx 0.22$ dex ($\ntwo$) and $0.20$ dex ($\otnt$), reflecting the intrinsic diversity in the NLRs of AGNs.

    \item Comparison between our calibrations shows good agreement between $\ntwo$ and $\otnt$ indices, which show tight correlations (Figure~\ref{fig_6}). However, a systematic offset is observed, with the $\otnt$ diagnostic yielding metallicities that are, on average, $\sim 0.07 \pm 0.10$ dex higher than those derived from the $\ntwo$ diagnostic. This difference is consistent with the distinct sensitivities of these line ratios to ionization conditions and abundance ratios. However, we explicitly linked the vertical stratification observed in the calibration grids (Figure~\ref{fig_4}) to the substantial systematic deviations ($\sim 0.50$ dex) found at the luminosity extremes (Figure~\ref{fig_6}), demonstrating that the magnitude of systematic error for any given source is causally determined by its specific position within the luminosity-stratified diagnostic plane.
    
\item Finally, we compared our results with the literature. Our calibrations yield metallicities systematically higher ($\sim 0.25-0.33$ dex) than the empirical $T_{\rm e}$--based \citetalias{Dors2021b} method and show varied offsets from early theoretical models (\citetalias{StorchiBergmann1998}), ranging from $\sim -0.35$ to $-0.42$ dex for \SBcalib{F1} and $\sim 0.10$ to $0.18$ dex for \SBcalib{F2}. However, the $\ntwo$ calibration shows good agreement with the semi-empirical \citetalias{Carvalho2020} calibration ($D \approx -0.04$ dex).
 \end{enumerate}

 This work presents a fundamental shift in AGN photoionization modelling by replacing the unobservable ionization parameter ($\log U$) with the directly observable X-ray luminosity ($\log \Lx$), thereby connecting diagnostic models to a measurable physical property that drives spectral hardness and heating. We demonstrate that neglecting $\Lx$ overlooks the systematic offsets intrinsic to the diagnostics, leading to metallicity errors up to $\sim0.50$ dex, and we provide robust, luminosity-dependent calibrations for the $\ntwo$ and $\otnt$ diagnostics to mitigate these biases. This multi-parameter framework, which will be extended to additional diagnostics in a future companion paper, offers a more reliable method for investigating the chemical evolution of AGN hosts and the connection between supermassive black holes and their environments.

\section*{Acknowledgements}
We thank the anonymous referee for their thorough review, insightful feedback, and highly constructive suggestions, which have substantially improved the overall clarity of this paper. MA gratefully acknowledges support from Fundação de Amparo à Pesquisa do Estado de São Paulo
(FAPESP, Processo: 2024/03727-3). OLD is grateful to Fundação de Amparo à Pesquisa do Estado de
São Paulo (FAPESP) and Conselho Nacional de Desenvolvimento Científico e Tecnológico (CNPq). RAR acknowledges support from the Conselho Nacional de Desenvolvimento Cient\'ifico e Tecnol\'ogico (CNPq; Proj. 303450/2022-3, 403398/2023-1, \& 441722/2023-7) and Coordena\c c\~ao de Aperfei\c coamento de Pessoal de N\'ivel Superior (CAPES;  Proj. 88887.894973/2023-00). 
RR acknowledges support from  Conselho Nacional de Desenvolvimento Cient\'{i}fico e Tecnol\'ogico  (CNPq, Proj. CNPq-445231/2024-6,311223/2020-6, 404238/2021-1, and 310413/2025-7), Funda\c{c}\~ao de amparo \`{a} pesquisa do Rio Grande do Sul (FAPERGS, Proj. 19/1750-2 and 24/2551-0001282-6) and Coordena\c{c}\~ao de Aperfei\c{c}oamento de Pessoal de N\'{i}vel Superior (CAPES, 88881.109987/2025-01). JVM acknowledges support from the Spanish grants: PID2022-136598NB-C32 and Severo Ochoa CEX2021-001131-S funded by MICIU/AEI/ 10.13039/501100011033.

\section*{Data Availability}
The observational data used in this study are publicly available from the \textit{Swift} Burst Alert Telescope (BAT) AGN Spectroscopic Survey (BASS) Data Release 2, accessible at \url{https://www.bass-survey.com/dr2.html}. The data products and analysis results generated specifically for this paper are available from the corresponding author upon reasonable request.

\bibliographystyle{mnras}
\bibliography{ref}

@ARTICLE{StorchiBergmann1998,
       author = {{Storchi-Bergmann}, Thaisa and {Schmitt}, Henrique R. and
         {Calzetti}, Daniela and {Kinney}, Anne L.},
        title = "{Chemical Abundance Calibrations for the Narrow-Line Region of Active Galaxies}",
      journal = {\aj},
         year = "1998",
        month = "Mar",
       volume = {115},
       number = {3},
        pages = {909 (SB98)-914},
          doi = {10.1086/300242},
archivePrefix = {arXiv},
       eprint = {astro-ph/9711302},
 primaryClass = {astro-ph},
       adsurl = {https://ui.adsabs.harvard.edu/abs/1998AJ....115..909S},
}

@ARTICLE{2000AJ....120.1579Y,
       author = {{York}, Donald G. and {Adelman}, J. and {Anderson}, Jr., John E. and {Anderson}, Scott F. and {Annis}, James and {Bahcall}, Neta A. and {Bakken}, J.~A. and {Barkhouser}, Robert and {Bastian}, Steven and {Berman}, Eileen and {Boroski}, William N. and {Bracker}, Steve and {Briegel}, Charlie and {Briggs}, John W. and {Brinkmann}, J. and {Brunner}, Robert and {Burles}, Scott and {Carey}, Larry and {Carr}, Michael A. and {Castander}, Francisco J. and {Chen}, Bing and {Colestock}, Patrick L. and {Connolly}, A.~J. and {Crocker}, J.~H. and {Csabai}, Istv{\'a}n and {Czarapata}, Paul C. and {Davis}, John Eric and {Doi}, Mamoru and {Dombeck}, Tom and {Eisenstein}, Daniel and {Ellman}, Nancy and {Elms}, Brian R. and {Evans}, Michael L. and {Fan}, Xiaohui and {Federwitz}, Glenn R. and {Fiscelli}, Larry and {Friedman}, Scott and {Frieman}, Joshua A. and {Fukugita}, Masataka and {Gillespie}, Bruce and {Gunn}, James E. and {Gurbani}, Vijay K. and {de Haas}, Ernst and {Haldeman}, Merle and {Harris}, Frederick H. and {Hayes}, J. and {Heckman}, Timothy M. and {Hennessy}, G.~S. and {Hindsley}, Robert B. and {Holm}, Scott and {Holmgren}, Donald J. and {Huang}, Chi-hao and {Hull}, Charles and {Husby}, Don and {Ichikawa}, Shin-Ichi and {Ichikawa}, Takashi and {Ivezi{\'c}}, {\v{Z}}eljko and {Kent}, Stephen and {Kim}, Rita S.~J. and {Kinney}, E. and {Klaene}, Mark and {Kleinman}, A.~N. and {Kleinman}, S. and {Knapp}, G.~R. and {Korienek}, John and {Kron}, Richard G. and {Kunszt}, Peter Z. and {Lamb}, D.~Q. and {Lee}, B. and {Leger}, R. French and {Limmongkol}, Siriluk and {Lindenmeyer}, Carl and {Long}, Daniel C. and {Loomis}, Craig and {Loveday}, Jon and {Lucinio}, Rich and {Lupton}, Robert H. and {MacKinnon}, Bryan and {Mannery}, Edward J. and {Mantsch}, P.~M. and {Margon}, Bruce and {McGehee}, Peregrine and {McKay}, Timothy A. and {Meiksin}, Avery and {Merelli}, Aronne and {Monet}, David G. and {Munn}, Jeffrey A. and {Narayanan}, Vijay K. and {Nash}, Thomas and {Neilsen}, Eric and {Neswold}, Rich and {Newberg}, Heidi Jo and {Nichol}, R.~C. and {Nicinski}, Tom and {Nonino}, Mario and {Okada}, Norio and {Okamura}, Sadanori and {Ostriker}, Jeremiah P. and {Owen}, Russell and {Pauls}, A. George and {Peoples}, John and {Peterson}, R.~L. and {Petravick}, Donald and {Pier}, Jeffrey R. and {Pope}, Adrian and {Pordes}, Ruth and {Prosapio}, Angela and {Rechenmacher}, Ron and {Quinn}, Thomas R. and {Richards}, Gordon T. and {Richmond}, Michael W. and {Rivetta}, Claudio H. and {Rockosi}, Constance M. and {Ruthmansdorfer}, Kurt and {Sandford}, Dale and {Schlegel}, David J. and {Schneider}, Donald P. and {Sekiguchi}, Maki and {Sergey}, Gary and {Shimasaku}, Kazuhiro and {Siegmund}, Walter A. and {Smee}, Stephen and {Smith}, J. Allyn and {Snedden}, S. and {Stone}, R. and {Stoughton}, Chris and {Strauss}, Michael A. and {Stubbs}, Christopher and {SubbaRao}, Mark and {Szalay}, Alexander S. and {Szapudi}, Istvan and {Szokoly}, Gyula P. and {Thakar}, Anirudda R. and {Tremonti}, Christy and {Tucker}, Douglas L. and {Uomoto}, Alan and {Vanden Berk}, Dan and {Vogeley}, Michael S. and {Waddell}, Patrick and {Wang}, Shu-i. and {Watanabe}, Masaru and {Weinberg}, David H. and {Yanny}, Brian and {Yasuda}, Naoki and {SDSS Collaboration}},
        title = "{The Sloan Digital Sky Survey: Technical Summary}",
      journal = {\aj},
         year = 2000,
        month = sep,
       volume = {120},
       number = {3},
        pages = {1579-1587},
          doi = {10.1086/301513},
archivePrefix = {arXiv},
       eprint = {astro-ph/0006396},
 primaryClass = {astro-ph},
       adsurl = {https://ui.adsabs.harvard.edu/abs/2000AJ....120.1579Y}
}

@ARTICLE{2020MNRAS.492..468D,
       author = {{Dors}, O.~L. and {Freitas-Lemes}, P. and {Am{\^o}res}, E.~B. and {P{\'e}rez-Montero}, E. and {Cardaci}, M.~V. and {H{\"a}gele}, G.~F. and {Armah}, M. and {Krabbe}, A.~C. and {Fa{\'u}ndez-Abans}, M.},
        title = "{Chemical abundances of Seyfert 2 AGNs - I. Comparing oxygen abundances from distinct methods using SDSS}",
      journal = {\mnras},
         year = 2020,
        month = feb,
       volume = {492},
       number = {1},
        pages = {468-479},
          doi = {10.1093/mnras/stz3492},
archivePrefix = {arXiv},
       eprint = {1912.04236},
 primaryClass = {astro-ph.GA},
       adsurl = {https://ui.adsabs.harvard.edu/abs/2020MNRAS.492..468D}
}

@ARTICLE{2020MNRAS.496.2191F,
       author = {{Flury}, Sophia R. and {Moran}, Edward C.},
        title = "{Chemical abundances in active galaxies}",
      journal = {\mnras},
         year = 2020,
        month = aug,
       volume = {496},
       number = {2},
        pages = {2191-2203},
          doi = {10.1093/mnras/staa1563},
archivePrefix = {arXiv},
       eprint = {2006.01113},
 primaryClass = {astro-ph.GA},
       adsurl = {https://ui.adsabs.harvard.edu/abs/2020MNRAS.496.2191F}
}

@ARTICLE{2003A&A...399.1003P,
       author = {{Pilyugin}, L.~S.},
        title = "{Abundance determinations in H II regions. Model fitting versus T$_{e}$-method}",
      journal = {\aap},
         year = 2003,
        month = mar,
       volume = {399},
        pages = {1003-1007},
          doi = {10.1051/0004-6361:20021669},
archivePrefix = {arXiv},
       eprint = {astro-ph/0211319},
 primaryClass = {astro-ph},
       adsurl = {https://ui.adsabs.harvard.edu/abs/2003A&A...399.1003P}
}

@ARTICLE{Cardelli1989,
       author = {{Cardelli}, Jason A. and {Clayton}, Geoffrey C. and {Mathis}, John S.},
        title = "{The Relationship between Infrared, Optical, and Ultraviolet Extinction}",
      journal = {\apj},
         year = "1989",
        month = "Oct",
       volume = {345},
        pages = {245},
          doi = {10.1086/167900},
       adsurl = {https://ui.adsabs.harvard.edu/abs/1989ApJ...345..245C}
}

@ARTICLE{2015ApJS..217...12D,
       author = {{Dopita}, Michael A. and {Shastri}, Prajval and {Davies}, Rebecca and {Kewley}, Lisa and {Hampton}, Elise and {Scharw{\"a}chter}, Julia and {Sutherland}, Ralph and {Kharb}, Preeti and {Jose}, Jessy and {Bhatt}, Harish and {Ramya}, S. and {Jin}, Chichuan and {Banfield}, Julie and {Zaw}, Ingyin and {Juneau}, St{\'e}phanie and {James}, Bethan and {Srivastava}, Shweta},
        title = "{Probing the Physics of Narrow Line Regions in Active Galaxies. II. The Siding Spring Southern Seyfert Spectroscopic Snapshot Survey (S7)}",
      journal = {\apjs},
         year = 2015,
        month = mar,
       volume = {217},
       number = {1},
          eid = {12},
        pages = {12},
          doi = {10.1088/0067-0049/217/1/12},
archivePrefix = {arXiv},
       eprint = {1501.02022},
 primaryClass = {astro-ph.GA},
       adsurl = {https://ui.adsabs.harvard.edu/abs/2015ApJS..217...12D}
}

@ARTICLE{2017MNRAS.467.3759T,
       author = {{Toribio San Cipriano}, L. and {Dom{\'\i}nguez-Guzm{\'a}n}, G. and {Esteban}, C. and {Garc{\'\i}a-Rojas}, J. and {Mesa-Delgado}, A. and {Bresolin}, F. and {Rodr{\'\i}guez}, M. and {Sim{\'o}n-D{\'\i}az}, S.},
        title = "{Carbon and oxygen in H II regions of the Magellanic Clouds: abundance discrepancy and chemical evolution}",
      journal = {\mnras},
         year = 2017,
        month = may,
       volume = {467},
       number = {3},
        pages = {3759-3774},
          doi = {10.1093/mnras/stx328},
archivePrefix = {arXiv},
       eprint = {1702.01120},
 primaryClass = {astro-ph.GA},
       adsurl = {https://ui.adsabs.harvard.edu/abs/2017MNRAS.467.3759T}
}

@ARTICLE{2008ApJ...681.1183K,
       author = {{Kewley}, Lisa J. and {Ellison}, Sara L.},
        title = "{Metallicity Calibrations and the Mass-Metallicity Relation for Star-forming Galaxies}",
      journal = {\apj},
         year = 2008,
        month = jul,
       volume = {681},
       number = {2},
        pages = {1183-1204},
          doi = {10.1086/587500},
archivePrefix = {arXiv},
       eprint = {0801.1849},
 primaryClass = {astro-ph},
       adsurl = {https://ui.adsabs.harvard.edu/abs/2008ApJ...681.1183K}
}

@ARTICLE{2021MNRAS.506L..11R,
       author = {{Riffel}, Rogemar A. and {Dors}, Oli L. and {Krabbe}, Angela C. and {Esteban}, C{\'e}sar},
        title = "{Electron temperature fluctuations in Seyfert galaxies}",
      journal = {\mnras},
         year = 2021,
        month = sep,
       volume = {506},
       number = {1},
        pages = {L11-L15},
          doi = {10.1093/mnrasl/slab064},
archivePrefix = {arXiv},
       eprint = {2106.03623},
 primaryClass = {astro-ph.GA},
       adsurl = {https://ui.adsabs.harvard.edu/abs/2021MNRAS.506L..11R}
}

@ARTICLE{Carvalho2020,
       author = {{Carvalho}, S.~P. and {Dors}, O.~L. and {Cardaci}, M.~V. and {H{\"a}gele}, G.~F. and {Krabbe}, A.~C. and {P{\'e}rez-Montero}, E. and {Monteiro}, A.~F. and {Armah}, M. and {Freitas-Lemes}, P.},
        title = "{Chemical abundances of Seyfert 2 AGNs - II. N2 metallicity calibration based on SDSS}",
      journal = {\mnras},
         year = 2020,
        month = mar,
       volume = {492},
       number = {4},
        pages = {5675 (C20)-5683},
          doi = {10.1093/mnras/staa193},
archivePrefix = {arXiv},
       eprint = {2001.07126},
 primaryClass = {astro-ph.GA},
       adsurl = {https://ui.adsabs.harvard.edu/abs/2020MNRAS.492.5675C},
     note = {(C20)}
}

@ARTICLE{2025A&A...696A.229P,
       author = {{P{\'e}rez-Montero}, E. and {Fern{\'a}ndez-Ontiveros}, J.~A. and {P{\'e}rez-D{\'\i}az}, B. and {V{\'\i}lchez}, J.~M. and {Amor{\'\i}n}, R.},
        title = "{Exploring the hardness of the ionizing radiation with the infrared softness diagram: II. Bimodal distributions in both the ionizing continuum slope and the excitation in active galactic nuclei}",
      journal = {\aap},
         year = 2025,
        month = apr,
       volume = {696},
          eid = {A229},
        pages = {A229},
          doi = {10.1051/0004-6361/202453276},
archivePrefix = {arXiv},
       eprint = {2503.09267},
 primaryClass = {astro-ph.GA},
       adsurl = {https://ui.adsabs.harvard.edu/abs/2025A&A...696A.229P}
}

@ARTICLE{Carr2023,
       author = {{Carr}, David J. and {Salzer}, John J. and {Gronwall}, Caryl and {Williams}, Anna L.},
        title = "{Metal Abundances of Intermediate-redshift Active Galactic Nuclei: Evidence for a Population of Lower-metallicity Seyfert 2 Galaxies at z = 0.3-0.4}",
      journal = {\apj},
         year = 2023,
        month = oct,
       volume = {955},
       number = {2},
          eid = {141},
        pages = {141},
          doi = {10.3847/1538-4357/aced91},
archivePrefix = {arXiv},
       eprint = {2308.06824},
 primaryClass = {astro-ph.GA},
       adsurl = {https://ui.adsabs.harvard.edu/abs/2023ApJ...955..141C}
}

@ARTICLE{2017MNRAS.470.1218P,
       author = {{Pereira-Santaella}, M. and {Rigopoulou}, D. and {Farrah}, D. and {Lebouteiller}, V. and {Li}, J.},
        title = "{Far-infrared metallicity diagnostics: application to local ultraluminous infrared galaxies}",
      journal = {\mnras},
         year = 2017,
        month = sep,
       volume = {470},
       number = {1},
        pages = {1218-1232},
          doi = {10.1093/mnras/stx1284},
archivePrefix = {arXiv},
       eprint = {1705.08367},
 primaryClass = {astro-ph.GA},
       adsurl = {https://ui.adsabs.harvard.edu/abs/2017MNRAS.470.1218P}
}

@ARTICLE{Richardson2022,
       author = {{Richardson}, Chris T. and {Simpson}, Connor and {Polimera}, Mugdha S. and {Kannappan}, Sheila J. and {Bellovary}, Jillian M. and {Greene}, Christopher and {Jenkins}, Sam},
        title = "{Optical and JWST Mid-IR Emission Line Diagnostics for Simultaneous IMBH and Stellar Excitation in z 0 Dwarf Galaxies}",
      journal = {\apj},
         year = 2022,
        month = mar,
       volume = {927},
       number = {2},
          eid = {165},
        pages = {165},
          doi = {10.3847/1538-4357/ac510c},
archivePrefix = {arXiv},
       eprint = {2202.01330},
 primaryClass = {astro-ph.GA},
       adsurl = {https://ui.adsabs.harvard.edu/abs/2022ApJ...927..165R}
}

@ARTICLE{2011ApJ...726...20M,
       author = {{Miller}, B.~P. and {Brandt}, W.~N. and {Schneider}, D.~P. and {Gibson}, R.~R. and {Steffen}, A.~T. and {Wu}, Jianfeng},
        title = "{X-ray Emission from Optically Selected Radio-intermediate and Radio-loud Quasars}",
      journal = {\apj},
         year = 2011,
        month = jan,
       volume = {726},
       number = {1},
          eid = {20},
        pages = {20},
          doi = {10.1088/0004-637X/726/1/20},
archivePrefix = {arXiv},
       eprint = {1010.4804},
 primaryClass = {astro-ph.CO},
       adsurl = {https://ui.adsabs.harvard.edu/abs/2011ApJ...726...20M}
}

@ARTICLE{2021RNAAS...5..101T,
       author = {{Timlin}, John and {Zhu}, Shifu and {Brandt}, W.~N. and {Laor}, Ari},
        title = "{The {\ensuremath{\alpha}}$_{ox}$-He II EW Connection in Radio-loud Quasars}",
      journal = {Research Notes of the American Astronomical Society},
         year = 2021,
        month = apr,
       volume = {5},
       number = {4},
          eid = {101},
        pages = {101},
          doi = {10.3847/2515-5172/abfbe5},
archivePrefix = {arXiv},
       eprint = {2104.14407},
 primaryClass = {astro-ph.HE},
       adsurl = {https://ui.adsabs.harvard.edu/abs/2021RNAAS...5..101T}
}

@ARTICLE{2021RAA....21....4Z,
       author = {{Zhou}, Min-Hua and {Gu}, Min-Feng},
        title = "{The composite X-ray spectra of radio-loud and radio-quiet SDSS quasars}",
      journal = {Research in Astronomy and Astrophysics},
         year = 2021,
        month = jan,
       volume = {21},
       number = {1},
          eid = {004},
        pages = {004},
          doi = {10.1088/1674-4527/21/1/4},
archivePrefix = {arXiv},
       eprint = {2007.01049},
 primaryClass = {astro-ph.HE},
       adsurl = {https://ui.adsabs.harvard.edu/abs/2021RAA....21....4Z}
}

@ARTICLE{1994ApJS...92...53W,
       author = {{Wilkes}, Belinda J. and {Tananbaum}, Harvey and {Worrall}, D.~M. and {Avni}, Yoram and {Oey}, M.~S. and {Flanagan}, Joan},
        title = "{The Einstein Database of IPC X-Ray Observations of Optically Selected and Radio-selected Quasars. I.}",
      journal = {\apjs},
         year = 1994,
        month = may,
       volume = {92},
        pages = {53},
          doi = {10.1086/191959},
       adsurl = {https://ui.adsabs.harvard.edu/abs/1994ApJS...92...53W}
}

@ARTICLE{2005AJ....130..387S,
       author = {{Strateva}, Iskra V. and {Brandt}, W.~N. and {Schneider}, Donald P. and {Vanden Berk}, Daniel G. and {Vignali}, Cristian},
        title = "{Soft X-Ray and Ultraviolet Emission Relations in Optically Selected AGN Samples}",
      journal = {\aj},
         year = 2005,
        month = aug,
       volume = {130},
       number = {2},
        pages = {387-405},
          doi = {10.1086/431247},
archivePrefix = {arXiv},
       eprint = {astro-ph/0503009},
 primaryClass = {astro-ph},
       adsurl = {https://ui.adsabs.harvard.edu/abs/2005AJ....130..387S}
}

@ARTICLE{2019A&A...630A.118V,
       author = {{Vito}, F. and {Brandt}, W.~N. and {Bauer}, F.~E. and {Calura}, F. and {Gilli}, R. and {Luo}, B. and {Shemmer}, O. and {Vignali}, C. and {Zamorani}, G. and {Brusa}, M. and {Civano}, F. and {Comastri}, A. and {Nanni}, R.},
        title = "{The X-ray properties of z > 6 quasars: no evident evolution of accretion physics in the first Gyr of the Universe}",
      journal = {\aap},
         year = 2019,
        month = oct,
       volume = {630},
          eid = {A118},
        pages = {A118},
          doi = {10.1051/0004-6361/201936217},
archivePrefix = {arXiv},
       eprint = {1908.09849},
 primaryClass = {astro-ph.GA},
       adsurl = {https://ui.adsabs.harvard.edu/abs/2019A&A...630A.118V}
}

@ARTICLE{2025arXiv250805397D,
       author = {{Dors}, O.~L. and {Oliveira}, C.~B. and {Cardaci}, M.~V. and {H{\"a}gele}, G.~F. and {Armah}, Mark and {Riffel}, R.~A. and {Ramos Vieira}, L. and {Almeida}, G.~C. and {Morais}, I.~N. and {Santos}, P.~C.},
        title = "{Metallicity of Active Galactic Nuclei from ultraviolet and optical emission lines-II. Revisiting the $C43$ metallicity calibration and its implications}",
      journal = {MNRAS},
    volume = {542},
    number = {4},
    pages = {3181-3197},
    year = {2025},
    month = sep,
    issn = {0035-8711},
    doi = {10.1093/mnras/staf1407},
archivePrefix = {arXiv},
       eprint = {2508.05397},
 primaryClass = {astro-ph.GA},
       adsurl = {https://ui.adsabs.harvard.edu/abs/2025MNRAS.tmp.1416D}
}

@ARTICLE{2024ApJ...977..187Z,
       author = {{Zhu}, Peixin and {Kewley}, Lisa J. and {Sutherland}, Ralph S.},
        title = "{Theoretical Diagnostics for Narrow-line Regions of Active Galactic Nuclei}",
      journal = {\apj},
         year = 2024,
        month = dec,
       volume = {977},
       number = {2},
          eid = {187},
        pages = {187},
          doi = {10.3847/1538-4357/ad8f37},
archivePrefix = {arXiv},
       eprint = {2411.04103},
 primaryClass = {astro-ph.GA},
       adsurl = {https://ui.adsabs.harvard.edu/abs/2024ApJ...977..187Z}
}

@ARTICLE{2024PASA...41...99O,
       author = {{Oliveira}, Celso B. and {Dors}, Oli and {Zinchenko}, Igor and {Cardaci}, Monica and {H{\"a}gele}, Guillermo and {Morais}, Istenio and {Santos}, Pedro and {Almeida}, Gleicy},
        title = "{Semi-empirical calibration of the oxygen abundance for LINER galaxies based on SDSS-IV MaNGA - The case for strong and weak AGN}",
      journal = {\pasa},
         year = 2024,
        month = dec,
       volume = {41},
          eid = {e099},
        pages = {e099},
          doi = {10.1017/pasa.2024.110},
archivePrefix = {arXiv},
       eprint = {2411.02043},
 primaryClass = {astro-ph.GA},
       adsurl = {https://ui.adsabs.harvard.edu/abs/2024PASA...41...99O}
}

@ARTICLE{2024MNRAS.534.3040D,
       author = {{Dors}, O.~L. and {Cardaci}, M.~V. and {H{\"a}gele}, G.~F. and {Valerdi}, M. and {Ilha}, G.~S. and {Oliveira}, C.~B. and {Riffel}, R.~A. and {Flury}, S.~R. and {Arellano-C{\'o}rdova}, K.~Z. and {Storchi-Bergmann}, T. and {Riffel}, R. and {Almeida}, G.~C. and {Morais}, I.~N.},
        title = "{Direct estimates of nitrogen abundance for Seyfert 2 nuclei}",
      journal = {\mnras},
         year = 2024,
        month = nov,
       volume = {534},
       number = {4},
        pages = {3040-3054},
          doi = {10.1093/mnras/stae2253},
archivePrefix = {arXiv},
       eprint = {2405.13906},
 primaryClass = {astro-ph.GA},
       adsurl = {https://ui.adsabs.harvard.edu/abs/2024MNRAS.534.3040D}
}

@ARTICLE{2007ApJ...671.1736I,
       author = {{Indriolo}, Nick and {Geballe}, Thomas R. and {Oka}, Takeshi and {McCall}, Benjamin J.},
        title = "{H$^{+}$$_{3}$ in Diffuse Interstellar Clouds: A Tracer for the Cosmic-Ray Ionization Rate}",
      journal = {\apj},
         year = 2007,
        month = dec,
       volume = {671},
       number = {2},
        pages = {1736-1747},
          doi = {10.1086/523036},
archivePrefix = {arXiv},
       eprint = {0709.1114},
 primaryClass = {astro-ph},
       adsurl = {https://ui.adsabs.harvard.edu/abs/2007ApJ...671.1736I}
}

@ARTICLE{2025A&A...693A.215K,
       author = {{Koutsoumpou}, E. and {Fern{\'a}ndez-Ontiveros}, J.~A. and {Dasyra}, K.~M. and {Spinoglio}, L.},
        title = "{Cosmic-ray ionization of low-excitation lines in active galactic nuclei and starburst galaxies}",
      journal = {\aap},
         year = 2025,
        month = jan,
       volume = {693},
          eid = {A215},
        pages = {A215},
          doi = {10.1051/0004-6361/202452232},
archivePrefix = {arXiv},
       eprint = {2411.17811},
 primaryClass = {astro-ph.GA},
       adsurl = {https://ui.adsabs.harvard.edu/abs/2025A&A...693A.215K}
}

@ARTICLE{Kraemer2000,
       author = {{Kraemer}, Steven B. and {Crenshaw}, D. Michael},
        title = "{Resolved Spectroscopy of the Narrow-Line Region in NGC 1068. III. Physical Conditions in the Emission-Line Gas}",
      journal = {\apj},
         year = 2000,
        month = dec,
       volume = {544},
       number = {2},
        pages = {763-779},
          doi = {10.1086/317246},
archivePrefix = {arXiv},
       eprint = {astro-ph/0007018},
 primaryClass = {astro-ph},
       adsurl = {https://ui.adsabs.harvard.edu/abs/2000ApJ...544..763K}
}

@ARTICLE{Stern2014,
       author = {{Stern}, Jonathan and {Laor}, Ari and {Baskin}, Alexei},
        title = "{Radiation pressure confinement - I. Ionized gas in the ISM of AGN hosts}",
      journal = {\mnras},
         year = 2014,
        month = feb,
       volume = {438},
       number = {2},
        pages = {901-921},
          doi = {10.1093/mnras/stt1843},
archivePrefix = {arXiv},
       eprint = {1309.7825},
 primaryClass = {astro-ph.CO},
       adsurl = {https://ui.adsabs.harvard.edu/abs/2014MNRAS.438..901S}
}

@ARTICLE{Dopita1996,
       author = {{Dopita}, Michael A. and {Sutherland}, Ralph S.},
        title = "{Spectral Signatures of Fast Shocks. I. Low-Density Model Grid}",
      journal = {\apjs},
         year = 1996,
        month = jan,
       volume = {102},
        pages = {161},
          doi = {10.1086/192255},
       adsurl = {https://ui.adsabs.harvard.edu/abs/1996ApJS..102..161D}
}

@ARTICLE{Contini2004,
       author = {{Contini}, Marcella},
        title = "{The complex structure of low-luminosity active galactic nuclei: NGC 4579}",
      journal = {\mnras},
         year = 2004,
        month = nov,
       volume = {354},
       number = {3},
        pages = {675-683},
          doi = {10.1111/j.1365-2966.2004.08222.x},
archivePrefix = {arXiv},
       eprint = {astro-ph/0407379},
 primaryClass = {astro-ph},
       adsurl = {https://ui.adsabs.harvard.edu/abs/2004MNRAS.354..675C}
}

@ARTICLE{Micelotta2010,
       author = {{Micelotta}, E.~R. and {Jones}, A.~P. and {Tielens}, A.~G.~G.~M.},
        title = "{Polycyclic aromatic hydrocarbon processing in interstellar shocks}",
      journal = {\aap},
         year = 2010,
        month = feb,
       volume = {510},
          eid = {A36},
        pages = {A36},
          doi = {10.1051/0004-6361/200911682},
archivePrefix = {arXiv},
       eprint = {0910.2461},
 primaryClass = {astro-ph.GA},
       adsurl = {https://ui.adsabs.harvard.edu/abs/2010A&A...510A..36M}
}

@ARTICLE{Binette1996,
       author = {{Binette}, L. and {Wilson}, A.~S. and {Storchi-Bergmann}, T.},
        title = "{Excitation and temperature of extended gas in active galaxies. II. Photoionization models with matter-bounded clouds.}",
      journal = {\aap},
         year = 1996,
        month = aug,
       volume = {312},
        pages = {365-379},
       adsurl = {https://ui.adsabs.harvard.edu/abs/1996A&A...312..365B}
}

@ARTICLE{Nagao2006a,
       author = {{Nagao}, T. and {Maiolino}, R. and {Marconi}, A.},
        title = "{Gas metallicity in the narrow-line regions of high-redshift active galactic nuclei}",
      journal = {\aap},
         year = 2006,
        month = mar,
       volume = {447},
       number = {3},
        pages = {863-876},
          doi = {10.1051/0004-6361:20054127},
archivePrefix = {arXiv},
       eprint = {astro-ph/0508652},
 primaryClass = {astro-ph},
       adsurl = {https://ui.adsabs.harvard.edu/abs/2006A&A...447..863N}
}

@ARTICLE{1981PASP...93....5B,
   author = {{Baldwin}, J.~A. and {Phillips}, M.~M. and {Terlevich}, R.},
    title = "{Classification parameters for the emission-line spectra of extragalactic objects}",
  journal = {\pasp},
     year = 1981,
    month = feb,
   volume = 93,
    pages = {5-19},
      doi = {10.1086/130766},
   adsurl = {http://adsabs.harvard.edu/abs/1981PASP...93....5B}
}

@ARTICLE{2015ApJ...815L..13R,
       author = {{Ricci}, C. and {Ueda}, Y. and {Koss}, M.~J. and {Trakhtenbrot}, B. and {Bauer}, F.~E. and {Gandhi}, P.},
        title = "{Compton-thick Accretion in the Local Universe}",
      journal = {\apjl},
         year = 2015,
        month = dec,
       volume = {815},
       number = {1},
          eid = {L13},
        pages = {L13},
          doi = {10.1088/2041-8205/815/1/L13},
archivePrefix = {arXiv},
       eprint = {1603.04852},
 primaryClass = {astro-ph.HE},
       adsurl = {https://ui.adsabs.harvard.edu/abs/2015ApJ...815L..13R}
}

@ARTICLE{2016ApJ...825...85K,
       author = {{Koss}, Michael J. and {Assef}, R. and {Balokovi{\'c}}, M. and {Stern}, D. and {Gandhi}, P. and {Lamperti}, I. and {Alexander}, D.~M. and {Ballantyne}, D.~R. and {Bauer}, F.~E. and {Berney}, S. and {Brandt}, W.~N. and {Comastri}, A. and {Gehrels}, N. and {Harrison}, F.~A. and {Lansbury}, G. and {Markwardt}, C. and {Ricci}, C. and {Rivers}, E. and {Schawinski}, K. and {Trakhtenbrot}, B. and {Treister}, E. and {Urry}, C. Megan},
        title = "{A New Population of Compton-thick AGNs Identified Using the Spectral Curvature above 10 keV}",
      journal = {\apj},
         year = 2016,
        month = jul,
       volume = {825},
       number = {2},
          eid = {85},
        pages = {85},
          doi = {10.3847/0004-637X/825/2/85},
archivePrefix = {arXiv},
       eprint = {1604.07825},
 primaryClass = {astro-ph.HE},
       adsurl = {https://ui.adsabs.harvard.edu/abs/2016ApJ...825...85K}
}

@ARTICLE{1994ApJ...429..572S,
       author = {{Storchi-Bergmann}, Thaisa and {Calzetti}, Daniela and {Kinney}, Anne L.},
        title = "{Ultraviolet to Near-Infrared Spectral Distributions of Star-forming Galaxies: Metallicity and Age Effects}",
      journal = {\apj},
         year = 1994,
        month = jul,
       volume = {429},
        pages = {572},
          doi = {10.1086/174345},
       adsurl = {https://ui.adsabs.harvard.edu/abs/1994ApJ...429..572S}
}

@ARTICLE{2011MNRAS.415.3616D,
       author = {{Dors}, Jr., O.~L. and {Krabbe}, Angela and {H{\"a}gele}, Guillermo F. and {P{\'e}rez-Montero}, Enrique},
        title = "{Analysing derived metallicities and ionization parameters from model-based determinations in ionized gaseous nebulae}",
      journal = {\mnras},
         year = 2011,
        month = aug,
       volume = {415},
       number = {4},
        pages = {3616-3626},
          doi = {10.1111/j.1365-2966.2011.18978.x},
archivePrefix = {arXiv},
       eprint = {1104.5460},
 primaryClass = {astro-ph.CO},
       adsurl = {https://ui.adsabs.harvard.edu/abs/2011MNRAS.415.3616D}
}

@ARTICLE{2014MNRAS.443.1291D,
       author = {{Dors}, Oli L. and {Cardaci}, M{\'o}nica V. and {H{\"a}gele}, Guillermo F. and {Krabbe}, {\^A}ngela C.},
        title = "{Metallicity evolution of AGNs from UV emission lines based on a new index}",
      journal = {\mnras},
         year = 2014,
        month = sep,
       volume = {443},
       number = {2},
        pages = {1291-1300},
          doi = {10.1093/mnras/stu1218},
archivePrefix = {arXiv},
       eprint = {1406.4832},
 primaryClass = {astro-ph.GA},
       adsurl = {https://ui.adsabs.harvard.edu/abs/2014MNRAS.443.1291D}
}

@ARTICLE{Calabro2023,
       author = {{Calabr{\`o}}, Antonello and {Pentericci}, Laura and {Feltre}, Anna and {Arrabal Haro}, Pablo and {Radovich}, Mario and {Seill{\'e}}, Lise-Marie and {Oliva}, Ernesto and {Daddi}, Emanuele and {Amor{\'\i}n}, Ricardo and {Bagley}, Micaela B. and {Bisigello}, Laura and {Buat}, V{\'e}ronique and {Castellano}, Marco and {Cleri}, Nikko J. and {Dickinson}, Mark and {Fern{\'a}ndez}, Vital and {Finkelstein}, Steven L. and {Giavalisco}, Mauro and {Grazian}, Andrea and {Hathi}, Nimish P. and {Hirschmann}, Michaela and {Juneau}, St{\'e}phanie and {Kartaltepe}, Jeyhan S. and {Koekemoer}, Anton M. and {Lucas}, Ray A. and {Papovich}, Casey and {P{\'e}rez-Gonz{\'a}lez}, Pablo G. and {Pirzkal}, Nor and {Santini}, Paola and {Trump}, Jonathan and {de la Vega}, Alexander and {Wilkins}, Stephen M. and {Yung}, L.~Y. Aaron and {Cassata}, Paolo and {Gobat}, Raphael A.~S. and {Mascia}, Sara and {Napolitano}, Lorenzo and {Vulcani}, Benedetta},
        title = "{Near-infrared emission line diagnostics for AGN from the local Universe to z {\ensuremath{\sim}} 3}",
      journal = {\aap},
         year = 2023,
        month = nov,
       volume = {679},
          eid = {A80},
        pages = {A80},
          doi = {10.1051/0004-6361/202347190},
archivePrefix = {arXiv},
       eprint = {2306.08605},
 primaryClass = {astro-ph.GA},
       adsurl = {https://ui.adsabs.harvard.edu/abs/2023A&A...679A..80C}
}

@article{Alloin1979,
  author  = {Alloin, D. and Collin-Souffrin, S. and Joly, M. and Vigroux, L.},
  year    = {1979},
  title   = {{Nitrogen and oxygen abundances in HII regions}},
  journal = {\aap},
  volume  = {78},
  pages   = {200},
  adsurl  = {https://ui.adsabs.harvard.edu/abs/1979A&A....78..200A}
}

@ARTICLE{VilaCostas1993,
       author = {{Vila-Costas}, M.~B. and {Edmunds}, M.~G.},
        title = "{The nitrogen-to-oxygen ratio in galaxies and its implications for the origin of nitrogen.}",
      journal = {\mnras},
         year = 1993,
        month = nov,
       volume = {265},
        pages = {199-212},
          doi = {10.1093/mnras/265.1.199},
       adsurl = {https://ui.adsabs.harvard.edu/abs/1993MNRAS.265..199V}
}

@ARTICLE{Castro2017,
       author = {{Castro}, C.~S. and {Dors}, O.~L. and {Cardaci}, M.~V. and {H{\"a}gele}, G.~F.},
        title = "{New metallicity calibration for Seyfert 2 galaxies based on the N2O2 index}",
      journal = {\mnras},
         year = 2017,
        month = may,
       volume = {467},
       number = {2},
        pages = {1507 (C17)-1514},
          doi = {10.1093/mnras/stx150},
archivePrefix = {arXiv},
       eprint = {1701.04997},
 primaryClass = {astro-ph.GA},
       adsurl = {https://ui.adsabs.harvard.edu/abs/2017MNRAS.467.1507C}
}

@ARTICLE{Contini1998,
       author = {{Contini}, Marcella and {Prieto}, M. Almudena and {Viegas}, Sueli M.},
        title = "{Gas and Dust Emission from the Nuclear Region of the Circinus Galaxy}",
      journal = {\apj},
         year = 1998,
        month = oct,
       volume = {505},
       number = {2},
        pages = {621-633},
          doi = {10.1086/306196},
archivePrefix = {arXiv},
       eprint = {astro-ph/9805029},
 primaryClass = {astro-ph},
       adsurl = {https://ui.adsabs.harvard.edu/abs/1998ApJ...505..621C}
}

@ARTICLE{Kewley2013,
       author = {{Kewley}, Lisa J. and {Dopita}, Michael A. and {Leitherer}, Claus and {Dav{\'e}}, Romeel and {Yuan}, Tiantian and {Allen}, Mark and {Groves}, Brent and {Sutherland}, Ralph},
        title = "{Theoretical Evolution of Optical Strong Lines across Cosmic Time}",
      journal = {\apj},
         year = 2013,
        month = sep,
       volume = {774},
       number = {2},
          eid = {100},
        pages = {100},
          doi = {10.1088/0004-637X/774/2/100},
archivePrefix = {arXiv},
       eprint = {1307.0508},
 primaryClass = {astro-ph.CO},
       adsurl = {https://ui.adsabs.harvard.edu/abs/2013ApJ...774..100K}
}

@ARTICLE{DAgostino2019b,
       author = {{D'Agostino}, Joshua J. and {Kewley}, Lisa J. and {Groves}, Brent A. and {Medling}, Anne M. and {Di Teodoro}, Enrico and {Dopita}, Michael A. and {Thomas}, Adam D. and {Sutherland}, Ralph S. and {Garcia-Burillo}, Santiago},
        title = "{Separating line emission from star formation, shocks, and AGN ionization in NGC 1068}",
      journal = {\mnras},
         year = 2019,
        month = aug,
       volume = {487},
       number = {3},
        pages = {4153-4168},
          doi = {10.1093/mnras/stz1611},
archivePrefix = {arXiv},
       eprint = {1906.07907},
 primaryClass = {astro-ph.GA},
       adsurl = {https://ui.adsabs.harvard.edu/abs/2019MNRAS.487.4153D}
}

@ARTICLE{Dors2021a,
       author = {{Dors}, O.~L. and {Contini}, M. and {Riffel}, R.~A. and {P{\'e}rez-Montero}, E. and {Krabbe}, A.~C. and {Cardaci}, M.~V. and {H{\"a}gele}, G.~F.},
        title = "{Chemical abundances of Seyfert 2 AGNs - IV. Composite models calculated by photoionization + shocks}",
      journal = {\mnras},
         year = 2021,
        month = feb,
       volume = {501},
       number = {1},
        pages = {1370-1383},
          doi = {10.1093/mnras/staa3707},
archivePrefix = {arXiv},
       eprint = {2011.12103},
 primaryClass = {astro-ph.GA},
       adsurl = {https://ui.adsabs.harvard.edu/abs/2021MNRAS.501.1370D}
}

@ARTICLE{Dopita2013,
       author = {{Dopita}, Michael A. and {Sutherland}, Ralph S. and {Nicholls}, David C. and {Kewley}, Lisa J. and {Vogt}, Fr{\'e}d{\'e}ric P.~A.},
        title = "{New Strong-line Abundance Diagnostics for H II Regions: Effects of {\ensuremath{\kappa}}-distributed Electron Energies and New Atomic Data}",
      journal = {\apjs},
         year = 2013,
        month = sep,
       volume = {208},
       number = {1},
          eid = {10},
        pages = {10},
          doi = {10.1088/0067-0049/208/1/10},
archivePrefix = {arXiv},
       eprint = {1307.5950},
 primaryClass = {astro-ph.CO},
       adsurl = {https://ui.adsabs.harvard.edu/abs/2013ApJS..208...10D}
}

@ARTICLE{Dopita2016,
       author = {{Dopita}, Michael A. and {Kewley}, Lisa J. and {Sutherland}, Ralph S. and {Nicholls}, David C.},
        title = "{Chemical abundances in high-redshift galaxies: a powerful new emission line diagnostic}",
      journal = {\apss},
         year = 2016,
        month = feb,
       volume = {361},
          eid = {61},
        pages = {61},
          doi = {10.1007/s10509-016-2657-8},
archivePrefix = {arXiv},
       eprint = {1601.01337},
 primaryClass = {astro-ph.GA},
       adsurl = {https://ui.adsabs.harvard.edu/abs/2016Ap&SS.361...61D}
}

@ARTICLE{Dors2015,
       author = {{Dors}, O.~L. and {Cardaci}, M.~V. and {H{\"a}gele}, G.~F. and {Rodrigues}, I. and {Grebel}, E.~K. and {Pilyugin}, L.~S. and {Freitas-Lemes}, P. and {Krabbe}, A.~C.},
        title = "{On the central abundances of active galactic nuclei and star-forming galaxies}",
      journal = {\mnras},
         year = 2015,
        month = nov,
       volume = {453},
       number = {4},
        pages = {4102-4111},
          doi = {10.1093/mnras/stv1916},
archivePrefix = {arXiv},
       eprint = {1508.07802},
 primaryClass = {astro-ph.GA},
       adsurl = {https://ui.adsabs.harvard.edu/abs/2015MNRAS.453.4102D}
}

@ARTICLE{Dors2020c,
       author = {{Dors}, O.~L. and {Maiolino}, R. and {Cardaci}, M.~V. and {H{\"a}gele}, G.~F. and {Krabbe}, A.~C. and {P{\'e}rez-Montero}, E. and {Armah}, M.},
        title = "{Chemical abundances of Seyfert 2 AGNs - III. Reducing the oxygen abundance discrepancy}",
      journal = {\mnras},
         year = 2020,
        month = aug,
       volume = {496},
       number = {3},
        pages = {3209-3221},
          doi = {10.1093/mnras/staa1781},
archivePrefix = {arXiv},
       eprint = {2006.09152},
 primaryClass = {astro-ph.GA},
       adsurl = {https://ui.adsabs.harvard.edu/abs/2020MNRAS.496.3209D}
}

@article{Ferland2017,
  author        = {Ferland, G. J. and Chatzikos, M. and Guzmán, F. and Lykins, M. L. and van Hoof, P. A. M. and Williams, R. J. R. and Abel, N. P. and Badnell, N. R. and Keenan, F. P. and Porter, R. L. and Stancil, P. C.},
  year          = {2017},
  title         = {{The 2017 Release of Cloudy}},
  journal       = {\rmxaa},
  volume        = {53},
  pages         = {385},
  eprint        = {1705.10877},
  archiveprefix = {arXiv},
  primaryclass  = {astro-ph.GA},
  adsurl        = {https://ui.adsabs.harvard.edu/abs/2017RMxAA..53..385F}
}

@ARTICLE{Ferland1998,
       author = {{Ferland}, G.~J. and {Korista}, K.~T. and {Verner}, D.~A. and {Ferguson}, J.~W. and {Kingdon}, J.~B. and {Verner}, E.~M.},
        title = "{CLOUDY 90: Numerical Simulation of Plasmas and Their Spectra}",
      journal = {\pasp},
         year = 1998,
        month = jul,
       volume = {110},
       number = {749},
        pages = {761-778},
          doi = {10.1086/316190},
       adsurl = {https://ui.adsabs.harvard.edu/abs/1998PASP..110..761F}
}

@ARTICLE{Ferland2013,
       author = {{Ferland}, G.~J. and {Porter}, R.~L. and {van Hoof}, P.~A.~M. and {Williams}, R.~J.~R. and {Abel}, N.~P. and {Lykins}, M.~L. and {Shaw}, G. and {Henney}, W.~J. and {Stancil}, P.~C.},
        title = "{The 2013 Release of Cloudy}",
      journal = {\rmxaa},
         year = 2013,
        month = apr,
       volume = {49},
        pages = {137-163},
          doi = {10.48550/arXiv.1302.4485},
archivePrefix = {arXiv},
       eprint = {1302.4485},
 primaryClass = {astro-ph.GA},
       adsurl = {https://ui.adsabs.harvard.edu/abs/2013RMxAA..49..137F}
}

@ARTICLE{Binette1985,
       author = {{Binette}, L. and {Dopita}, M.~A. and {Tuohy}, I.~R.},
        title = "{Radiative shock-wave theory. II. High-velocity shocks and thermal instabilities.}",
      journal = {\apj},
         year = 1985,
        month = oct,
       volume = {297},
        pages = {476-491},
          doi = {10.1086/163544},
       adsurl = {https://ui.adsabs.harvard.edu/abs/1985ApJ...297..476B}
}

@ARTICLE{Sutherland1993,
       author = {{Sutherland}, Ralph S. and {Dopita}, M.~A.},
        title = "{Cooling Functions for Low-Density Astrophysical Plasmas}",
      journal = {\apjs},
         year = 1993,
        month = sep,
       volume = {88},
        pages = {253},
          doi = {10.1086/191823},
       adsurl = {https://ui.adsabs.harvard.edu/abs/1993ApJS...88..253S}
}

@ARTICLE{Sutherland2017,
       author = {{Sutherland}, Ralph S. and {Dopita}, Michael A.},
        title = "{Effects of Preionization in Radiative Shocks. I. Self-consistent Models}",
      journal = {\apjs},
         year = 2017,
        month = apr,
       volume = {229},
       number = {2},
          eid = {34},
        pages = {34},
          doi = {10.3847/1538-4365/aa6541},
archivePrefix = {arXiv},
       eprint = {1702.07453},
 primaryClass = {astro-ph.IM},
       adsurl = {https://ui.adsabs.harvard.edu/abs/2017ApJS..229...34S}
}

@ARTICLE{Chatzikos2023,
       author = {{Gunasekera}, Chamani M. and {van Hoof}, Peter A.~M. and {Chatzikos}, Marios and {Ferland}, Gary J.},
        title = "{The 23.01 Release of Cloudy}",
      journal = {Research Notes of the American Astronomical Society},
         year = 2023,
        month = nov,
       volume = {7},
       number = {11},
          eid = {246},
        pages = {246},
          doi = {10.3847/2515-5172/ad0e75},
archivePrefix = {arXiv},
       eprint = {2311.10163},
 primaryClass = {astro-ph.GA},
       adsurl = {https://ui.adsabs.harvard.edu/abs/2023RNAAS...7..246G}
}

@ARTICLE{Shakura1973,
       author = {{Shakura}, N.~I. and {Sunyaev}, R.~A.},
        title = "{Black holes in binary systems. Observational appearance.}",
      journal = {\aap},
         year = 1973,
        month = jan,
       volume = {24},
        pages = {337-355},
       adsurl = {https://ui.adsabs.harvard.edu/abs/1973A&A....24..337S}
}

@ARTICLE{Tananbaum1979,
       author = {{Tananbaum}, H. and {Avni}, Y. and {Branduardi}, G. and {Elvis}, M. and {Fabbiano}, G. and {Feigelson}, E. and {Giacconi}, R. and {Henry}, J.~P. and {Pye}, J.~P. and {Soltan}, A. and {Zamorani}, G.},
        title = "{X-ray studies of quasars with the Einstein Observatory.}",
      journal = {\apjl},
         year = 1979,
        month = nov,
       volume = {234},
        pages = {L9-L13},
          doi = {10.1086/183100},
       adsurl = {https://ui.adsabs.harvard.edu/abs/1979ApJ...234L...9T}
}

@ARTICLE{Marconi2004,
       author = {{Marconi}, A. and {Risaliti}, G. and {Gilli}, R. and {Hunt}, L.~K. and {Maiolino}, R. and {Salvati}, M.},
        title = "{Local supermassive black holes, relics of active galactic nuclei and the X-ray background}",
      journal = {\mnras},
         year = 2004,
        month = jun,
       volume = {351},
       number = {1},
        pages = {169-185},
          doi = {10.1111/j.1365-2966.2004.07765.x},
archivePrefix = {arXiv},
       eprint = {astro-ph/0311619},
 primaryClass = {astro-ph},
       adsurl = {https://ui.adsabs.harvard.edu/abs/2004MNRAS.351..169M}
}

@ARTICLE{Ichikawa2017,
       author = {{Ichikawa}, Kohei and {Ricci}, Claudio and {Ueda}, Yoshihiro and {Matsuoka}, Kenta and {Toba}, Yoshiki and {Kawamuro}, Taiki and {Trakhtenbrot}, Benny and {Koss}, Michael J.},
        title = "{The Complete Infrared View of Active Galactic Nuclei from the 70 Month Swift/BAT Catalog}",
      journal = {\apj},
         year = 2017,
        month = jan,
       volume = {835},
       number = {1},
          eid = {74},
        pages = {74},
          doi = {10.3847/1538-4357/835/1/74},
archivePrefix = {arXiv},
       eprint = {1611.09858},
 primaryClass = {astro-ph.GA},
       adsurl = {https://ui.adsabs.harvard.edu/abs/2017ApJ...835...74I}
}

@ARTICLE{Brandt2015,
       author = {{Brandt}, W.~N. and {Alexander}, D.~M.},
        title = "{Cosmic X-ray surveys of distant active galaxies. The demographics, physics, and ecology of growing supermassive black holes}",
      journal = {\aapr},
         year = 2015,
        month = jan,
       volume = {23},
          eid = {1},
        pages = {1},
          doi = {10.1007/s00159-014-0081-z},
archivePrefix = {arXiv},
       eprint = {1501.01982},
 primaryClass = {astro-ph.HE},
       adsurl = {https://ui.adsabs.harvard.edu/abs/2015A&ARv..23....1B}
}

@ARTICLE{Just2007,
       author = {{Just}, D.~W. and {Brandt}, W.~N. and {Shemmer}, O. and {Steffen}, A.~T. and {Schneider}, D.~P. and {Chartas}, G. and {Garmire}, G.~P.},
        title = "{The X-Ray Properties of the Most Luminous Quasars from the Sloan Digital Sky Survey}",
      journal = {\apj},
         year = 2007,
        month = aug,
       volume = {665},
       number = {2},
        pages = {1004-1022},
          doi = {10.1086/519990},
archivePrefix = {arXiv},
       eprint = {0705.3059},
 primaryClass = {astro-ph},
       adsurl = {https://ui.adsabs.harvard.edu/abs/2007ApJ...665.1004J}
}

@ARTICLE{2006AJ....131.2826S,
       author = {{Steffen}, A.~T. and {Strateva}, I. and {Brandt}, W.~N. and {Alexander}, D.~M. and {Koekemoer}, A.~M. and {Lehmer}, B.~D. and {Schneider}, D.~P. and {Vignali}, C.},
        title = "{The X-Ray-to-Optical Properties of Optically Selected Active Galaxies over Wide Luminosity and Redshift Ranges}",
      journal = {\aj},
         year = 2006,
        month = jun,
       volume = {131},
       number = {6},
        pages = {2826-2842},
          doi = {10.1086/503627},
archivePrefix = {arXiv},
       eprint = {astro-ph/0602407},
 primaryClass = {astro-ph},
       adsurl = {https://ui.adsabs.harvard.edu/abs/2006AJ....131.2826S}
}

@ARTICLE{Lusso2010,
       author = {{Lusso}, E. and {Comastri}, A. and {Vignali}, C. and {Zamorani}, G. and {Brusa}, M. and {Gilli}, R. and {Iwasawa}, K. and {Salvato}, M. and {Civano}, F. and {Elvis}, M. and {Merloni}, A. and {Bongiorno}, A. and {Trump}, J.~R. and {Koekemoer}, A.~M. and {Schinnerer}, E. and {Le Floc'h}, E. and {Cappelluti}, N. and {Jahnke}, K. and {Sargent}, M. and {Silverman}, J. and {Mainieri}, V. and {Fiore}, F. and {Bolzonella}, M. and {Le F{\`e}vre}, O. and {Garilli}, B. and {Iovino}, A. and {Kneib}, J.~P. and {Lamareille}, F. and {Lilly}, S. and {Mignoli}, M. and {Scodeggio}, M. and {Vergani}, D.},
        title = "{The X-ray to optical-UV luminosity ratio of X-ray selected type 1 AGN in XMM-COSMOS}",
      journal = {\aap},
         year = 2010,
        month = mar,
       volume = {512},
          eid = {A34},
        pages = {A34},
          doi = {10.1051/0004-6361/200913298},
archivePrefix = {arXiv},
       eprint = {0912.4166},
 primaryClass = {astro-ph.CO},
       adsurl = {https://ui.adsabs.harvard.edu/abs/2010A&A...512A..34L}
}

@ARTICLE{Lusso2017,
       author = {{Lusso}, E. and {Risaliti}, G.},
        title = "{Quasars as standard candles. I. The physical relation between disc and coronal emission}",
      journal = {\aap},
         year = 2017,
        month = jun,
       volume = {602},
          eid = {A79},
        pages = {A79},
          doi = {10.1051/0004-6361/201630079},
archivePrefix = {arXiv},
       eprint = {1703.05299},
 primaryClass = {astro-ph.HE},
       adsurl = {https://ui.adsabs.harvard.edu/abs/2017A&A...602A..79L}
}

@ARTICLE{Gupta2018,
       author = {{Gupta}, Maitrayee and {Sikora}, Marek and {Rusinek}, Katarzyna and {Madejski}, Greg M.},
        title = "{Comparison of hard X-ray spectra of luminous radio galaxies and their radio-quiet counterparts}",
      journal = {\mnras},
         year = 2018,
        month = nov,
       volume = {480},
       number = {3},
        pages = {2861-2871},
          doi = {10.1093/mnras/sty2043},
archivePrefix = {arXiv},
       eprint = {1808.07170},
 primaryClass = {astro-ph.HE},
       adsurl = {https://ui.adsabs.harvard.edu/abs/2018MNRAS.480.2861G}
}

@ARTICLE{Gupta2020,
       author = {{Gupta}, Maitrayee and {Sikora}, Marek and {Rusinek}, Katarzyna},
        title = "{Comparison of SEDs of very massive radio-loud and radio-quiet AGN}",
      journal = {\mnras},
         year = 2020,
        month = feb,
       volume = {492},
       number = {1},
        pages = {315-325},
          doi = {10.1093/mnras/stz3384},
archivePrefix = {arXiv},
       eprint = {1904.04684},
 primaryClass = {astro-ph.HE},
       adsurl = {https://ui.adsabs.harvard.edu/abs/2020MNRAS.492..315G}
}

@ARTICLE{Nagao2006b,
       author = {{Nagao}, T. and {Maiolino}, R. and {Marconi}, A.},
        title = "{Gas metallicity diagnostics in star-forming galaxies}",
      journal = {\aap},
         year = 2006,
        month = nov,
       volume = {459},
       number = {1},
        pages = {85-101},
          doi = {10.1051/0004-6361:20065216},
archivePrefix = {arXiv},
       eprint = {astro-ph/0603580},
 primaryClass = {astro-ph},
       adsurl = {https://ui.adsabs.harvard.edu/abs/2006A&A...459...85N}
}

@ARTICLE{Groves2004a,
       author = {{Groves}, Brent A. and {Dopita}, Michael A. and {Sutherland}, Ralph S.},
        title = "{Dusty, Radiation Pressure-Dominated Photoionization. I. Model Description, Structure, and Grids}",
      journal = {\apjs},
         year = 2004,
        month = jul,
       volume = {153},
       number = {1},
        pages = {9-73},
          doi = {10.1086/421113},
archivePrefix = {arXiv},
       eprint = {astro-ph/0404175},
 primaryClass = {astro-ph},
       adsurl = {https://ui.adsabs.harvard.edu/abs/2004ApJS..153....9G}
}

@ARTICLE{1999ARA&A..37..487H,
       author = {{Hamann}, Fred and {Ferland}, Gary},
        title = "{Elemental Abundances in Quasistellar Objects: Star Formation and Galactic Nuclear Evolution at High Redshifts}",
      journal = {\araa},
         year = 1999,
        month = jan,
       volume = {37},
        pages = {487-531},
          doi = {10.1146/annurev.astro.37.1.487},
archivePrefix = {arXiv},
       eprint = {astro-ph/9904223},
 primaryClass = {astro-ph},
       adsurl = {https://ui.adsabs.harvard.edu/abs/1999ARA&A..37..487H}
}

@ARTICLE{Kewley2001,
       author = {{Kewley}, L.~J. and {Dopita}, M.~A. and {Sutherland}, R.~S. and {Heisler}, C.~A. and {Trevena}, J.},
        title = "{Theoretical Modeling of Starburst Galaxies}",
      journal = {\apj},
         year = 2001,
        month = jul,
       volume = {556},
       number = {1},
        pages = {121-140},
          doi = {10.1086/321545},
archivePrefix = {arXiv},
       eprint = {astro-ph/0106324},
 primaryClass = {astro-ph},
       adsurl = {https://ui.adsabs.harvard.edu/abs/2001ApJ...556..121K}
}

@ARTICLE{CidFernandes2011,
       author = {{Cid Fernandes}, R. and {Stasi{\'n}ska}, G. and {Mateus}, A. and {Vale Asari}, N.},
        title = "{A comprehensive classification of galaxies in the Sloan Digital Sky Survey: how to tell true from fake AGN?}",
      journal = {\mnras},
         year = 2011,
        month = may,
       volume = {413},
       number = {3},
        pages = {1687-1699},
          doi = {10.1111/j.1365-2966.2011.18244.x},
archivePrefix = {arXiv},
       eprint = {1012.4426},
 primaryClass = {astro-ph.CO},
       adsurl = {https://ui.adsabs.harvard.edu/abs/2011MNRAS.413.1687C}
}

@ARTICLE{Davies2014,
       author = {{Davies}, Rebecca L. and {Kewley}, Lisa J. and {Ho}, I.-Ting and {Dopita}, Michael A.},
        title = "{Starburst-AGN mixing - II. Optically selected active galaxies}",
      journal = {\mnras},
         year = 2014,
        month = nov,
       volume = {444},
       number = {4},
        pages = {3961-3974},
          doi = {10.1093/mnras/stu1740},
archivePrefix = {arXiv},
       eprint = {1408.5888},
 primaryClass = {astro-ph.GA},
       adsurl = {https://ui.adsabs.harvard.edu/abs/2014MNRAS.444.3961D}
}

@ARTICLE{Kewley2019,
       author = {{Kewley}, Lisa J. and {Nicholls}, David C. and {Sutherland}, Ralph S.},
        title = "{Understanding Galaxy Evolution Through Emission Lines}",
      journal = {\araa},
         year = 2019,
        month = aug,
       volume = {57},
        pages = {511-570},
          doi = {10.1146/annurev-astro-081817-051832},
archivePrefix = {arXiv},
       eprint = {1910.09730},
 primaryClass = {astro-ph.GA},
       adsurl = {https://ui.adsabs.harvard.edu/abs/2019ARA&A..57..511K}
}

@ARTICLE{Koss2017,
       author = {{Koss}, Michael and {Trakhtenbrot}, Benny and {Ricci}, Claudio and {Lamperti}, Isabella and {Oh}, Kyuseok and {Berney}, Simon and {Schawinski}, Kevin and {Balokovi{\'c}}, Mislav and {Baronchelli}, Linda and {Crenshaw}, D. Michael and {Fischer}, Travis and {Gehrels}, Neil and {Harrison}, Fiona and {Hashimoto}, Yasuhiro and {Hogg}, Drew and {Ichikawa}, Kohei and {Masetti}, Nicola and {Mushotzky}, Richard and {Sartori}, Lia and {Stern}, Daniel and {Treister}, Ezequiel and {Ueda}, Yoshihiro and {Veilleux}, Sylvain and {Winter}, Lisa},
        title = "{BAT AGN Spectroscopic Survey. I. Spectral Measurements, Derived Quantities, and AGN Demographics}",
      journal = {\apj},
         year = 2017,
        month = nov,
       volume = {850},
       number = {1},
          eid = {74},
        pages = {74},
          doi = {10.3847/1538-4357/aa8ec9},
archivePrefix = {arXiv},
       eprint = {1707.08123},
 primaryClass = {astro-ph.HE},
       adsurl = {https://ui.adsabs.harvard.edu/abs/2017ApJ...850...74K}
}

@ARTICLE{Koss2022a,
       author = {{Koss}, Michael J. and {Trakhtenbrot}, Benny and {Ricci}, Claudio and {Bauer}, Franz E. and {Treister}, Ezequiel and {Mushotzky}, Richard and {Urry}, C. Megan and {Ananna}, Tonima T. and {Balokovi{\'c}}, Mislav and {den Brok}, Jakob S. and {Cenko}, S. Bradley and {Harrison}, Fiona and {Ichikawa}, Kohei and {Lamperti}, Isabella and {Lein}, Amy and {Mej{\'\i}a-Restrepo}, Julian E. and {Oh}, Kyuseok and {Pacucci}, Fabio and {Pfeifle}, Ryan W. and {Powell}, Meredith C. and {Privon}, George C. and {Ricci}, Federica and {Salvato}, Mara and {Schawinski}, Kevin and {Shimizu}, Taro and {Smith}, Krista L. and {Stern}, Daniel},
        title = "{BASS. XXI. The Data Release 2 Overview}",
      journal = {\apjs},
         year = 2022,
        month = jul,
       volume = {261},
       number = {1},
          eid = {1},
        pages = {1},
          doi = {10.3847/1538-4365/ac6c8f},
archivePrefix = {arXiv},
       eprint = {2207.12428},
 primaryClass = {astro-ph.GA},
       adsurl = {https://ui.adsabs.harvard.edu/abs/2022ApJS..261....1K}
}

@ARTICLE{Oh2022,
       author = {{Oh}, Kyuseok and {Koss}, Michael J. and {Ueda}, Yoshihiro and {Stern}, Daniel and {Ricci}, Claudio and {Trakhtenbrot}, Benny and {Powell}, Meredith C. and {den Brok}, Jakob S. and {Lamperti}, Isabella and {Mushotzky}, Richard and {Ricci}, Federica and {B{\"a}r}, Rudolf E. and {Rojas}, Alejandra F. and {Ichikawa}, Kohei and {Riffel}, Rog{\'e}rio and {Treister}, Ezequiel and {Harrison}, Fiona and {Urry}, C. Megan and {Bauer}, Franz E. and {Schawinski}, Kevin},
        title = "{BASS. XXIV. The BASS DR2 Spectroscopic Line Measurements and AGN Demographics}",
      journal = {\apjs},
         year = 2022,
        month = jul,
       volume = {261},
       number = {1},
          eid = {4},
        pages = {4},
          doi = {10.3847/1538-4365/ac5b68},
archivePrefix = {arXiv},
       eprint = {2203.00017},
 primaryClass = {astro-ph.GA},
       adsurl = {https://ui.adsabs.harvard.edu/abs/2022ApJS..261....4O}
}

@ARTICLE{Luridiana2015,
       author = {{Luridiana}, V. and {Morisset}, C. and {Shaw}, R.~A.},
        title = "{PyNeb: a new tool for analyzing emission lines. I. Code description and validation of results}",
      journal = {\aap},
         year = 2015,
        month = jan,
       volume = {573},
          eid = {A42},
        pages = {A42},
          doi = {10.1051/0004-6361/201323152},
archivePrefix = {arXiv},
       eprint = {1410.6662},
 primaryClass = {astro-ph.IM},
       adsurl = {https://ui.adsabs.harvard.edu/abs/2015A&A...573A..42L}
}

@ARTICLE{Binette2024,
       author = {{Binette}, Luc and {Zovaro}, Henry R.~M. and {Villar Mart{\'\i}n}, Montserrat and {Dors}, Oli L. and {Krongold}, Yair and {Morisset}, Christophe and {Revalski}, Mitchell and {Alarie}, Alexandre and {Riffel}, Rogemar A. and {Dopita}, Michael A.},
        title = "{Constraints on the densities and temperature of the Seyfert 2 narrow line region}",
      journal = {\aap},
         year = 2024,
        month = apr,
       volume = {684},
          eid = {A53},
        pages = {A53},
          doi = {10.1051/0004-6361/202245754},
archivePrefix = {arXiv},
       eprint = {2401.06972},
 primaryClass = {astro-ph.GA},
       adsurl = {https://ui.adsabs.harvard.edu/abs/2024A&A...684A..53B}
}

@ARTICLE{Peimbert1967,
       author = {{Peimbert}, Manuel},
        title = "{Temperature Determinations of H II Regions}",
      journal = {\apj},
         year = "1967",
        month = "Dec",
       volume = {150},
        pages = {825},
          doi = {10.1086/149385},
       adsurl = {https://ui.adsabs.harvard.edu/abs/1967ApJ...150..825P}
}

@ARTICLE{Garnett1992,
       author = {{Garnett}, Donald R.},
        title = "{Electron Temperature Variations and the Measurement of Nebular Abundances}",
      journal = {\aj},
         year = 1992,
        month = apr,
       volume = {103},
        pages = {1330},
          doi = {10.1086/116146},
       adsurl = {https://ui.adsabs.harvard.edu/abs/1992AJ....103.1330G}
}

@ARTICLE{2022ApJ...930...14R,
       author = {{Revalski}, Mitchell and {Crenshaw}, D. Michael and {Rafelski}, Marc and {Kraemer}, Steven B. and {Polack}, Garrett E. and {Falc{\~a}o}, Anna Trindade and {Fischer}, Travis C. and {Meena}, Beena and {Martinez}, Francisco and {Schmitt}, Henrique R. and {Collins}, Nicholas R. and {Falcone}, Julia},
        title = "{Quantifying Feedback from Narrow Line Region Outflows in Nearby Active Galaxies. IV. The Effects of Different Density Estimates on the Ionized Gas Masses and Outflow Rates}",
      journal = {\apj},
         year = 2022,
        month = may,
       volume = {930},
       number = {1},
          eid = {14},
        pages = {14},
          doi = {10.3847/1538-4357/ac5f3d},
archivePrefix = {arXiv},
       eprint = {2203.07387},
 primaryClass = {astro-ph.GA},
       adsurl = {https://ui.adsabs.harvard.edu/abs/2022ApJ...930...14R}
}

@ARTICLE{2021ApJ...910..139R,
       author = {{Revalski}, Mitchell and {Meena}, Beena and {Martinez}, Francisco and {Polack}, Garrett E. and {Crenshaw}, D. Michael and {Kraemer}, Steven B. and {Collins}, Nicholas R. and {Fischer}, Travis C. and {Schmitt}, Henrique R. and {Schmidt}, Judy and {Maksym}, W. Peter and {Rafelski}, Marc},
        title = "{Quantifying Feedback from Narrow Line Region Outflows in Nearby Active Galaxies. III. Results for the Seyfert 2 Galaxies Markarian 3, Markarian 78, and NGC 1068}",
      journal = {\apj},
         year = 2021,
        month = apr,
       volume = {910},
       number = {2},
          eid = {139},
        pages = {139},
          doi = {10.3847/1538-4357/abdcad},
archivePrefix = {arXiv},
       eprint = {2101.06270},
 primaryClass = {astro-ph.GA},
       adsurl = {https://ui.adsabs.harvard.edu/abs/2021ApJ...910..139R}
}

@ARTICLE{2018ApJ...856...46R,
       author = {{Revalski}, M. and {Crenshaw}, D.~M. and {Kraemer}, S.~B. and {Fischer}, T.~C. and {Schmitt}, H.~R. and {Machuca}, C.},
        title = "{Quantifying Feedback from Narrow Line Region Outflows in Nearby Active Galaxies. I. Spatially Resolved Mass Outflow Rates for the Seyfert 2 Galaxy Markarian 573}",
      journal = {\apj},
         year = 2018,
        month = mar,
       volume = {856},
       number = {1},
          eid = {46},
        pages = {46},
          doi = {10.3847/1538-4357/aab107},
archivePrefix = {arXiv},
       eprint = {1802.07734},
 primaryClass = {astro-ph.GA},
       adsurl = {https://ui.adsabs.harvard.edu/abs/2018ApJ...856...46R}
}

@ARTICLE{2024MNRAS.527.8193D,
       author = {{Dors}, Oli L. and {Cardaci}, M.~V. and {H{\"a}gele}, G.~F. and {Ilha}, G.~S. and {Oliveira}, C.~B. and {Riffel}, R.~A. and {Riffel}, R. and {Krabbe}, A.~C.},
        title = "{Cosmic metallicity evolution of Active Galactic Nuclei: implications for optical diagnostic diagrams}",
      journal = {\mnras},
         year = 2024,
        month = jan,
       volume = {527},
       number = {3},
        pages = {8193-8212},
          doi = {10.1093/mnras/stad3667},
archivePrefix = {arXiv},
       eprint = {2311.14026},
 primaryClass = {astro-ph.GA},
       adsurl = {https://ui.adsabs.harvard.edu/abs/2024MNRAS.527.8193D}
}

@ARTICLE{Asplund2021,
       author = {{Asplund}, M. and {Amarsi}, A.~M. and {Grevesse}, N.},
        title = "{The chemical make-up of the Sun: A 2020 vision}",
      journal = {\aap},
         year = 2021,
        month = sep,
       volume = {653},
          eid = {A141},
        pages = {A141},
          doi = {10.1051/0004-6361/202140445},
archivePrefix = {arXiv},
       eprint = {2105.01661},
 primaryClass = {astro-ph.SR},
       adsurl = {https://ui.adsabs.harvard.edu/abs/2021A&A...653A.141A}
}

@ARTICLE{Komossa2008,
       author = {{Komossa}, S. and {Xu}, D. and {Zhou}, H. and {Storchi-Bergmann}, T. and {Binette}, L.},
        title = "{On the Nature of Seyfert Galaxies with High [O III] {\ensuremath{\lambda}}5007 Blueshifts}",
      journal = {\apj},
         year = 2008,
        month = jun,
       volume = {680},
       number = {2},
        pages = {926-938},
          doi = {10.1086/587932},
archivePrefix = {arXiv},
       eprint = {0803.0240},
 primaryClass = {astro-ph},
       adsurl = {https://ui.adsabs.harvard.edu/abs/2008ApJ...680..926K}
}

@ARTICLE{Groves2004b,
       author = {{Groves}, Brent A. and {Dopita}, Michael A. and {Sutherland}, Ralph S.},
        title = "{Dust in Photoionized Nebulae. I. The Effect on Emission-Line Ratios and the DUSTY Code}",
      journal = {\apjs},
         year = 2004,
        month = jul,
       volume = {153},
       number = {1},
        pages = {75-91},
          doi = {10.1086/421113},
       adsurl = {https://ui.adsabs.harvard.edu/abs/2004ApJS..153...75G}
}

@ARTICLE{Groves2006b,
       author = {{Groves}, B. and {Dopita}, M. and {Sutherland}, R.},
        title = "{The infrared emission from the narrow line region}",
      journal = {\aap},
         year = 2006,
        month = nov,
       volume = {458},
       number = {2},
        pages = {405-416},
          doi = {10.1051/0004-6361:20065097},
archivePrefix = {arXiv},
       eprint = {astro-ph/0608057},
 primaryClass = {astro-ph},
       adsurl = {https://ui.adsabs.harvard.edu/abs/2006A&A...458..405G}
}

@ARTICLE{Ferguson1997,
       author = {{Ferguson}, Jason W. and {Korista}, Kirk T. and {Baldwin}, Jack A. and {Ferland}, Gary J.},
        title = "{Locally Optimally Emitting Clouds and the Narrow Emission Lines in Seyfert Galaxies}",
      journal = {\apj},
         year = 1997,
        month = sep,
       volume = {487},
       number = {1},
        pages = {122-141},
          doi = {10.1086/304611},
archivePrefix = {arXiv},
       eprint = {astro-ph/9705083},
 primaryClass = {astro-ph},
       adsurl = {https://ui.adsabs.harvard.edu/abs/1997ApJ...487..122F}
}

@ARTICLE{Netzer2015,
       author = {{Netzer}, Hagai},
        title = "{Revisiting the Unified Model of Active Galactic Nuclei}",
      journal = {\araa},
         year = 2015,
        month = aug,
       volume = {53},
        pages = {365-408},
          doi = {10.1146/annurev-astro-082214-122302},
archivePrefix = {arXiv},
       eprint = {1505.00811},
 primaryClass = {astro-ph.GA},
       adsurl = {https://ui.adsabs.harvard.edu/abs/2015ARA&A..53..365N}
}

@ARTICLE{Rose2018,
       author = {{Rose}, Marvin and {Tadhunter}, Clive and {Ramos Almeida}, Cristina and {Rodr{\'\i}guez Zaur{\'\i}n}, Javier and {Santoro}, Francesco and {Spence}, Robert},
        title = "{Quantifying the AGN-driven outflows in ULIRGs (QUADROS) - I: VLT/Xshooter observations of nine nearby objects}",
      journal = {\mnras},
         year = 2018,
        month = feb,
       volume = {474},
       number = {1},
        pages = {128-156},
          doi = {10.1093/mnras/stx2590},
archivePrefix = {arXiv},
       eprint = {1710.06600},
 primaryClass = {astro-ph.GA},
       adsurl = {https://ui.adsabs.harvard.edu/abs/2018MNRAS.474..128R}
}

@ARTICLE{Baron2019,
       author = {{Baron}, Dalya and {Netzer}, Hagai},
        title = "{Discovering AGN-driven winds through their infrared emission - II. Mass outflow rate and energetics}",
      journal = {\mnras},
         year = 2019,
        month = jul,
       volume = {486},
       number = {3},
        pages = {4290-4303},
          doi = {10.1093/mnras/stz1070},
archivePrefix = {arXiv},
       eprint = {1903.11076},
 primaryClass = {astro-ph.GA},
       adsurl = {https://ui.adsabs.harvard.edu/abs/2019MNRAS.486.4290B}
}

@ARTICLE{Davies2020,
       author = {{Davies}, R. and {Baron}, D. and {Shimizu}, T. and {Netzer}, H. and {Burtscher}, L. and {de Zeeuw}, P.~T. and {Genzel}, R. and {Hicks}, E.~K.~S. and {Koss}, M. and {Lin}, M. -Y. and {Lutz}, D. and {Maciejewski}, W. and {M{\"u}ller-S{\'a}nchez}, F. and {Orban de Xivry}, G. and {Ricci}, C. and {Riffel}, R. and {Riffel}, R.~A. and {Rosario}, D. and {Schartmann}, M. and {Schnorr-M{\"u}ller}, A. and {Shangguan}, J. and {Sternberg}, A. and {Sturm}, E. and {Storchi-Bergmann}, T. and {Tacconi}, L. and {Veilleux}, S.},
        title = "{Ionized outflows in local luminous AGN: what are the real densities and outflow rates?}",
      journal = {\mnras},
         year = 2020,
        month = nov,
       volume = {498},
       number = {3},
        pages = {4150-4177},
          doi = {10.1093/mnras/staa2413},
archivePrefix = {arXiv},
       eprint = {2003.06153},
 primaryClass = {astro-ph.GA},
       adsurl = {https://ui.adsabs.harvard.edu/abs/2020MNRAS.498.4150D}
}

@ARTICLE{Planck2021,
       author = {{Planck Collaboration} and {Aghanim}, N. and {Akrami}, Y. and {Ashdown}, M. and {Aumont}, J. and {Baccigalupi}, C. and {Ballardini}, M. and {Banday}, A.~J. and {Barreiro}, R.~B. and {Bartolo}, N. and {Basak}, S. and {Battye}, R. and {Benabed}, K. and {Bernard}, J. -P. and {Bersanelli}, M. and {Bielewicz}, P. and {Bock}, J.~J. and {Bond}, J.~R. and {Borrill}, J. and {Bouchet}, F.~R. and {Boulanger}, F. and {Bucher}, M. and {Burigana}, C. and {Butler}, R.~C. and {Calabrese}, E. and {Cardoso}, J. -F. and {Carron}, J. and {Challinor}, A. and {Chiang}, H.~C. and {Chluba}, J. and {Colombo}, L.~P.~L. and {Combet}, C. and {Contreras}, D. and {Crill}, B.~P. and {Cuttaia}, F. and {de Bernardis}, P. and {de Zotti}, G. and {Delabrouille}, J. and {Delouis}, J. -M. and {Di Valentino}, E. and {Diego}, J.~M. and {Dor{\'e}}, O. and {Douspis}, M. and {Ducout}, A. and {Dupac}, X. and {Dusini}, S. and {Efstathiou}, G. and {Elsner}, F. and {En{\ss}lin}, T.~A. and {Eriksen}, H.~K. and {Fantaye}, Y. and {Farhang}, M. and {Fergusson}, J. and {Fernandez-Cobos}, R. and {Finelli}, F. and {Forastieri}, F. and {Frailis}, M. and {Fraisse}, A.~A. and {Franceschi}, E. and {Frolov}, A. and {Galeotta}, S. and {Galli}, S. and {Ganga}, K. and {G{\'e}nova-Santos}, R.~T. and {Gerbino}, M. and {Ghosh}, T. and {Gonz{\'a}lez-Nuevo}, J. and {G{\'o}rski}, K.~M. and {Gratton}, S. and {Gruppuso}, A. and {Gudmundsson}, J.~E. and {Hamann}, J. and {Handley}, W. and {Hansen}, F.~K. and {Herranz}, D. and {Hildebrandt}, S.~R. and {Hivon}, E. and {Huang}, Z. and {Jaffe}, A.~H. and {Jones}, W.~C. and {Karakci}, A. and {Keih{\"a}nen}, E. and {Keskitalo}, R. and {Kiiveri}, K. and {Kim}, J. and {Kisner}, T.~S. and {Knox}, L. and {Krachmalnicoff}, N. and {Kunz}, M. and {Kurki-Suonio}, H. and {Lagache}, G. and {Lamarre}, J. -M. and {Lasenby}, A. and {Lattanzi}, M. and {Lawrence}, C.~R. and {Le Jeune}, M. and {Lemos}, P. and {Lesgourgues}, J. and {Levrier}, F. and {Lewis}, A. and {Liguori}, M. and {Lilje}, P.~B. and {Lilley}, M. and {Lindholm}, V. and {L{\'o}pez-Caniego}, M. and {Lubin}, P.~M. and {Ma}, Y. -Z. and {Mac{\'\i}as-P{\'e}rez}, J.~F. and {Maggio}, G. and {Maino}, D. and {Mandolesi}, N. and {Mangilli}, A. and {Marcos-Caballero}, A. and {Maris}, M. and {Martin}, P.~G. and {Martinelli}, M. and {Mart{\'\i}nez-Gonz{\'a}lez}, E. and {Matarrese}, S. and {Mauri}, N. and {McEwen}, J.~D. and {Meinhold}, P.~R. and {Melchiorri}, A. and {Mennella}, A. and {Migliaccio}, M. and {Millea}, M. and {Mitra}, S. and {Miville-Desch{\^e}nes}, M. -A. and {Molinari}, D. and {Montier}, L. and {Morgante}, G. and {Moss}, A. and {Natoli}, P. and {N{\o}rgaard-Nielsen}, H.~U. and {Pagano}, L. and {Paoletti}, D. and {Partridge}, B. and {Patanchon}, G. and {Peiris}, H.~V. and {Perrotta}, F. and {Pettorino}, V. and {Piacentini}, F. and {Polastri}, L. and {Polenta}, G. and {Puget}, J. -L. and {Rachen}, J.~P. and {Reinecke}, M. and {Remazeilles}, M. and {Renzi}, A. and {Rocha}, G. and {Rosset}, C. and {Roudier}, G. and {Rubi{\~n}o-Mart{\'\i}n}, J.~A. and {Ruiz-Granados}, B. and {Salvati}, L. and {Sandri}, M. and {Savelainen}, M. and {Scott}, D. and {Shellard}, E.~P.~S. and {Sirignano}, C. and {Sirri}, G. and {Spencer}, L.~D. and {Sunyaev}, R. and {Suur-Uski}, A. -S. and {Tauber}, J.~A. and {Tavagnacco}, D. and {Tenti}, M. and {Toffolatti}, L. and {Tomasi}, M. and {Trombetti}, T. and {Valenziano}, L. and {Valiviita}, J. and {Van Tent}, B. and {Vibert}, L. and {Vielva}, P. and {Villa}, F. and {Vittorio}, N. and {Wandelt}, B.~D. and {Wehus}, I.~K. and {White}, M. and {White}, S.~D.~M. and {Zacchei}, A. and {Zonca}, A.},
        title = "{Planck 2018 results. VI. Cosmological parameters (Corrigendum)}",
      journal = {\aap},
         year = 2021,
        month = aug,
       volume = {652},
          eid = {C4},
        pages = {C4},
          doi = {10.1051/0004-6361/201833910e},
       adsurl = {https://ui.adsabs.harvard.edu/abs/2021A&A...652C...4P}
}

@ARTICLE{Ricci2017BASS,
       author = {{Ricci}, C. and {Trakhtenbrot}, B. and {Koss}, M.~J. and {Ueda}, Y. and {Del Vecchio}, I. and {Treister}, E. and {Schawinski}, K. and {Paltani}, S. and {Oh}, K. and {Lamperti}, I. and {Berney}, S. and {Gandhi}, P. and {Ichikawa}, K. and {Bauer}, F.~E. and {Ho}, L.~C. and {Asmus}, D. and {Beckmann}, V. and {Soldi}, S. and {Balokovi{\'c}}, M. and {Gehrels}, N. and {Markwardt}, C.~B.},
        title = "{BAT AGN Spectroscopic Survey. V. X-Ray Properties of the Swift/BAT 70-month AGN Catalog}",
      journal = {\apjs},
         year = 2017,
        month = dec,
       volume = {233},
       number = {2},
          eid = {17},
        pages = {17},
          doi = {10.3847/1538-4365/aa96ad},
archivePrefix = {arXiv},
       eprint = {1709.03989},
 primaryClass = {astro-ph.HE},
       adsurl = {https://ui.adsabs.harvard.edu/abs/2017ApJS..233...17R}
}

@ARTICLE{Maiolino2008,
       author = {{Maiolino}, R. and {Nagao}, T. and {Grazian}, A. and {Cocchia}, F. and {Marconi}, A. and {Mannucci}, F. and {Cimatti}, A. and {Pipino}, A. and {Ballero}, S. and {Calura}, F. and {Chiappini}, C. and {Fontana}, A. and {Granato}, G.~L. and {Matteucci}, F. and {Pastorini}, G. and {Pentericci}, L. and {Risaliti}, G. and {Salvati}, M. and {Silva}, L.},
        title = "{AMAZE. I. The evolution of the mass-metallicity relation at z > 3}",
      journal = {\aap},
         year = 2008,
        month = sep,
       volume = {488},
       number = {2},
        pages = {463-479},
          doi = {10.1051/0004-6361:200809678},
archivePrefix = {arXiv},
       eprint = {0806.2410},
 primaryClass = {astro-ph},
       adsurl = {https://ui.adsabs.harvard.edu/abs/2008A&A...488..463M}
}

@ARTICLE{Maiolino2019Rev,
       author = {{Maiolino}, R. and {Mannucci}, F.},
        title = "{De re metallica: the cosmic chemical evolution of galaxies}",
      journal = {\aapr},
         year = 2019,
        month = feb,
       volume = {27},
       number = {1},
          eid = {3},
        pages = {3},
          doi = {10.1007/s00159-018-0112-2},
archivePrefix = {arXiv},
       eprint = {1811.09642},
 primaryClass = {astro-ph.GA},
       adsurl = {https://ui.adsabs.harvard.edu/abs/2019A&ARv..27....3M}
}

@ARTICLE{2010A&A...517A..85L,
       author = {{L{\'o}pez-S{\'a}nchez}, {\'A}. R. and {Esteban}, C.},
        title = "{Massive star formation in Wolf-Rayet galaxies. IV. Colours, chemical-composition analysis and metallicity-luminosity relations}",
      journal = {\aap},
         year = 2010,
        month = jul,
       volume = {517},
          eid = {A85},
        pages = {A85},
          doi = {10.1051/0004-6361/201014156},
archivePrefix = {arXiv},
       eprint = {1004.0626},
 primaryClass = {astro-ph.CO},
       adsurl = {https://ui.adsabs.harvard.edu/abs/2010A&A...517A..85L}
}

@ARTICLE{Marino2013,
       author = {{Marino}, R.~A. and {Rosales-Ortega}, F.~F. and {S{\'a}nchez}, S.~F. and {Gil de Paz}, A. and {V{\'\i}lchez}, J. and {Miralles-Caballero}, D. and {Kehrig}, C. and {P{\'e}rez-Montero}, E. and {Stanishev}, V. and {Iglesias-P{\'a}ramo}, J. and {D{\'\i}az}, A.~I. and {Castillo-Morales}, A. and {Kennicutt}, R. and {L{\'o}pez-S{\'a}nchez}, A.~R. and {Galbany}, L. and {Garc{\'\i}a-Benito}, R. and {Mast}, D. and {Mendez-Abreu}, J. and {Monreal-Ibero}, A. and {Husemann}, B. and {Walcher}, C.~J. and {Garc{\'\i}a-Lorenzo}, B. and {Masegosa}, J. and {Del Olmo Orozco}, A. and {Mour{\~a}o}, A.~M. and {Ziegler}, B. and {Moll{\'a}}, M. and {Papaderos}, P. and {S{\'a}nchez-Bl{\'a}zquez}, P. and {Gonz{\'a}lez Delgado}, R.~M. and {Falc{\'o}n-Barroso}, J. and {Roth}, M.~M. and {van de Ven}, G. and {CALIFA Team}},
        title = "{The O3N2 and N2 abundance indicators revisited: improved calibrations based on CALIFA and T$_{e}$-based literature data}",
      journal = {\aap},
         year = 2013,
        month = nov,
       volume = {559},
          eid = {A114},
        pages = {A114},
          doi = {10.1051/0004-6361/201321956},
archivePrefix = {arXiv},
       eprint = {1307.5316},
 primaryClass = {astro-ph.CO},
       adsurl = {https://ui.adsabs.harvard.edu/abs/2013A&A...559A.114M}
}

@ARTICLE{Curti2017,
       author = {{Curti}, M. and {Cresci}, G. and {Mannucci}, F. and {Marconi}, A. and {Maiolino}, R. and {Esposito}, S.},
        title = "{New fully empirical calibrations of strong-line metallicity indicators in star-forming galaxies}",
      journal = {\mnras},
         year = 2017,
        month = feb,
       volume = {465},
       number = {2},
        pages = {1384-1400},
          doi = {10.1093/mnras/stw2766},
archivePrefix = {arXiv},
       eprint = {1610.06939},
 primaryClass = {astro-ph.GA},
       adsurl = {https://ui.adsabs.harvard.edu/abs/2017MNRAS.465.1384C}
}

@ARTICLE{McGaugh1991,
       author = {{McGaugh}, Stacy S.},
        title = "{H II Region Abundances: Model Oxygen Line Ratios}",
      journal = {\apj},
         year = 1991,
        month = oct,
       volume = {380},
        pages = {140},
          doi = {10.1086/170569},
       adsurl = {https://ui.adsabs.harvard.edu/abs/1991ApJ...380..140M}
}

@ARTICLE{Kewley2002,
       author = {{Kewley}, L.~J. and {Dopita}, M.~A.},
        title = "{Using Strong Lines to Estimate Abundances in Extragalactic H II Regions and Starburst Galaxies}",
      journal = {\apjs},
         year = 2002,
        month = sep,
       volume = {142},
       number = {1},
        pages = {35-52},
          doi = {10.1086/341326},
archivePrefix = {arXiv},
       eprint = {astro-ph/0206495},
 primaryClass = {astro-ph},
       adsurl = {https://ui.adsabs.harvard.edu/abs/2002ApJS..142...35K}
}

@ARTICLE{Tremonti2004,
       author = {{Tremonti}, Christy A. and {Heckman}, Timothy M. and {Kauffmann}, Guinevere and {Brinchmann}, Jarle and {Charlot}, St{\'e}phane and {White}, Simon D.~M. and {Seibert}, Mark and {Peng}, Eric W. and {Schlegel}, David J. and {Uomoto}, Alan and {Fukugita}, Masataka and {Brinkmann}, Jon},
        title = "{The Origin of the Mass-Metallicity Relation: Insights from 53,000 Star-forming Galaxies in the Sloan Digital Sky Survey}",
      journal = {\apj},
         year = 2004,
        month = oct,
       volume = {613},
       number = {2},
        pages = {898-913},
          doi = {10.1086/423264},
archivePrefix = {arXiv},
       eprint = {astro-ph/0405537},
 primaryClass = {astro-ph},
       adsurl = {https://ui.adsabs.harvard.edu/abs/2004ApJ...613..898T}
}

@ARTICLE{Denicol2002,
       author = {{Denicol{\'o}}, Glenda and {Terlevich}, Roberto and {Terlevich}, Elena},
        title = "{New light on the search for low-metallicity galaxies - I. The N2 calibrator}",
      journal = {\mnras},
         year = 2002,
        month = feb,
       volume = {330},
       number = {1},
        pages = {69-74},
          doi = {10.1046/j.1365-8711.2002.05041.x},
archivePrefix = {arXiv},
       eprint = {astro-ph/0110356},
 primaryClass = {astro-ph},
       adsurl = {https://ui.adsabs.harvard.edu/abs/2002MNRAS.330...69D}
}

@ARTICLE{1981ApJ...245..357Z,
       author = {{Zamorani}, G. and {Henry}, J.~P. and {Maccacaro}, T. and {Tananbaum}, H. and {Soltan}, A. and {Avni}, Y. and {Liebert}, J. and {Stocke}, J. and {Strittmatter}, P.~A. and {Weymann}, R.~J. and {Smith}, M.~G. and {Condon}, J.~J.},
        title = "{X-ray studies of quasars with the Einstein Observatory II.}",
      journal = {\apj},
         year = 1981,
        month = apr,
       volume = {245},
        pages = {357-374},
          doi = {10.1086/158815},
       adsurl = {https://ui.adsabs.harvard.edu/abs/1981ApJ...245..357Z}
}

@ARTICLE{2024ApJ...960..108Z,
       author = {{Zhang}, XueGuang},
        title = "{Are There Higher Electron Densities in Narrow Emission Line Regions of Type-1 AGNs than in Type-2 AGNs?}",
      journal = {\apj},
         year = 2024,
        month = jan,
       volume = {960},
       number = {2},
          eid = {108},
        pages = {108},
          doi = {10.3847/1538-4357/ad029a},
archivePrefix = {arXiv},
       eprint = {2309.00852},
 primaryClass = {astro-ph.GA},
       adsurl = {https://ui.adsabs.harvard.edu/abs/2024ApJ...960..108Z}
}

@ARTICLE{1992ApJ...387...95F,
       author = {{Ferland}, G.~J. and {Peterson}, B.~M. and {Horne}, K. and {Welsh}, W.~F. and {Nahar}, S.~N.},
        title = "{Anisotropic Line Emission and the Geometry of the Broad-Line Region in Active Galactic Nuclei}",
      journal = {\apj},
         year = 1992,
        month = mar,
       volume = {387},
        pages = {95},
          doi = {10.1086/171063},
       adsurl = {https://ui.adsabs.harvard.edu/abs/1992ApJ...387...95F}
}

@ARTICLE{Dors2021b,
       author = {{Dors}, Oli L.},
        title = "{Chemical abundances in Seyfert galaxies - VI. Empirical abundance calibration}",
      journal = {\mnras},
         year = 2021,
        month = oct,
       volume = {507},
       number = {1},
        pages = {466 (D21)-474},
          doi = {10.1093/mnras/stab2166},
       adsurl = {https://ui.adsabs.harvard.edu/abs/2021MNRAS.507..466D}
}

@ARTICLE{Oliveira2022,
       author = {{Oliveira}, C.~B. and {Krabbe}, A.~C. and {Hernandez-Jimenez}, J.~A. and {Dors}, O.~L. and {Zinchenko}, I.~A. and {H{\"a}gele}, G.~F. and {Cardaci}, M.~V. and {Monteiro}, A.~F.},
        title = "{Chemical abundance of LINER galaxies - metallicity calibrations based on SDSS-IV MaNGA}",
      journal = {\mnras},
         year = 2022,
        month = oct,
       volume = {515},
       number = {4},
        pages = {6093-6108},
          doi = {10.1093/mnras/stac2118},
archivePrefix = {arXiv},
       eprint = {2207.10260},
 primaryClass = {astro-ph.GA},
       adsurl = {https://ui.adsabs.harvard.edu/abs/2022MNRAS.515.6093O}
}

@ARTICLE{Blanc2015,
       author = {{Blanc}, Guillermo A. and {Kewley}, Lisa and {Vogt}, Fr{\'e}d{\'e}ric P.~A. and {Dopita}, Michael A.},
        title = "{IZI: Inferring the Gas Phase Metallicity (Z) and Ionization Parameter (q) of Ionized Nebulae Using Bayesian Statistics}",
      journal = {\apj},
         year = 2015,
        month = jan,
       volume = {798},
       number = {2},
          eid = {99},
        pages = {99},
          doi = {10.1088/0004-637X/798/2/99},
archivePrefix = {arXiv},
       eprint = {1410.8146},
 primaryClass = {astro-ph.GA},
       adsurl = {https://ui.adsabs.harvard.edu/abs/2015ApJ...798...99B}
}

@ARTICLE{ValeAsari2016,
       author = {{Vale Asari}, N. and {Stasi{\'n}ska}, G. and {Morisset}, C. and {Cid Fernandes}, R.},
        title = "{BOND: Bayesian Oxygen and Nitrogen abundance Determinations in giant H II regions using strong and semistrong lines}",
      journal = {\mnras},
         year = 2016,
        month = aug,
       volume = {460},
       number = {2},
        pages = {1739-1757},
          doi = {10.1093/mnras/stw971},
archivePrefix = {arXiv},
       eprint = {1605.01057},
 primaryClass = {astro-ph.GA},
       adsurl = {https://ui.adsabs.harvard.edu/abs/2016MNRAS.460.1739V}
}

@ARTICLE{Bian2017,
       author = {{Bian}, Fuyan and {Kewley}, Lisa J. and {Dopita}, Michael A. and {Blanc}, Guillermo A.},
        title = "{Mass-Metallicity Relation for Local Analogs of High-redshift galaxies: Implications for the Evolution of the Mass-Metallicity Relations}",
      journal = {\apj},
         year = 2017,
        month = jan,
       volume = {834},
       number = {1},
          eid = {51},
        pages = {51},
          doi = {10.3847/1538-4357/834/1/51},
archivePrefix = {arXiv},
       eprint = {1611.08595},
 primaryClass = {astro-ph.GA},
       adsurl = {https://ui.adsabs.harvard.edu/abs/2017ApJ...834...51B}
}

@ARTICLE{Chevallard2016,
       author = {{Chevallard}, Jacopo and {Charlot}, St{\'e}phane},
        title = "{Modelling and interpreting spectral energy distributions of galaxies with BEAGLE}",
      journal = {\mnras},
         year = 2016,
        month = oct,
       volume = {462},
       number = {2},
        pages = {1415-1443},
          doi = {10.1093/mnras/stw1756},
archivePrefix = {arXiv},
       eprint = {1603.03037},
 primaryClass = {astro-ph.GA},
       adsurl = {https://ui.adsabs.harvard.edu/abs/2016MNRAS.462.1415C}
}

@ARTICLE{VidalGarcia2024,
       author = {{Vidal-Garc{\'\i}a}, A. and {Plat}, A. and {Curtis-Lake}, E. and {Feltre}, A. and {Hirschmann}, M. and {Chevallard}, J. and {Charlot}, S.},
        title = "{BEAGLE-AGN I: simultaneous constraints on the properties of gas in star-forming and AGN narrow-line regions in galaxies}",
      journal = {\mnras},
         year = 2024,
        month = jan,
       volume = {527},
       number = {3},
        pages = {7217-7241},
          doi = {10.1093/mnras/stad3252},
archivePrefix = {arXiv},
       eprint = {2211.13648},
 primaryClass = {astro-ph.GA},
       adsurl = {https://ui.adsabs.harvard.edu/abs/2024MNRAS.527.7217V}
}

@ARTICLE{Carnall2019,
       author = {{Carnall}, A.~C. and {McLure}, R.~J. and {Dunlop}, J.~S. and {Cullen}, F. and {McLeod}, D.~J. and {Wild}, V. and {Johnson}, B.~D. and {Appleby}, S. and {Dav{\'e}}, R. and {Amorin}, R. and {Bolzonella}, M. and {Castellano}, M. and {Cimatti}, A. and {Cucciati}, O. and {Gargiulo}, A. and {Garilli}, B. and {Marchi}, F. and {Pentericci}, L. and {Pozzetti}, L. and {Schreiber}, C. and {Talia}, M. and {Zamorani}, G.},
        title = "{The VANDELS survey: the star-formation histories of massive quiescent galaxies at 1.0 < z < 1.3}",
      journal = {\mnras},
         year = 2019,
        month = nov,
       volume = {490},
       number = {1},
        pages = {417-439},
          doi = {10.1093/mnras/stz2544},
archivePrefix = {arXiv},
       eprint = {1903.11082},
 primaryClass = {astro-ph.GA},
       adsurl = {https://ui.adsabs.harvard.edu/abs/2019MNRAS.490..417C}
}

@ARTICLE{Thomas2018,
       author = {{Thomas}, Adam D. and {Dopita}, Michael A. and {Kewley}, Lisa J. and {Groves}, Brent A. and {Sutherland}, Ralph S. and {Hopkins}, Andrew M. and {Blanc}, Guillermo A.},
        title = "{Interrogating Seyferts with NebulaBayes: Spatially Probing the Narrow-line Region Radiation Fields and Chemical Abundances}",
      journal = {\apj},
         year = 2018,
        month = apr,
       volume = {856},
       number = {2},
          eid = {89},
        pages = {89},
          doi = {10.3847/1538-4357/aab3db},
archivePrefix = {arXiv},
       eprint = {1803.00740},
 primaryClass = {astro-ph.GA},
       adsurl = {https://ui.adsabs.harvard.edu/abs/2018ApJ...856...89T}
}

@ARTICLE{PerezMonteiro2019,
       author = {{P{\'e}rez-Montero}, E. and {Dors}, O.~L. and {V{\'\i}lchez}, J.~M. and {Garc{\'\i}a-Benito}, R. and {Cardaci}, M.~V. and {H{\"a}gele}, G.~F.},
        title = "{A bayesian-like approach to derive chemical abundances in type-2 active galactic nuclei based on photoionization models}",
      journal = {\mnras},
         year = 2019,
        month = oct,
       volume = {489},
       number = {2},
        pages = {2652-2668},
          doi = {10.1093/mnras/stz2278},
archivePrefix = {arXiv},
       eprint = {1908.04827},
 primaryClass = {astro-ph.GA},
       adsurl = {https://ui.adsabs.harvard.edu/abs/2019MNRAS.489.2652P}
}

@ARTICLE{Peimbert2017,
       author = {{Peimbert}, Manuel and {Peimbert}, Antonio and {Delgado-Inglada}, Gloria},
        title = "{Nebular Spectroscopy: A Guide on Hii Regions and Planetary Nebulae}",
      journal = {\pasp},
         year = 2017,
        month = aug,
       volume = {129},
       number = {978},
        pages = {082001},
          doi = {10.1088/1538-3873/aa72c3},
archivePrefix = {arXiv},
       eprint = {1705.06323},
 primaryClass = {astro-ph.GA},
       adsurl = {https://ui.adsabs.harvard.edu/abs/2017PASP..129h2001P}
}

@ARTICLE{Matsuoka2011,
       author = {{Matsuoka}, K. and {Nagao}, T. and {Marconi}, A. and {Maiolino}, R. and {Taniguchi}, Y.},
        title = "{The mass-metallicity relation of SDSS quasars}",
      journal = {\aap},
         year = 2011,
        month = mar,
       volume = {527},
          eid = {A100},
        pages = {A100},
          doi = {10.1051/0004-6361/201015584},
archivePrefix = {arXiv},
       eprint = {1011.5811},
 primaryClass = {astro-ph.CO},
       adsurl = {https://ui.adsabs.harvard.edu/abs/2011A&A...527A.100M}
}

@ARTICLE{Oh2017,
       author = {{Oh}, Kyuseok and {Schawinski}, Kevin and {Koss}, Michael and {Trakhtenbrot}, Benny and {Lamperti}, Isabella and {Ricci}, Claudio and {Mushotzky}, Richard and {Veilleux}, Sylvain and {Berney}, Simon and {Crenshaw}, D. Michael and {Gehrels}, Neil and {Harrison}, Fiona and {Masetti}, Nicola and {Soto}, Kurt T. and {Stern}, Daniel and {Treister}, Ezequiel and {Ueda}, Yoshihiro},
        title = "{BAT AGN Spectroscopic Survey - III. An observed link between AGN Eddington ratio and narrow-emission-line ratios}",
      journal = {\mnras},
         year = 2017,
        month = jan,
       volume = {464},
       number = {2},
        pages = {1466-1473},
          doi = {10.1093/mnras/stw2467},
archivePrefix = {arXiv},
       eprint = {1609.08625},
 primaryClass = {astro-ph.GA},
       adsurl = {https://ui.adsabs.harvard.edu/abs/2017MNRAS.464.1466O}
}

@ARTICLE{Groves2006a,
       author = {{Groves}, Brent A. and {Heckman}, Timothy M. and {Kauffmann}, Guinevere},
        title = "{Emission-line diagnostics of low-metallicity active galactic nuclei}",
      journal = {\mnras},
         year = 2006,
        month = oct,
       volume = {371},
       number = {4},
        pages = {1559-1569},
          doi = {10.1111/j.1365-2966.2006.10812.x},
archivePrefix = {arXiv},
       eprint = {astro-ph/0607311},
 primaryClass = {astro-ph},
       adsurl = {https://ui.adsabs.harvard.edu/abs/2006MNRAS.371.1559G}
}

@book{Osterbrock2006,
       author = {{Osterbrock}, Donald E. and {Ferland}, Gary J.},
        title     = {{Astrophysics of Gaseous Nebulae and Active Galactic Nuclei}},
  edition   = {2nd},
  publisher = {University Science Books},
  address   = {Sausalito, California, USA},
         year = 2006,
       adsurl = {https://ui.adsabs.harvard.edu/abs/2006agna.book.....O}
}

@ARTICLE{2014A&A...561A..10P,
       author = {{Proxauf}, B. and {{\"O}ttl}, S. and {Kimeswenger}, S.},
        title = "{Upgrading electron temperature and electron density diagnostic diagrams of forbidden line emission}",
      journal = {\aap},
         year = 2014,
        month = jan,
       volume = {561},
          eid = {A10},
        pages = {A10},
          doi = {10.1051/0004-6361/201322581},
archivePrefix = {arXiv},
       eprint = {1311.5041},
 primaryClass = {astro-ph.IM},
       adsurl = {https://ui.adsabs.harvard.edu/abs/2014A&A...561A..10P}
}

@ARTICLE{Sanders2016,
       author = {{Sanders}, Ryan L. and {Shapley}, Alice E. and {Kriek}, Mariska and {Reddy}, Naveen A. and {Freeman}, William R. and {Coil}, Alison L. and {Siana}, Brian and {Mobasher}, Bahram and {Shivaei}, Irene and {Price}, Sedona H. and {de Groot}, Laura},
        title = "{The MOSDEF Survey: Electron Density and Ionization Parameter at z \raisebox{-0.5ex}\textasciitilde 2.3}",
      journal = {\apj},
         year = 2016,
        month = jan,
       volume = {816},
       number = {1},
          eid = {23},
        pages = {23},
          doi = {10.3847/0004-637X/816/1/23},
archivePrefix = {arXiv},
       eprint = {1509.03636},
 primaryClass = {astro-ph.GA},
       adsurl = {https://ui.adsabs.harvard.edu/abs/2016ApJ...816...23S}
}

@article{Keenan1997,
  title={Nebular and Auroral emission lines of [Ar iv] in the optical spectra of planetary nebulae},
  author={Keenan, FP and McKenna, FC and Bell, KL and Ramsbottom, CA and Wickstead, AW and Aller, LH and Hyung, S},
  journal={The Astrophysical Journal},
  volume={487},
  number={1},
  pages={457},
  year={1997},
  publisher={IOP Publishing}
}

@ARTICLE{Matsuoka2009,
       author = {{Matsuoka}, K. and {Nagao}, T. and {Maiolino}, R. and {Marconi}, A. and {Taniguchi}, Y.},
        title = "{Chemical evolution of high-redshift radio galaxies}",
      journal = {\aap},
         year = 2009,
        month = sep,
       volume = {503},
       number = {3},
        pages = {721-730},
          doi = {10.1051/0004-6361/200811478},
archivePrefix = {arXiv},
       eprint = {0905.1581},
 primaryClass = {astro-ph.CO},
       adsurl = {https://ui.adsabs.harvard.edu/abs/2009A&A...503..721M}
}

@ARTICLE{Vaona2012,
       author = {{Vaona}, L. and {Ciroi}, S. and {Di Mille}, F. and {Cracco}, V. and {La Mura}, G. and {Rafanelli}, P.},
        title = "{Spectral properties of the narrow-line region in Seyfert galaxies selected from the SDSS-DR7}",
      journal = {\mnras},
         year = 2012,
        month = dec,
       volume = {427},
       number = {2},
        pages = {1266-1283},
          doi = {10.1111/j.1365-2966.2012.22060.x},
archivePrefix = {arXiv},
       eprint = {1210.5201},
 primaryClass = {astro-ph.CO},
       adsurl = {https://ui.adsabs.harvard.edu/abs/2012MNRAS.427.1266V}
}

@ARTICLE{Feltre2016,
       author = {{Feltre}, A. and {Charlot}, S. and {Gutkin}, J.},
        title = "{Nuclear activity versus star formation: emission-line diagnostics at ultraviolet and optical wavelengths}",
      journal = {\mnras},
         year = 2016,
        month = mar,
       volume = {456},
       number = {3},
        pages = {3354-3374},
          doi = {10.1093/mnras/stv2794},
archivePrefix = {arXiv},
       eprint = {1511.08217},
 primaryClass = {astro-ph.GA},
       adsurl = {https://ui.adsabs.harvard.edu/abs/2016MNRAS.456.3354F}
}

@ARTICLE{Armah2023,
       author = {{Armah}, Mark and {Riffel}, Rog{\'e}rio and {Dors}, O.~L. and {Oh}, Kyuseok and {Koss}, Michael J. and {Ricci}, Claudio and {Trakhtenbrot}, Benny and {Valerdi}, Mabel and {Riffel}, Rogemar A. and {Krabbe}, Angela C.},
        title = "{Oxygen abundances in the narrow line regions of Seyfert galaxies and the metallicity-luminosity relation}",
      journal = {\mnras},
         year = 2023,
        month = apr,
       volume = {520},
       number = {2},
        pages = {1687-1703},
          doi = {10.1093/mnras/stad217},
archivePrefix = {arXiv},
       eprint = {2301.07596},
 primaryClass = {astro-ph.GA},
       adsurl = {https://ui.adsabs.harvard.edu/abs/2023MNRAS.520.1687A}
}

@ARTICLE{Armah2024,
       author = {{Armah}, Mark and {Riffel}, Rog{\'e}rio and {Dahmer-Hahn}, L.~G. and {Davies}, R.~I. and {Dors}, O.~L. and {Kakkad}, Darshan and {Riffel}, Rogemar A. and {Rodr{\'\i}guez-Ardila}, A. and {Ruschel-Dutra}, D. and {Storchi-Bergmann}, T.},
        title = "{Spatially resolved gas-phase metallicity in Seyfert galaxies}",
      journal = {\mnras},
         year = 2024,
        month = nov,
       volume = {534},
       number = {3},
        pages = {2723-2757},
          doi = {10.1093/mnras/stae2263},
archivePrefix = {arXiv},
       eprint = {2409.20465},
 primaryClass = {astro-ph.GA},
       adsurl = {https://ui.adsabs.harvard.edu/abs/2024MNRAS.534.2723A}
}

@ARTICLE{Ellison2008,
       author = {{Ellison}, Sara L. and {Patton}, David R. and {Simard}, Luc and {McConnachie}, Alan W.},
        title = "{Galaxy Pairs in the Sloan Digital Sky Survey. I. Star Formation, Active Galactic Nucleus Fraction, and the Mass-Metallicity Relation}",
      journal = {\aj},
         year = 2008,
        month = may,
       volume = {135},
       number = {5},
        pages = {1877-1899},
          doi = {10.1088/0004-6256/135/5/1877},
archivePrefix = {arXiv},
       eprint = {0803.0161},
 primaryClass = {astro-ph},
       adsurl = {https://ui.adsabs.harvard.edu/abs/2008AJ....135.1877E}
}

@ARTICLE{Hagele2008,
       author = {{H{\"a}gele}, Guillermo F. and {D{\'\i}az}, {\'A}ngeles I. and {Terlevich}, Elena and {Terlevich}, Roberto and {P{\'e}rez-Montero}, Enrique and {Cardaci}, M{\'o}nica V.},
        title = "{Precision abundance analysis of bright HII galaxies}",
      journal = {\mnras},
         year = 2008,
        month = jan,
       volume = {383},
       number = {1},
        pages = {209-229},
          doi = {10.1111/j.1365-2966.2007.12527.x},
archivePrefix = {arXiv},
       eprint = {0710.1828},
 primaryClass = {astro-ph},
       adsurl = {https://ui.adsabs.harvard.edu/abs/2008MNRAS.383..209H}
}

@ARTICLE{Peimbert1969,
       author = {{Peimbert}, M. and {Costero}, R.},
        title = "{Chemical Abundances in Galactic HII Regions}",
      journal = {Boletin de los Observatorios Tonantzintla y Tacubaya},
         year = 1969,
        month = may,
       volume = {5},
        pages = {3-22},
       adsurl = {https://ui.adsabs.harvard.edu/abs/1969BOTT....5....3P}
}

@ARTICLE{PerezMontero2017,
       author = {{P{\'e}rez-Montero}, Enrique},
        title = "{Ionized Gaseous Nebulae Abundance Determination from the Direct Method}",
      journal = {\pasp},
         year = 2017,
        month = apr,
       volume = {129},
       number = {974},
        pages = {043001},
          doi = {10.1088/1538-3873/aa5abb},
archivePrefix = {arXiv},
       eprint = {1702.04255},
 primaryClass = {astro-ph.GA},
       adsurl = {https://ui.adsabs.harvard.edu/abs/2017PASP..129d3001P}
}

@ARTICLE{Pettini2004,
       author = {{Pettini}, Max and {Pagel}, Bernard E.~J.},
        title = "{[OIII]/[NII] as an abundance indicator at high redshift}",
      journal = {\mnras},
         year = 2004,
        month = mar,
       volume = {348},
       number = {3},
        pages = {L59-L63},
          doi = {10.1111/j.1365-2966.2004.07591.x},
archivePrefix = {arXiv},
       eprint = {astro-ph/0401128},
 primaryClass = {astro-ph},
       adsurl = {https://ui.adsabs.harvard.edu/abs/2004MNRAS.348L..59P}
}

@ARTICLE{GarciaRojas2007,
       author = {{Garc{\'\i}a-Rojas}, Jorge and {Esteban}, C{\'e}sar},
        title = "{On the Abundance Discrepancy Problem in H II Regions}",
      journal = {\apj},
         year = 2007,
        month = nov,
       volume = {670},
       number = {1},
        pages = {457-470},
          doi = {10.1086/521871},
archivePrefix = {arXiv},
       eprint = {0707.3518},
 primaryClass = {astro-ph},
       adsurl = {https://ui.adsabs.harvard.edu/abs/2007ApJ...670..457G}
}

@ARTICLE{Riffel2017,
       author = {{Riffel}, Rogemar A. and {Storchi-Bergmann}, Thaisa and {Riffel}, Rogerio and {Dahmer-Hahn}, Luis G. and {Diniz}, Marlon R. and {Sch{\"o}nell}, Astor J. and {Dametto}, Natacha Z.},
        title = "{Gemini NIFS survey of feeding and feedback processes in nearby active galaxies - I. Stellar kinematics}",
      journal = {\mnras},
         year = 2017,
        month = sep,
       volume = {470},
       number = {1},
        pages = {992-1016},
          doi = {10.1093/mnras/stx1308},
archivePrefix = {arXiv},
       eprint = {1705.06949},
 primaryClass = {astro-ph.GA},
       adsurl = {https://ui.adsabs.harvard.edu/abs/2017MNRAS.470..992R}
}

@ARTICLE{Riffel2021,
       author = {{Riffel}, Rogemar A. and {Bianchin}, Marina and {Riffel}, Rog{\'e}rio and {Storchi-Bergmann}, Thaisa and {Sch{\"o}nell}, Astor J. and {Dahmer-Hahn}, Luis Gabriel and {Dametto}, Natacha Z. and {Diniz}, Marlon R.},
        title = "{Gemini NIFS survey of feeding and feedback in nearby active galaxies - IV. Excitation}",
      journal = {\mnras},
         year = 2021,
        month = may,
       volume = {503},
       number = {4},
        pages = {5161-5178},
          doi = {10.1093/mnras/stab788},
archivePrefix = {arXiv},
       eprint = {2103.08736},
 primaryClass = {astro-ph.GA},
       adsurl = {https://ui.adsabs.harvard.edu/abs/2021MNRAS.503.5161R}
}

@ARTICLE{Audibert2017,
       author = {{Audibert}, Anelise and {Riffel}, Rog{\'e}rio and {Sales}, Dinalva A. and {Pastoriza}, Miriani G. and {Ruschel-Dutra}, Daniel},
        title = "{Probing the active galactic nucleus unified model torus properties in Seyfert galaxies}",
      journal = {\mnras},
         year = 2017,
        month = jan,
       volume = {464},
       number = {2},
        pages = {2139-2173},
          doi = {10.1093/mnras/stw2477},
archivePrefix = {arXiv},
       eprint = {1609.08972},
 primaryClass = {astro-ph.GA},
       adsurl = {https://ui.adsabs.harvard.edu/abs/2017MNRAS.464.2139A}
}

@ARTICLE{Shimizu2019,
       author = {{Shimizu}, T. Taro and {Davies}, R.~I. and {Lutz}, D. and {Burtscher}, L. and {Lin}, M. and {Baron}, D. and {Davies}, R.~L. and {Genzel}, R. and {Hicks}, E.~K.~S. and {Koss}, M. and {Maciejewski}, W. and {M{\"u}ller-S{\'a}nchez}, F. and {Orban de Xivry}, G. and {Price}, S.~H. and {Ricci}, C. and {Riffel}, R. and {Riffel}, R.~A. and {Rosario}, D. and {Schartmann}, M. and {Schnorr-M{\"u}ller}, A. and {Sternberg}, A. and {Sturm}, E. and {Storchi-Bergmann}, T. and {Tacconi}, L. and {Veilleux}, S.},
        title = "{The multiphase gas structure and kinematics in the circumnuclear region of NGC 5728}",
      journal = {\mnras},
         year = 2019,
        month = dec,
       volume = {490},
       number = {4},
        pages = {5860-5887},
          doi = {10.1093/mnras/stz2802},
archivePrefix = {arXiv},
       eprint = {1907.03801},
 primaryClass = {astro-ph.GA},
       adsurl = {https://ui.adsabs.harvard.edu/abs/2019MNRAS.490.5860S}
}

@ARTICLE{GRAVITY2024,
       author = {{Gravity Collaboration} and {Amorim}, A. and {Bourdarot}, G. and {Brandner}, W. and {Cao}, Y. and {Cl{\'e}net}, Y. and {Davies}, R. and {de Zeeuw}, P.~T. and {Dexter}, J. and {Drescher}, A. and {Eckart}, A. and {Eisenhauer}, F. and {Fabricius}, M. and {Feuchtgruber}, H. and {F{\"o}rster Schreiber}, N.~M. and {Garcia}, P.~J.~V. and {Genzel}, R. and {Gillessen}, S. and {Gratadour}, D. and {H{\"o}nig}, S. and {Kishimoto}, M. and {Lacour}, S. and {Lutz}, D. and {Millour}, F. and {Netzer}, H. and {Ott}, T. and {Perraut}, K. and {Perrin}, G. and {Peterson}, B.~M. and {Petrucci}, P.~O. and {Pfuhl}, O. and {Prieto}, A. and {Rabien}, S. and {Rouan}, D. and {Santos}, D.~J.~D. and {Shangguan}, J. and {Shimizu}, T. and {Sternberg}, A. and {Straubmeier}, C. and {Sturm}, E. and {Tacconi}, L.~J. and {Tristram}, K.~R.~W. and {Widmann}, F. and {Woillez}, J.},
        title = "{VLTI/GRAVITY interferometric measurements of the innermost dust structure sizes around active galactic nuclei}",
      journal = {\aap},
         year = 2024,
        month = oct,
       volume = {690},
          eid = {A76},
        pages = {A76},
          doi = {10.1051/0004-6361/202450746},
archivePrefix = {arXiv},
       eprint = {2407.13458},
 primaryClass = {astro-ph.GA},
       adsurl = {https://ui.adsabs.harvard.edu/abs/2024A&A...690A..76G}
}

@ARTICLE{2019MNRAS.488.5185N,
       author = {{Netzer}, Hagai},
        title = "{Bolometric correction factors for active galactic nuclei}",
      journal = {\mnras},
         year = 2019,
        month = oct,
       volume = {488},
       number = {4},
        pages = {5185-5191},
          doi = {10.1093/mnras/stz2016},
archivePrefix = {arXiv},
       eprint = {1907.09534},
 primaryClass = {astro-ph.GA},
       adsurl = {https://ui.adsabs.harvard.edu/abs/2019MNRAS.488.5185N}
}

@ARTICLE{2008ApJ...685..160N,
       author = {{Nenkova}, Maia and {Sirocky}, Matthew M. and {Nikutta}, Robert and {Ivezi{\'c}}, {\v{Z}}eljko and {Elitzur}, Moshe},
        title = "{AGN Dusty Tori. II. Observational Implications of Clumpiness}",
      journal = {\apj},
         year = 2008,
        month = sep,
       volume = {685},
       number = {1},
        pages = {160-180},
          doi = {10.1086/590483},
archivePrefix = {arXiv},
       eprint = {0806.0512},
 primaryClass = {astro-ph},
       adsurl = {https://ui.adsabs.harvard.edu/abs/2008ApJ...685..160N}
}

@ARTICLE{2020A&A...636A..73D,
       author = {{Duras}, F. and {Bongiorno}, A. and {Ricci}, F. and {Piconcelli}, E. and {Shankar}, F. and {Lusso}, E. and {Bianchi}, S. and {Fiore}, F. and {Maiolino}, R. and {Marconi}, A. and {Onori}, F. and {Sani}, E. and {Schneider}, R. and {Vignali}, C. and {La Franca}, F.},
        title = "{Universal bolometric corrections for active galactic nuclei over seven luminosity decades}",
      journal = {\aap},
         year = 2020,
        month = apr,
       volume = {636},
          eid = {A73},
        pages = {A73},
          doi = {10.1051/0004-6361/201936817},
archivePrefix = {arXiv},
       eprint = {2001.09984},
 primaryClass = {astro-ph.GA},
       adsurl = {https://ui.adsabs.harvard.edu/abs/2020A&A...636A..73D}
}

@ARTICLE{RuschelDutra2021,
       author = {{Ruschel-Dutra}, D. and {Storchi-Bergmann}, T. and {Schnorr-M{\"u}ller}, A. and {Riffel}, R.~A. and {Dall'Agnol de Oliveira}, B. and {Lena}, D. and {Robinson}, A. and {Nagar}, N. and {Elvis}, M.},
        title = "{AGNIFS survey of local AGN: GMOS-IFU data and outflows in 30 sources}",
      journal = {\mnras},
         year = 2021,
        month = oct,
       volume = {507},
       number = {1},
        pages = {74-89},
          doi = {10.1093/mnras/stab2058},
archivePrefix = {arXiv},
       eprint = {2107.07635},
 primaryClass = {astro-ph.GA},
       adsurl = {https://ui.adsabs.harvard.edu/abs/2021MNRAS.507...74R}
}

@ARTICLE{Kakkad2022,
       author = {{Kakkad}, D. and {Sani}, E. and {Rojas}, A.~F. and {Mallmann}, Nicolas D. and {Veilleux}, S. and {Bauer}, Franz E. and {Ricci}, F. and {Mushotzky}, R. and {Koss}, M. and {Ricci}, C. and {Treister}, E. and {Privon}, George C. and {Nguyen}, N. and {B{\"a}r}, R. and {Harrison}, F. and {Oh}, K. and {Powell}, M. and {Riffel}, R. and {Stern}, D. and {Trakhtenbrot}, B. and {Urry}, C.~M.},
        title = "{BASS XXXI: Outflow scaling relations in low redshift X-ray AGN host galaxies with MUSE}",
      journal = {\mnras},
         year = 2022,
        month = apr,
       volume = {511},
       number = {2},
        pages = {2105-2124},
          doi = {10.1093/mnras/stac103},
archivePrefix = {arXiv},
       eprint = {2201.04149},
 primaryClass = {astro-ph.GA},
       adsurl = {https://ui.adsabs.harvard.edu/abs/2022MNRAS.511.2105K}
}

@ARTICLE{Riffel2023,
       author = {{Riffel}, R.~A. and {Storchi-Bergmann}, T. and {Riffel}, R. and {Bianchin}, M. and {Zakamska}, N.~L. and {Ruschel-Dutra}, D. and {Bentz}, M.~C. and {Burtscher}, L. and {Crenshaw}, D.~M. and {Dahmer-Hahn}, L.~G. and {Dametto}, N.~Z. and {Davies}, R.~I. and {Diniz}, M.~R. and {Fischer}, T.~C. and {Harrison}, C.~M. and {Mainieri}, V. and {Revalski}, M. and {Rodriguez-Ardila}, A. and {Rosario}, D.~J. and {Sch{\"o}nell}, A.~J.},
        title = "{The AGNIFS survey: spatially resolved observations of hot molecular and ionized outflows in nearby active galaxies}",
      journal = {\mnras},
         year = 2023,
        month = may,
       volume = {521},
       number = {2},
        pages = {1832-1848},
          doi = {10.1093/mnras/stad599},
archivePrefix = {arXiv},
       eprint = {2302.11324},
 primaryClass = {astro-ph.GA},
       adsurl = {https://ui.adsabs.harvard.edu/abs/2023MNRAS.521.1832R}
}

@ARTICLE{2022MNRAS.514.5506D,
       author = {{Dors}, O.~L. and {Valerdi}, M. and {Freitas-Lemes}, P. and {Krabbe}, A.~C. and {Riffel}, R.~A. and {Am{\^o}res}, E.~B. and {Riffel}, R. and {Armah}, M. and {Monteiro}, A.~F. and {Oliveira}, C.~B.},
        title = "{Chemical abundances in Seyfert galaxies - IX. Helium abundance estimates}",
      journal = {\mnras},
         year = 2022,
        month = aug,
       volume = {514},
       number = {4},
        pages = {5506-5527},
          doi = {10.1093/mnras/stac1722},
archivePrefix = {arXiv},
       eprint = {2206.09836},
 primaryClass = {astro-ph.GA},
       adsurl = {https://ui.adsabs.harvard.edu/abs/2022MNRAS.514.5506D}
}

@ARTICLE{Storey1995,
       author = {{Storey}, P.~J. and {Hummer}, D.~G.},
        title = "{Recombination line intensities for hydrogenic ions-IV. Total recombination coefficients and machine-readable tables for Z=1 to 8}",
      journal = {\mnras},
         year = 1995,
        month = jan,
       volume = {272},
       number = {1},
        pages = {41-48},
          doi = {10.1093/mnras/272.1.41},
       adsurl = {https://ui.adsabs.harvard.edu/abs/1995MNRAS.272...41S}
}

\bsp
\label{lastpage}

\end{document}